\documentclass[a4paper,fleqn,usenatbib]{mnras}

\usepackage{newtxtext,newtxmath}

\usepackage[T1]{fontenc}
\usepackage{ae,aecompl}

\usepackage{graphicx}	
\usepackage{amsmath}	
\usepackage{amssymb}	
\usepackage{CJKutf8}	
\usepackage{enumitem}   

\newcommand{\hst}{{\it HST\/}}       
       
\newcommand{\chandra}{{\it Chandra\/}}

\newcommand{\xmm}{\hbox{\it XMM-Newton\/}}

\newcommand{\athena}{{\it Athena\/}}

\newcommand{\gaia}{{\it Gaia\/}}
\newcommand{\ep}{{\it Einstein Probe\/}}

\newcommand{\xray}{\hbox{X-ray}}  
\newcommand{\cdfs}{\hbox{CDF-S}}
\newcommand{\cdfn}{\hbox{CDF-N}}

\title[Fast extragalactic X-ray transients]{Searching for fast extragalactic X-ray transients in {\textit{Chandra}} surveys}

\author[G. Yang et al.]{
G. Yang\begin{CJK*}{UTF8}{gbsn} (杨光),\end{CJK*}$^{1,2}$\thanks{E-mail: gyang206265@gmail.com (GY)}
W. N. Brandt,$^{1,2,3}$
S. F. Zhu\begin{CJK*}{UTF8}{gbsn} (朱世甫),\end{CJK*}$^{1,2}$
F. E. Bauer,$^{4,5,6}$
\newauthor
B. Luo\begin{CJK*}{UTF8}{gbsn} (罗斌),\end{CJK*}$^{7}$
Y. Q. Xue\begin{CJK*}{UTF8}{gbsn} (薛永泉),\end{CJK*}$^{8,9}$
and
X. C. Zheng\begin{CJK*}{UTF8}{gbsn} (郑学琛)\end{CJK*}$^{10}$
\\
$^{1}$Department of Astronomy and Astrophysics, 525 Davey Lab, The Pennsylvania State University, University Park, PA 16802, USA\\
$^{2}$Institute for Gravitation and the Cosmos, The Pennsylvania State University, University Park, PA 16802, USA\\
$^{3}$Department of Physics, 104 Davey Laboratory, The Pennsylvania State University, University Park, PA 16802, USA\\
$^{4}$Instituto de Astrof{\'{\i}}sica and Centro de Astroingenier{\'{\i}}a, Facultad de F{\'{i}}sica, Pontificia Universidad Cat{\'{o}}lica de Chile, Casilla 306, Santiago, Chile\\
$^{5}$Millennium Institute of Astrophysics (MAS), Nuncio Monse{\~{n}}or S{\'{o}}tero Sanz 100, Providencia, Santiago, Chile\\
$^{6}$Space Science Institute, 4750 Walnut Street, Suite 205, Boulder, Colorado 80301, USA\\
$^{7}$School of Astronomy \& Space Science, Nanjing University, Nanjing 210093, China\\
$^{8}$CAS Key Laboratory for Research in Galaxies and Cosmology, Department of Astronomy, 
University of Science and Technology of China, Hefei 230026, China\\
$^{9}$School of Astronomy and Space Science, University of Science and Technology of China,
Hefei 230026, China\\
$^{10}$Leiden Observatory, Leiden University, PO Box 9513, NL-2300 RA Leiden, the Netherlands
}

\date{Accepted XXX. Received YYY; in original form ZZZ}

\pubyear{2018}

\begin{document}
\label{firstpage}
\pagerange{\pageref{firstpage}--\pageref{lastpage}}
\maketitle

\begin{abstract}
Recent works have discovered two fast ($\approx 10$~ks) extragalactic \xray\ transients 
in the \chandra\ Deep Field-South (\cdfs\ XT1 and XT2).
These findings suggest that a large population of {similar} 
extragalactic transients might exist in archival \xray\ observations.
We develop a method that can effectively detect {such} transients 
in a single \chandra\ exposure, and systematically apply it to \chandra\ 
surveys of \cdfs, \cdfn, DEEP2, UDS, COSMOS, and \hbox{E-CDF-S}, totaling 19~Ms of exposure.
We find {13 transient candidates}, including \cdfs\ XT1 and XT2.
With the aid of available excellent multiwavelength observations, we identify the 
physical nature of all these candidates.
Aside from \cdfs\ XT1 and XT2, the other {11} sources are all stellar objects, 
{and all of them have ${z}$-band magnitudes brighter than 20.}
We estimate an event rate of ${59^{+77}_{-38}\ \rm{evt\ yr^{-1} deg^{-2}}}$ for 
{\cdfs\ XT-like} transients with \hbox{0.5--7 keV} peak fluxes 
${\log F_{\rm peak} \gtrsim -12.6}$ (erg~cm$^{-2}$~s$^{-1}$).
This event rate translates to ${\approx 15^{+20}_{-10}}$ transients existing among
\chandra\ archival observations at Galactic latitudes $|b|>20^{\circ}$, which 
can be probed in future work.
Future missions such as \athena\ and the \ep\ with large 
{grasps (effective area $\times$ field of view)} are needed 
to discover a large sample ($\sim$~thousands) of fast extragalactic \xray\ transients.
\end{abstract}




\section{Introduction}\label{sec:intro}
X-ray observations can provide uniquely insightful views of many astronomical 
phenomena such as accretion and mergers of compact objects 
(e.g. \citealt{brandt15}; \citealt{pooley18}).
The \xray\ sky is variable.
Main-sequence stars (especially dwarfs) have strong flares powered 
by magnetic reconnection, generally lasting minutes to hours 
\citep[e.g.][]{haisch91, gudel09}.
\xray\ binaries have various variability behaviors such as pulsations,
bursts, and quasi-periodic oscillations \citep[e.g.][]{van_der_klis89, belloni14}.
Active galactic nuclei (AGNs) typically have red-noise {\xray} variability,   
with characteristic amplitudes being $\lesssim 0.5$~dex on timescales from
$\sim$~an hour to $\sim$~10 years 
\citep[e.g.][]{markowitz03b, markowitz03, yang16, paolillo17, zheng17}.
{However, some relatively rare AGN and related phenomena, 
e.g. tidal disruption events, changing-look AGNs, and narrow-line Seyfert~1s, 
can have larger {\xray} variability 
amplitudes (e.g. \hbox{\citealt{komossa15}}; {\hbox{\citealt{kara16}}}; 
\hbox{\citealt{ricci16}}; \hbox{\citealt{gallo18}}).}

Recently, a new type of \xray\ variability phenomenon has been revealed in 
the form of two relatively faint \xray\ transients found in the \chandra\ 
observations of the \chandra\ Deep Field-South (\cdfs\ XT1 and XT2; 
\hbox{\citealt{bauer17}}; \hbox{\citealt{xue19}}).
Both transients are fast ($T_{90} \approx 10$~ks,\footnote{$T_{90}$ 
is defined as the time interval between the arrival times of the \hbox{5\%-th} 
photon and the \hbox{95\%-th} photon.} observed-frame).
Their origins are found to be extragalactic, with optical/near-infrared (NIR) 
counterparts at $z\approx 2.1$ (photometric redshift) and $z=0.74$ 
(spectroscopic redshift), respectively.
Both transients have $\gtrsim 100$~counts detected, corresponding to enormous 
amounts of energy release ($\gtrsim 10^{48}$~erg, assuming isotropic emission).
Due to the lack of simultaneous multiwavelength observations and the small 
sample size of transients, the physical origins are not well determined with
some possibilities being off-axis gamma-ray bursts, tidal-disruption events, 
mergers of neutron stars, and shock-breakout events.
In this paper, we regard \cdfs\ XT1 and XT2 as the same ``type'' of 
transients considering their observational similarities in flux, timespan,
and extragalactic origin, although their physical causes might be different. 

Given the short timescales ($T_{90}\approx 10$~ks) and large numbers of 
counts ($\gtrsim 100$) for \cdfs\ XT1 and XT2, such transients should be 
easy to detect in any $\gtrsim 10$~ks \chandra\ exposure. 
The two transients are both detected in a small survey area 
($\approx 480$~arcmin$^2$) and relatively short timespan 
(2014 October and 2015 March),
indicating that a large population of \xray\ transients might exist.
\cite{bauer17} performed a preliminary transient search in the \chandra\ source 
catalog (CSC; \citealt{evans10}), which compiled \chandra\ observations 
before 2010 August 10.  
They did not find transients similar to the \cdfs\ transients.
However, this CSC search is not conclusive, because the CSC is not dedicated 
to discovering fast transients and thus potential transients might be missed 
or poorly/incorrectly characterized.
Also, many CSC sources have only a single short \chandra\ visit, making 
it difficult to ascertain the transient and quiescent levels.
The CSC sources also generally lack deep optical/NIR observations, preventing
further studies of the physical nature of potential transients. 

To mitigate the above issues, in this work, we search for similar 
transients in \chandra\ archival observations of \xray\ surveys.
We develop a method to identify {\cdfs\ XT-like} transients in a single 
\chandra\ exposure, which is applicable to any \chandra\ imaging observation. 
In the surveys, most \xray\ sources have been visited by two or more 
\chandra\ exposures, allowing us to inspect transients with multi-epoch 
\xray\ data and study their quiescent behaviors.
Deep multiwavelength data are critical in clarifying the physical origins 
of \xray\ transients.
\cdfs\ XT1 and XT2 have optical/NIR counterparts with $V\gtrsim 25$~mag
and $H\gtrsim 24$~mag (\hbox{\citealt{bauer17}}; \hbox{\citealt{xue19}}), well 
beyond the detection limit of wide-field surveys such as SDSS \citep{york00} 
and UKIDSS \citep{lawrence07}.
\cite{glennie15} discovered an \xray\ transient in one \chandra\ archival 
observation, but were not able to clarify its physical origin due to the 
lack of deep multiwavelength data.
Our selected \xray\ surveys are accompanied by deep multiwavelength 
observations, allowing identifications of optical/NIR counterparts for 
the selected transients.

{The main aim of this paper is to search for fast extragalactic \xray\ 
transients that are similar to \cdfs\ XT1 and XT2 rather than general 
\xray\ transients (although our search is effective for a fairly wide
range of transients; see Appendix~\ref{sec:oth_lc}).
The structure of this paper is organized as follows.
We detail our \xray\ transient-selection algorithm and assess its 
efficiency with simulations in \S\ref{sec:method}.
We describe our \xray\ data, selection of transient candidates, and 
optical/NIR counterparts in \S\ref{sec:analyses}.
We estimate the event rate of \cdfs\ XT-like transients based on our 
results and discuss the prospect of future missions in \S\ref{sec:evtR}.
We summarize our results in \S\ref{sec:sum}.}

Throughout this paper, we assume a cosmology with $H_0=70$~km~s$^{-1}$~Mpc$^{-1}$, 
$\Omega_M=0.3$, and $\Omega_{\Lambda}=0.7$.
Quoted uncertainties are at the $1\sigma$\ (68\%) confidence level, unless 
otherwise stated.
Quoted optical/infrared magnitudes are AB magnitudes.

\section{Methodology}\label{sec:method}
{In \S\ref{sec:alg}, we detail our algorithm for transient-candidate 
searching, which is designed to find \cdfs\ XT-like events within individual 
\chandra\ exposures.
Our algorithm is simple and fast, and can be easily implemented for any individual
\chandra\ observations. 
We perform intensive Monte~Carlo simulations and assess the sensitivity 
of our algorithm in \S\ref{sec:eff}.
}

\subsection{Algorithm for Transient-Candidate Selection}\label{sec:alg}
{Our algorithm works on an unbinned \chandra\ light curve, i.e. an array 
of photon arrival times of a source, for which the background has
been estimated.
Below, we denote ${N_{\rm tot}}$ (${N_{\rm bkg}}$) as 
the number of total (background) counts for the light curve.
We require that the source lies within an off-axis angle of $8'$, following 
previous \chandra\ studies (e.g. \hbox{\citealt{vito16}}; \hbox{\citealt{yang16}}).
This is because \chandra's performance (as measured by, e.g. effective area and 
PSF size) drops significantly beyond $8'$.
Additionally, we require that the light-curve length is shorter than 50~ks to 
avoid large numbers of accumulated background counts in long exposures. 
Exposures longer than 50~ks are chopped into a few parts to meet this 
requirement (\S\ref{sec:analyses}).
In \S\ref{sec:sim_res}, we show that our algorithm reaches a uniform sensitivity
for nearly all observations shorter than 50~ks.
{Note that the choice of 50~ks is somewhat subjective; the flux limit 
and the results of our transient search (\S\ref{sec:sim_res} and \S\ref{sec:cp}) 
do not change significantly if we adjust this value between $\approx 16$~ks and 
$\approx 100$~ks.
Choosing a value below $\approx 16$~ks could chop some observations into 
$\lesssim 8$~ks parts, which are ineffective in our selection of XT-like 
transients (see \S\ref{sec:eff}).
Choosing a value above $\approx 100$~ks could leave some long observations unchopped,
which have relatively high accumulated background, affecting transient 
detection.\footnote{If we do not chop the observations, the actually detected 
extragalactic transients among our data will be the same, although there will be four more 
stellar flares detected (\S\ref{sec:analyses}).}
}
}

{Our algorithm first calculates ${N_1}$ and ${N_2}$,} 
defined as the numbers of 
counts at $t=(t_{\rm s}, t_{\rm m})$ and $t=(t_{\rm m}, t_{\rm e})$, 
respectively, where $t_{\rm s}$ and $t_{\rm e}$ are the times when the 
exposure starts and ends, respectively, and 
$t_{\rm m}=(t_{\rm s}+t_{\rm e})/2$, i.e. the midpoint of exposure time.
Since typical \chandra\ observations are continuous and uninterrupted by 
background flares ($\approx 1\%$ of exposure time), our two-part division 
of the exposure is legitimate.

We select a source in an observation as a transient candidate if it 
satisfies all of the following criteria (Method~1):
{
\begin{enumerate}[wide=0pt, widest=99, leftmargin=\parindent, labelsep=*]
    \item[(A)] ${N_{\rm tot}}$ is larger than the 5${\sigma}$ 
        Poisson upper limit of ${N_{\rm bkg}}$;
    \item[(B)] {$N_1$ and $N_2$ are statistically different at
        a ${>4\sigma}$ significance level;}
    \item[(C)] ${N_1 > 5\times N_2}$ or ${N_2 > 5\times N_1}$.
\end{enumerate}
}
Criterion~A filters out faint sources that have low signal-to-noise ratios (S/N),
{and thus boosts the speed of the selection process.}
This criterion is also helpful in avoiding false detections caused by rare 
background flares, since flares can dominate the detected counts for faint sources.
{Criterion~B selects sources that have significantly different count rates in the 
first-half and second-half exposures.
{Technically, we realize Criterion~B with the $E$-test \citep{krishnamoorthy04}.
The $E$-test can test if two Poisson variables ($N_1$ and $N_2$ in our case) are 
drawn from the same distribution, and simultaneously considers the statistical 
fluctuations of both variables.}
Criteria~A and B are based on statistical significance, and 
they select high-S/N sources with significant variability.
However, these criteria are not sufficient, since they cannot rule out AGNs which 
have stochastic variability.
To deal with this AGN issue, we also add Criterion~C, which requires that the 
flux-variation amplitude is large.
Since the characteristic AGN variability amplitudes (on timescales from
$\sim$~an hour to $\sim$~10 years) are a factor of $\lesssim 3$ (\S\ref{sec:intro}), 
we choose the amplitude threshold as a factor of 5 to cleanly rule out AGN variability.
We note that the choice of amplitude threshold is empirical: a low value could not 
remove AGNs effectively; a high value could miss potential transients.  
We have tested adjusting the threshold slightly (e.g., by a factor of 
{3 or 4} instead of 5), 
and the number of extragalactic transients we found in \S\ref{sec:analyses} does 
not change.
}

Method~1 is not efficient in selecting transients that happen at 
$t\approx t_{\rm m}$, because these transients will have similar $N_1$ and $N_2$.
To overcome this defect, we also select transients with the following method.
We denote $N'_1$ as the number of counts at $t=(t_{\rm bgn}, t_{\rm q1})$
plus that at $t=(t_{\rm q3}, t_{\rm end})$, where $t_{\rm q1}$ and 
$t_{\rm q3}$ are the first and third quartiles of the observation time,
and $N'_2$ as the number of counts at $t=(t_{\rm q1}, t_{\rm q3})$. 
We also select a source as a transient candidate, if it satisfies (Method~2)
{
\begin{enumerate}[wide=0pt, widest=99, leftmargin=\parindent, labelsep=*]
    \item[(A$'$)] ${N_{\rm tot}}$ is larger than the 
        5${\sigma}$ Poisson upper limit of ${N_{\rm bkg}}$;
    \item[(B$'$)] {$N'_1$ and $N'_2$ are statistically different at
        a ${>4\sigma}$ significance level;}
    \item[(C$'$)] ${N'_1 > 5\times N'_2}$ or ${N'_2 > 5\times N'_1}$.
\end{enumerate}
In \S\ref{sec:eff}, we prove the necessity of adopting both Method~1 
and Method~2 for transient selection.}

\subsection{Efficiency of the Selection Algorithm}\label{sec:eff}
In this Section, we assess the efficiency of our transient-selection algorithm
(\S\ref{sec:select}) with Monte~Carlo simulations.
{
In \S\ref{sec:conf}, we detail our simulation configurations.
In \S\ref{sec:gau}, we define a ``gauge'' to measure the efficiency of our 
algorithm.
In \S\ref{sec:sim_res}, we present our simulation results, i.e. the performance
of our algorithm.
}

\subsubsection{Simulation Configurations}\label{sec:conf}
\begin{figure}
\includegraphics[width=\linewidth]{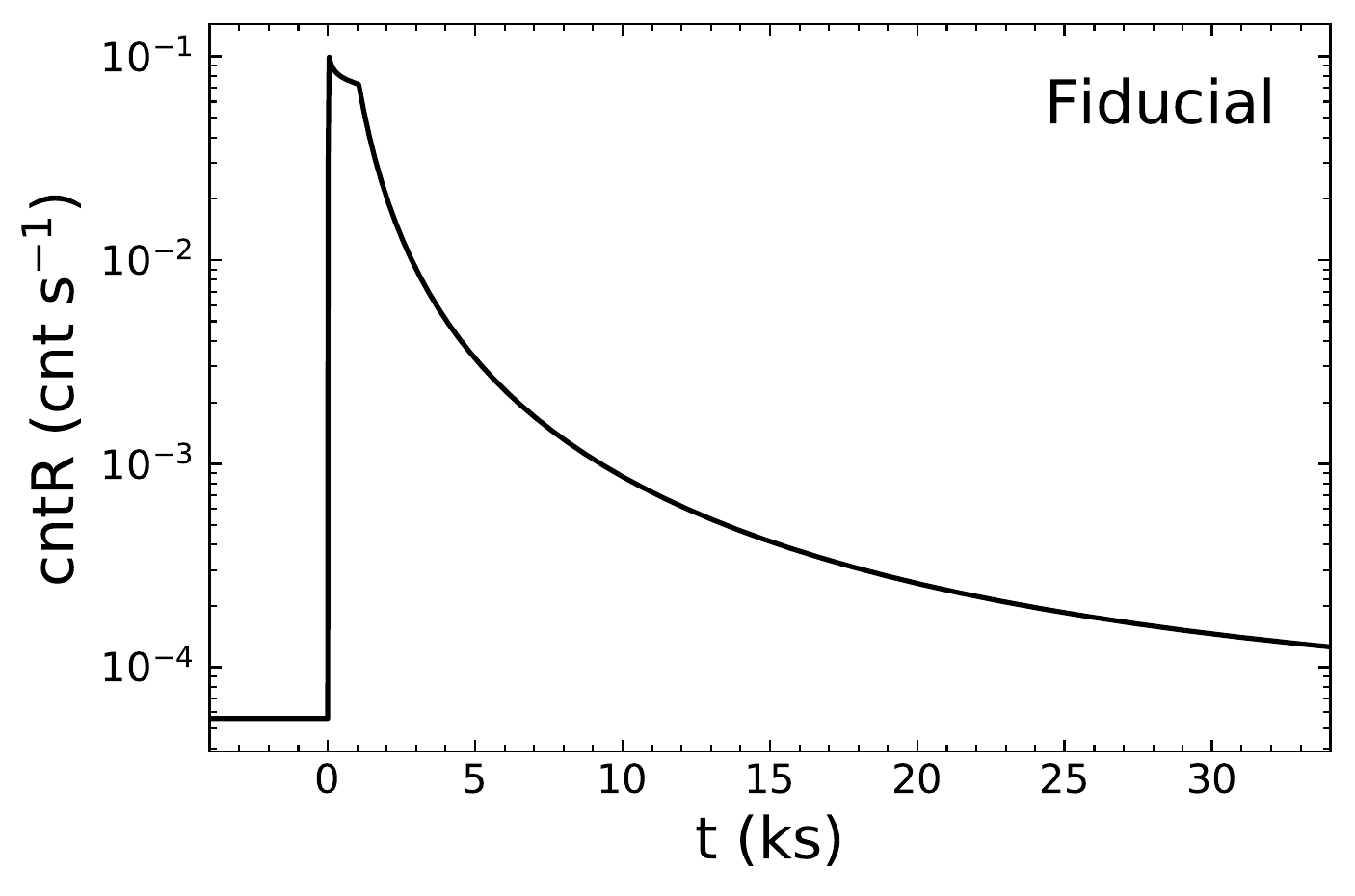}
\caption{The fiducial light-curve model adopted in our simulations.
The light-curve shape is similar to those of \cdfs\ XT1 and XT2.
The time ($x$-axis) zero point is chosen such that the transient 
starts at $t=0$. 
{The plot is generated with peak flux 
${\log\ F_{\rm peak} = -12.0}$ (cgs), 
for display purposes only.
In the simulations, we test different ${F_{\rm peak}}$ values 
(\S\ref{sec:sim_res}).
}
}
\label{fig:sim_lc_model}
\end{figure}

The simulations are based on a {fiducial light-curve model}.
Since our main goal is to search for fast extragalactic transients analogous to 
\cdfs\ XT1 and XT2, we adopt a light-curve model similar to the best-fit 
models of these two transients (\hbox{\citealt{bauer17}}; \hbox{\citealt{xue19}}). 
The light-curve shape in the model is described by 
\begin{equation}\label{eq:lc}
\rm{cntR}(t) \propto 
\begin{cases}
    0, &t<0 \\
    t, & 0\leq t< t_1 \\
    t^{\alpha_1}, & t_1\leq t< t_2 \\
    t^{\alpha_2}, & t\geq t_2
\end{cases}
\end{equation}
where cntR is the count rate in units of counts~s$^{-1}$.
Here, we follow the convention that the transient starts at $t=0$.
For $t$ between 0 and $t_1$, the cntR rises to the peak value. 
This time interval is very short ($\lesssim 100$~s for both \cdfs\ XT1 and XT2),
and thus the exact functional form is not important.
Here, we adopt a basic form of a linear rise and set $t_1= 50$~s.
For $t$ between $t_1$ and $t_2$, the light curve is roughly in a plateau with 
an index of $\alpha_1 = -0.1$.
This plateau only exists for XT2 (2.3~ks) but not for XT1, 
and we adopt $t_2 = t_1+ 1$~ks. 
For $t>t_2$, the adopted cntR is a power-law decline with an index
of $\alpha_2=-2$, which is between those of XT1 ($-1.5$) and XT2 ($-2.2$).
{We adopt a power-law spectral shape with photon index of 
${\Gamma=1.6}$ for the model, which is consistent {with} those measured 
for both XT1 and XT2.}
We note that changing the model parameters slightly (e.g.\ changing $t_1$
to 100~s and $\Gamma$ to 2.0) does not significantly affect our simulation 
results.
{In Appendix~\ref{sec:oth_lc}, we also perform simulations for 
some other types of transients that are significantly different from 
the \cdfs\ XTs, although these transients are not the main focus of 
this work; 
these simulations show that our algorithm can identify transients with
timescales $\lesssim$ exposure time while the details of the light-curve 
shapes do not affect the sensitivity significantly.
}
We plot the adopted light-curve model in Fig.~\ref{fig:sim_lc_model}.
The $T_{90}$ for this light-curve setting is 9.4~ks, similar to those of 
XT1 and XT2.
This similarity is expected, because our model in Eq.~\ref{eq:lc}
is based on the light-curve shapes of XT1 and XT2.
{
Under the fiducial-model configuration, the conversion between 
peak flux and total net counts is 
\begin{equation}\label{eq:Nnet}
{
    N_{\rm net} \approx 1.6\times 10^{14} F_{\rm peak}\ (\rm erg\ cm^{-2}\ s^{-1}).
}
\end{equation}
The conversion factor is calculated with 
{\sc pimms}, assuming a typical off-axis angle of 5$\arcmin$ when accounting 
for vignetting (i.e. the drop of photon-collecting area toward large off-axis 
angle; see Appendix~\ref{sec:oth_ang} for other off-axis 
angles).\footnote{{See http://cxc.harvard.edu/toolkit/pimms.jsp
for {\sc pimms}; see 
http://cxc.harvard.edu/proposer/POG/html/chap4.html for vignetting.}}
}

Background noise is also needed for the simulations.
{Here, background includes both detector background and
sky \xray\ background for \hbox{0.5--7 keV}. 
The background-extraction region is an annulus centered at the 
\xray\ source (see \S\ref{sec:xray_data} for details).}
The background level rises as a function of off-axis angle.
In the simulations, we assume a background of ${5.6\times 10^{-5}}$~cnt~s$^{-1}$,
which is the typical background level at an off-axis angle of 5$\arcmin$ 
(see Appendix~\ref{sec:oth_ang} for other off-axis angles).
The adopted background is also approximately the median value for all \xray\ 
sources in our studied surveys.
This background level only corresponds to ${\approx 3}$
background counts for a 50~ks light curve, which is the longest light 
curve analyzed (see \S\ref{sec:alg}).

\subsubsection{Efficiency Gauge}\label{sec:gau}
{For a given set of ${F_{\rm peak}}$ and 
${t_{\rm exp}}$ (exposure time), we can estimate the probability of transient 
detection (${P_{\rm det}}$) as a function of 
${t_{\rm m}}$ (observation midpoint; \S\ref{sec:alg}) with the 
simulation procedures described below.
}
Since the transient starts at $t=0$ (\S\ref{sec:conf}), $t_{\rm m}$ actually 
means the relative time between the exposure midpoint and the transient start time.

First, we simulate light curves in the time interval of 
$t=(-t_{\rm exp},\ t_{\rm exp})$.
We divide $t=(-t_{\rm exp},\ t_{\rm exp})$ into small bins with $\Delta t=5$~s.
We then calculate the expected total counts in each bin.
Using these values, we generate the counts in each bin with a Poisson 
distribution, which gives a simulated light curve.
We repeat the procedures and generate 1,000 light curves.
We apply both Method~1 and Method~2 (\S\ref{sec:select}) for these
light curves and calculate the fraction of successful detections.
We adopt this fraction as the detection probability ($P_{\rm det}$).

\begin{figure}
\includegraphics[width=\linewidth]{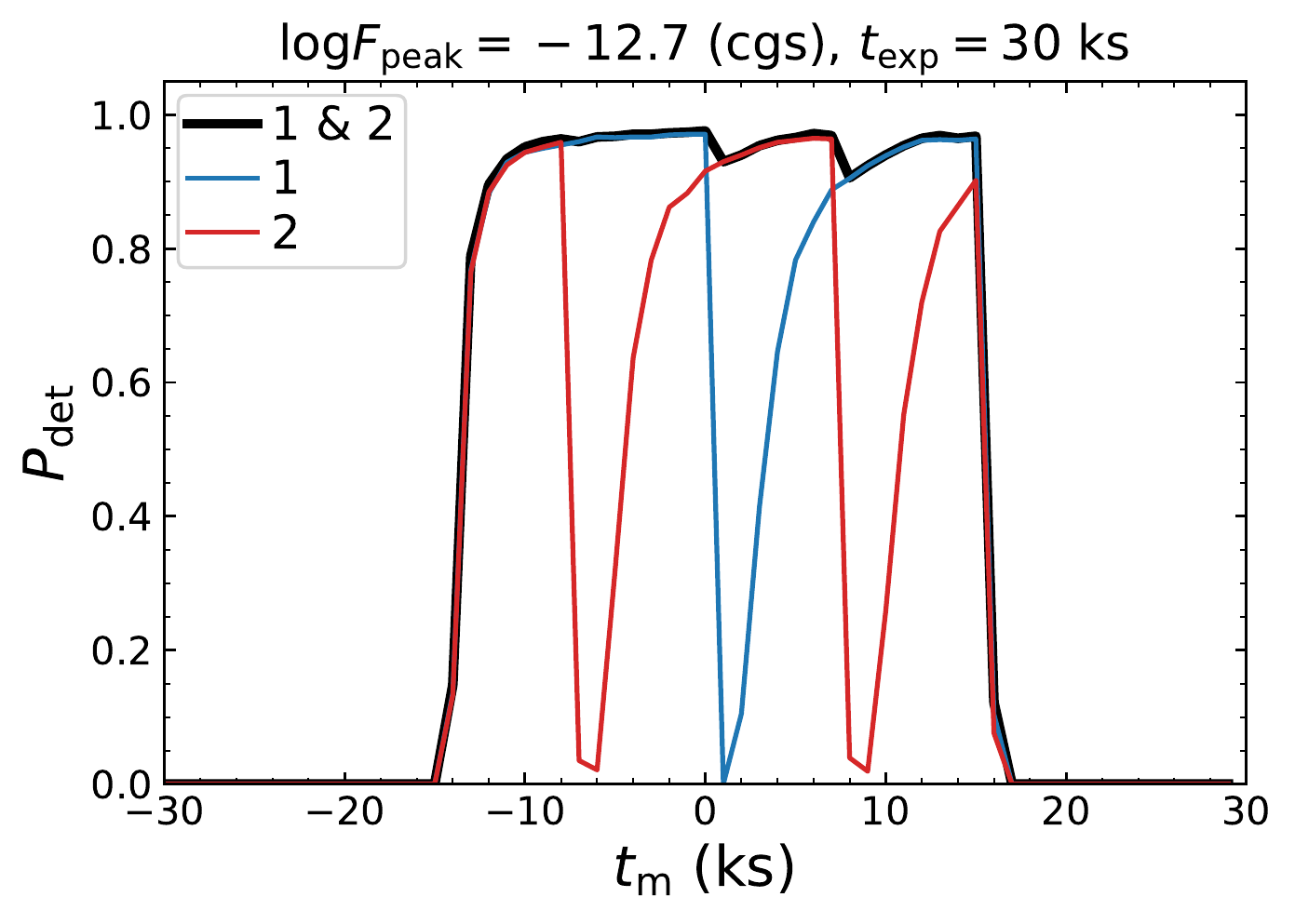}
\caption{$P_{\rm det}$ as a function of exposure 
midpoint (see \S\ref{sec:select}). 
The {black} curve represents the results of both Method~1 and Method~2;
the {blue} and {red} curves represent the results of Method~1 
and Method~2, respectively.
$P_{\rm det}$ is calculated based on simulations (\S\ref{sec:eff}).
The time ($x$-axis) zero point is chosen such that the transient 
starts at $t=0$, {and thus ${t_{\rm m}}$ means 
the relative time between the exposure midpoint and the transient 
start time.}
As labelled, different panels are for different net counts and 
exposure times.
There are some significant drops in the curves of Method~1 and Method~2, 
which are related to our transient-detection algorithm.
For example, when the transients happens at $t\approx t_{\rm m}$, 
the efficiency of Method~1 is low (see \S\ref{sec:alg}).
}
\label{fig:Pdet}
\end{figure}

{
Fig.~\ref{fig:Pdet} displays an example of ${P_{\rm det}}$ 
vs.\ ${t_{\rm m}}$
for ${\log F_{\rm peak}=-12.7}$ (cgs) and 
${t_{\rm exp}=30}$~ks.
Besides showing the ${P_{\rm det}}$ when using both Method~1 and Method~2 
(see \S\ref{sec:alg}), Fig.~\ref{fig:Pdet} also displays the ${P_{\rm det}}$ 
when using Method~1 and Method~2 separately.
Note that ${P_{\rm det}}$ drops significantly for some 
${t_{\rm m}}$ values when using Method~1 and Method~2 separately.
However, such drops are greatly alleviated when using both Methods, 
indicating the necessity of our combined method strategy.
}

{
From Fig.~\ref{fig:Pdet}, ${P_{\rm det}}$ (using both Methods) is not 
constant for different ${t_{\rm m}}$. 
This ${P_{\rm det}}$ variation makes it difficult to use ${P_{\rm det}}$ as 
a direct measure of algorithm performance as a function of ${F_{\rm peak}}$ and 
${t_{\rm exp}}$.
Therefore, we define an {``effective''} detection probability (${P_{\rm eff}}$) 
averaged over different ${t_{\rm m}}$ as a gauge to measure the efficiency, i.e.
\begin{equation}
{
P_{\rm eff} = \frac{\int^{+\infty}_{-\infty} P_{\rm det}(t_{\rm m}) 
d t_{\rm m} }{ t_{\rm exp} }.
}
\end{equation}
From this definition, ${P_{\rm eff}}$ ranges from 0 to 
${\approx 1}$ for a given set of ${F_{\rm peak}}$ and 
${t_{\rm exp}}$,\footnote{{${P_{\rm eff}}$ might 
be slightly greater than unity, because a transient may be detected even when
it is partially covered by the observations.}} 
with higher values indicating higher average detection efficiency.
}

\subsubsection{Simulation Results}\label{sec:sim_res}
\begin{figure}
\includegraphics[width=\linewidth]{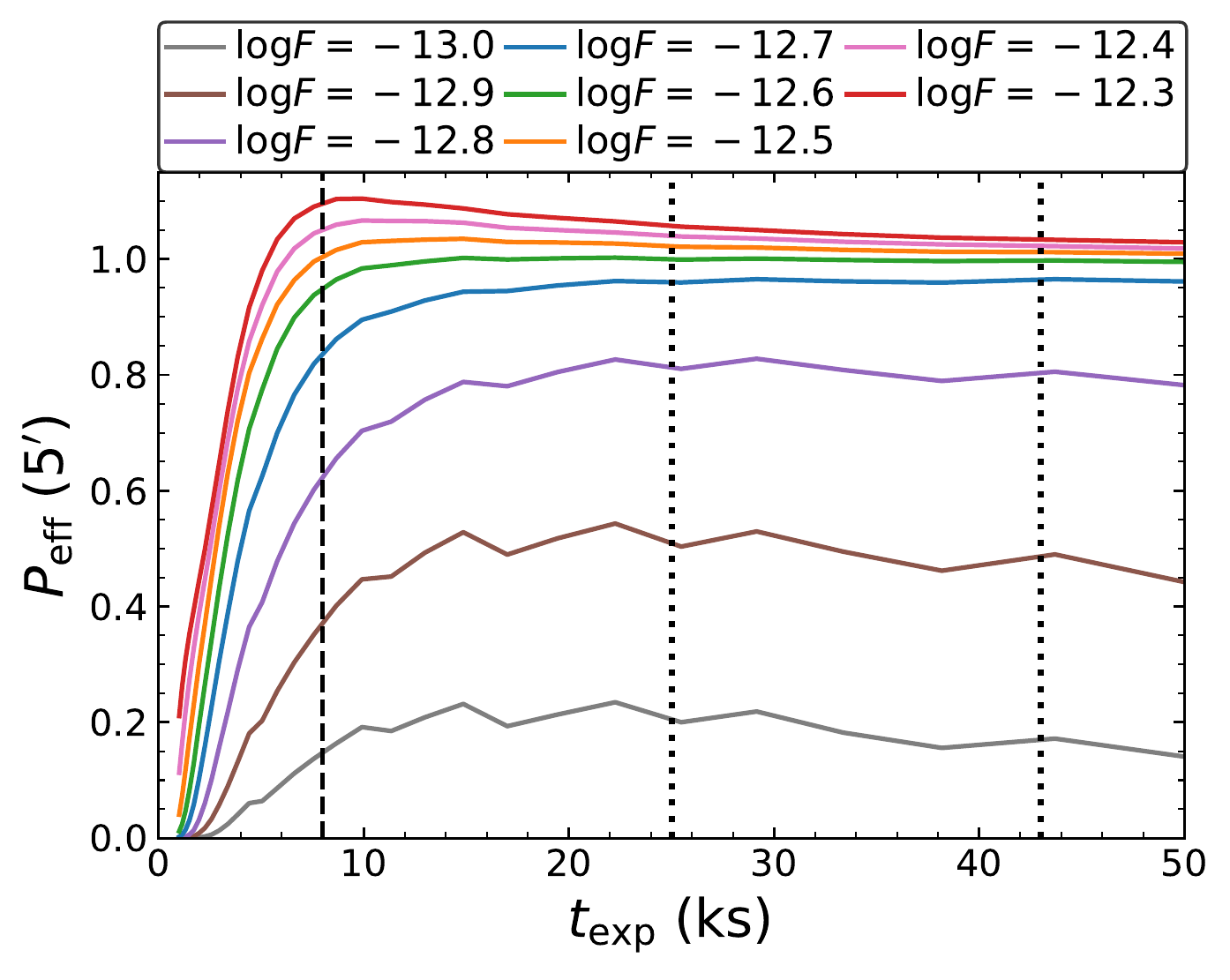}
\caption{{$P_{\rm eff}$ as a function of ${t_{\rm exp}}$
for a typical off-axis angle of ${5'}$ (see \S\ref{sec:conf}).
Different colors indicate different ${F_{\rm peak}}$ as labelled (cgs units).
The black vertical dashed line marks ${t_{\rm exp}=8}$~ks.
The black vertical dotted lines indicate the \hbox{20\%--80\%} percentile range of the 
exposure times of our data (\S\ref{sec:select}).
Above our selected flux limit (${\log F_{\rm peak}=-12.6}$; \S\ref{sec:sim_res})
${P_{\rm eff}}$ rises to ${\approx 1}$ for ${t_{\rm exp} \gtrsim 8}$~ks. 
}}
\label{fig:Peff}
\end{figure}

{We calculate ${P_{\rm eff}}$ for different ${t_{\rm exp}}$ 
and ${F_{\rm peak}}$ and show the results in Fig.~\ref{fig:Peff}.
As expected, ${P_{\rm eff}}$ rises toward high ${F_{\rm peak}}$ at a given 
${t_{\rm exp}}$, because brighter sources have higher S/N.
We choose ${\log F_{\rm peak} \approx -12.6}$~(cgs) as our detection limit, above 
which ${P_{\rm eff}\approx 1}$ for a wide range of ${t_{\rm exp}=8\text{--}50}$~ks.
Note that this flux limit is much lower than the peak fluxes of \cdfs\ XT1 and XT2 
(see Table~\ref{tab:xray_prop}).
The estimated flux limit is mainly used to estimate the event rate in 
\S\ref{sec:evtR}, we note that there are still non-zero probabilities to detect 
transients below this limit (see Fig.~\ref{fig:Peff}).
Here, we remind readers that ${t_{\rm exp}=50}$~ks is the maximum exposure time 
accepted by our algorithm (see \S\ref{sec:alg}).
We note that the simulation results above are calculated from our 
fiducial model which is similar to \cdfs\ XTs (\S\ref{sec:conf}; see Appendix~\ref{sec:oth_lc}
for some other transient models), since our main purpose is to find \cdfs\ XT-like transients. 
The simulation results are based on the instrumental response and background at a typical off-axis 
angle of ${5'}$ (\S\ref{sec:conf}), and we present the results at other off-axis angles 
in Appendix~\ref{sec:oth_ang}.
}

{
Below ${t_{\rm exp}=8}$~ks, ${P_{\rm eff}}$ drops significantly at a given 
${F_{\rm peak}}$ (see Fig.~\ref{fig:Peff}). 
This is because, when the exposure time becomes significantly shorter than the 
transient timescale, the observed light curve will be similar to a normal variable 
source, and thus may not be selected by our algorithm.
In our estimation of event rate (\S\ref{sec:evtR}), we do not include observations 
that are shorter than 8~ks, although we do not discard these observations in our 
transient search (\S\ref{sec:analyses}).
Only a negligible fraction of observation time (${\approx 0.1\%}$; 
see \S\ref{sec:evtR}) in our analyzed \xray\ data is from ${<8}$~ks 
exposures.
}


\section{Data and Analyses}\label{sec:analyses}
{The scope of this paper is to search for \cdfs\ XT-like extragalactic 
transients.
Utilizing the methodology detailed in \S\ref{sec:method}, we first perform an 
initial search for transient candidates in the \xray\ survey data 
(\S\ref{sec:xray_data} and \S\ref{sec:select}).
Since stellar objects can have strong \xray\ flares that might be selected by our 
algorithm, we need to exclude stars from our selected transient candidates.
We perform this task with the high-quality multiwavelength data available for the 
surveys (\S\ref{sec:cp}).
}

\subsection{X-ray Data and Processing}\label{sec:xray_data}
In this work, we analyze the \chandra\ survey data from the
\cdfs, \cdfn, DEEP2, UDS, COSMOS, and \hbox{E-CDF-S} regions.
The survey properties are summarized in Table~\ref{tab:surv_prop}.
DEEP2 includes the full field of EGS (\hbox{DEEP2-1}) and three other fields 
(\hbox{DEEP2-2}, \hbox{DEEP2-3}, and \hbox{DEEP2-4}) with shallower 
($\approx 10$~ks) exposures.
The total exposure time of these surveys is 19~Ms.
All the surveys are at high Galactic latitude ($|b|\gtrsim 40^\circ$), 
matching our main interest of searching for extragalactic transients.
Also, these surveys have deep multiwavelength coverage, allowing us to 
study the physical origins of the transients (\S\ref{sec:intro} 
and \S\ref{sec:cp}).

We download all the \chandra\ data products of observations related to the 
surveys, and run the {\sc chandra\_repro} script in 
{\sc ciao 4.10}.\footnote{{{\sc chandra\_repro} cannot process observation 
1431 (\cdfs), which consists of two separate exposures. 
For this observation, we use the data products from \cite{luo17}, who split the 
observation into two continuous exposures.
We perform transient searching for these two exposures independently 
(\S\ref{sec:select}), but do not find transient candidates in the two 
exposures.}}
The {\sc chandra\_repro} script performs standard cleaning and calibration 
processes,\footnote{http://cxc.harvard.edu/ciao/ahelp/chandra\_repro.html} and 
yields a clean event file for each observation.
{Based on the data products of {\sc chandra\_repro}, we produce the 
exposure map for each observation using the {\sc ciao} script {\sc fluximage}. 
The exposure maps denote the ``effective'' exposure times for different positions 
in the field of view, and instrumental factors such as bad pixels and vignetting
are taken into account.}

For each event file, we extract the \hbox{0.5--7~keV} photons of each \xray\ 
source presented in the \xray\ catalogs (Table~\ref{tab:surv_prop}). 
Since the \chandra\ background is extremely low, any sources with 
$\gtrsim 10$ net counts should be detected by the \xray\ surveys.
This level of counts is much lower than that of our transient-selection 
sensitivity (see below), and thus we should not miss any 
transients due to their absence in the \xray\ catalogs.
The total events are extracted from an aperture of $1.5\times R_{90}$, where 
$R_{90}$ is the radius encircling 90\% of the \xray\ counts.
We adopt $R_{90}$ as a function of off-axis angle from Table~A1 of \cite{vito16}.
{From simulations with the {\sc ciao} script {\sc simulate\_psf},
we find that this aperture size (${1.5\times R_{90}}$) encircles 
nearly all (${\gtrsim 98\%}$) \xray\ counts regardless of off-axis 
angle.}
The background events are extracted from an annulus with inner and outer 
radii of $1.5\times R_{90}$ and $1.5\times R_{90} + 20~\rm{pixels}$.
The background area is $9$~times larger than the source area for a typical 
source at an off-axis angle of $5\arcmin$.
If the background region covers a nearby \xray\ source, we mask the source
(also with radius of $1.5\times R_{90}$), and do not include the masked area 
when estimating the background.
We note that changing the source and background extraction regions slightly 
will not affect our qualitative results.
{We estimate the background counts in the source region 
(${N_{\rm bkg}}$; \S\ref{sec:alg}) by scaling the counts in the 
background region by a factor.
Here, the scaling factor is the sum of the exposure-map values 
in the source area divided by that in the background area.}

\begin{table*}
\centering
\caption{Properties of X-ray Surveys Analyzed in this Work}
\label{tab:surv_prop}
\begin{tabular}{cccccc}
\hline\hline
Survey & Area & Total Exp. & Obs.\ Num. & Src.\ Num. & Reference \\
(1) & (2) & (3) & (4) & (5) & (6) \\
\hline
CDF-S & 0.13 & 6.9 & 101 & 1008 & \cite{luo17} \\ 
CDF-N & 0.12 & 2.0 & 20 & 683 & \cite{xue16} \\ 
DEEP2 & 3.28 & 3.7 & 139 & 2976 & \cite{goulding12, nandra15} \\ 
UDS & 0.33 & 1.2 & 25 & 868 & \cite{kocevski18}; Suh et al.\ in prep. \\ 
COSMOS & 2.20 & 4.5 & 117 & 4016 & \cite{civano16, marchesi16} \\ 
E-CDF-S & 0.31 & 1.0 & 9 & 1003 & \cite{xue16} \\ 
\hline
All & 6.38 & 19.3 & 411 & 10554 & -- \\
\hline
\end{tabular}
\begin{flushleft}
{\sc Note.} ---
(1) \xray\ survey name.
(2) Survey area in deg$^2$.
(3) Total exposure time in Ms.
(4) Number of \chandra\ observations {(before chopping; see \S\ref{sec:select})}.
(5) \xray\ source number.
(6) References where the survey details and source catalog are presented.
Additional information about the \cdfs, \cdfn, and \hbox{E-CDF-S} can be found in \cite{xue17}. 
\end{flushleft}
\end{table*}

\subsection{Selection of Transient Candidates}\label{sec:select}
\begin{figure}
\includegraphics[width=\linewidth]{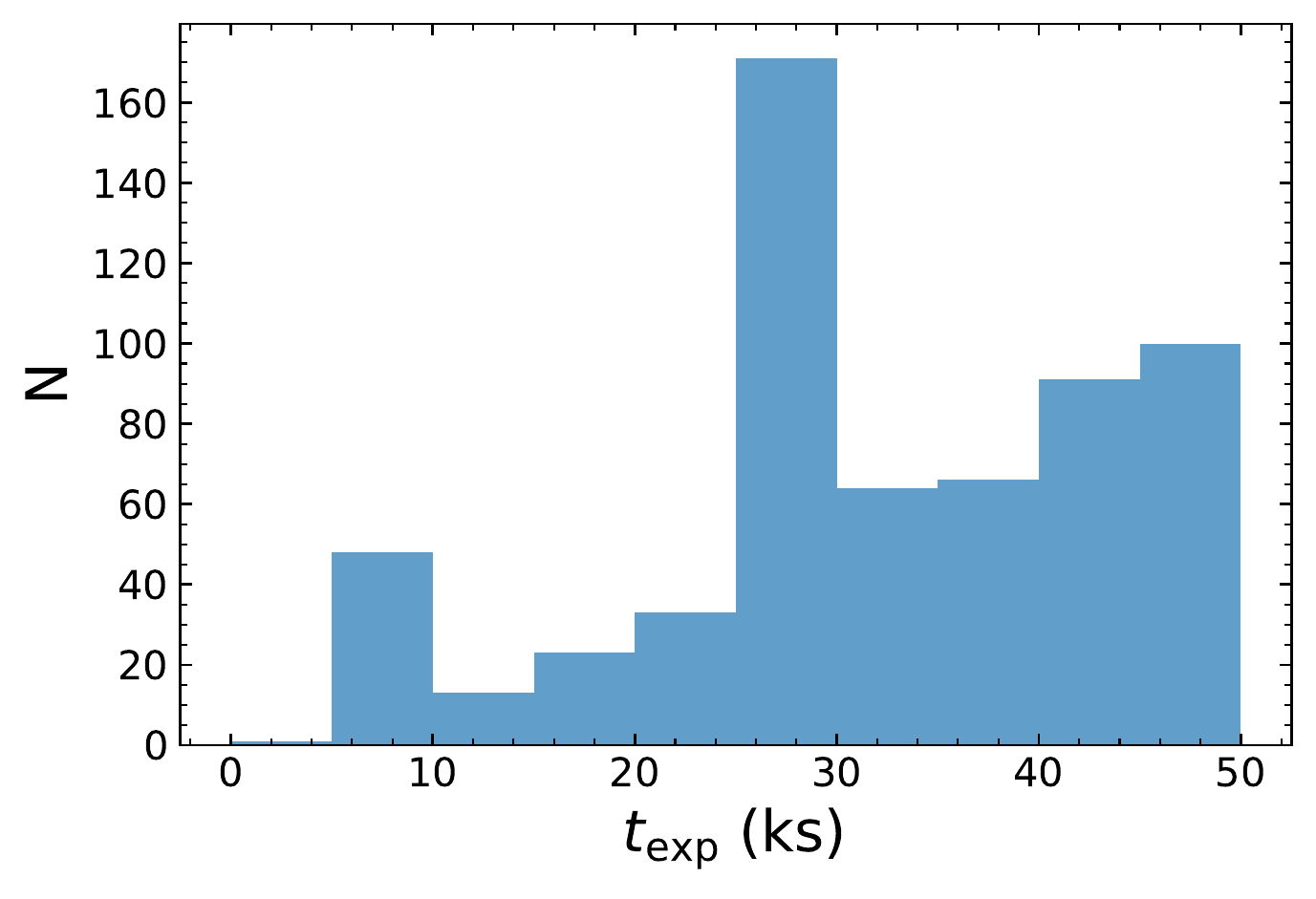}
\caption[]{{The $t_{\rm exp}$ distribution of the 610 exposures 
(after exposure chopping; \S\ref{sec:select}) analyzed in this work. 
There is a peak at the $t_{\rm exp}=25\text{--}30$~ks bin, because 
the original observation set has many (74) exposures of
$t_{\rm exp}=50\text{--}60$~ks, and these exposures are chopped 
to exposures of $t_{\rm exp}=25\text{--}30$~ks.
}}
\label{fig:texp_dis}
\end{figure}

{We apply the algorithm in \S\ref{sec:alg} to the light curves extracted in 
\S\ref{sec:xray_data}.
We note that the transient selection is only applied to sources with off-axis
angle of ${<8'}$ to avoid the low-quality \xray\ data beyond 
${8'}$ (\S\ref{sec:alg}).
If a light curve is longer than 50~ks (the maximum length accepted by our 
algorithm; \S\ref{sec:alg}), we chop it into several continuous parts with each 
having the same ${t_{\rm exp}}$ shorter than (or equal to) 50~ks 
(\S\ref{sec:method}). 
For example, for a 80~ks exposure, we divide it into two parts each having 
${t_{\rm exp}=40}$~ks.
We then perform transient selection for each chopped light curve independently.
After this observation-chopping process, we have 610 exposures with a median  
${t_{\rm exp}}$ of 30~ks and a \hbox{20\%--80\%} percentile range of 
\hbox{25--43 ks}.
{We show the $t_{\rm exp}$ distribution of these 610 exposures in 
Fig.~\ref{fig:texp_dis}.}
}

For Method~1 (2), Criterion~A (A$'$) selects a total of 9379 (9379) events
in the {610 exposures} analyzed.
Among these events, Criterion~B (B$'$) further selects {31 (24)} events. 
Finally, Criterion~C (C$'$) picks out 11 (5) events as the events selected by 
Method~1 (2).
{For the events filtered out by Criterion~C (C$'$), $\approx 70\%$ of 
them are stellar flares, identified with the methods detailed in \S\ref{sec:cp};
the other $\approx 30\%$ have extragalactic origins.
We have examined the light curves of these extragalactic sources and found 
all of them have significant non-zero quiescent fluxes, and thus they are 
likely AGNs rather than extragalactic transients.
This result demonstrates the capability of Criterion~C (C$'$) in removing AGN 
variability (\S\ref{sec:alg}).
}
We merge the events selected by Method~1 and Method~2, leading to a sample of 
13 unique transient candidates. 
Among these 13 candidates, 8 and 2 are uniquely selected by Method~1 and Method~2,
respectively, indicating the importance of using both Methods 
(see \S\ref{sec:method}).

We visually inspect the background light curves of these {transient candidates}, 
and do not find significant flares.
We have checked the \xray\ images of the transients in both sky and detector coordinates.
For each source, the events are concentrated and extended in sky and detector 
coordinates, respectively. 
This indicates that the {transient candidates} are physical \xray\ sources 
rather than hot pixels,
because hot pixels will lead to extended (concentrated) patterns in the sky (detector) 
coordinates caused by \chandra\ dithering.

\begin{figure}
\includegraphics[width=\linewidth]{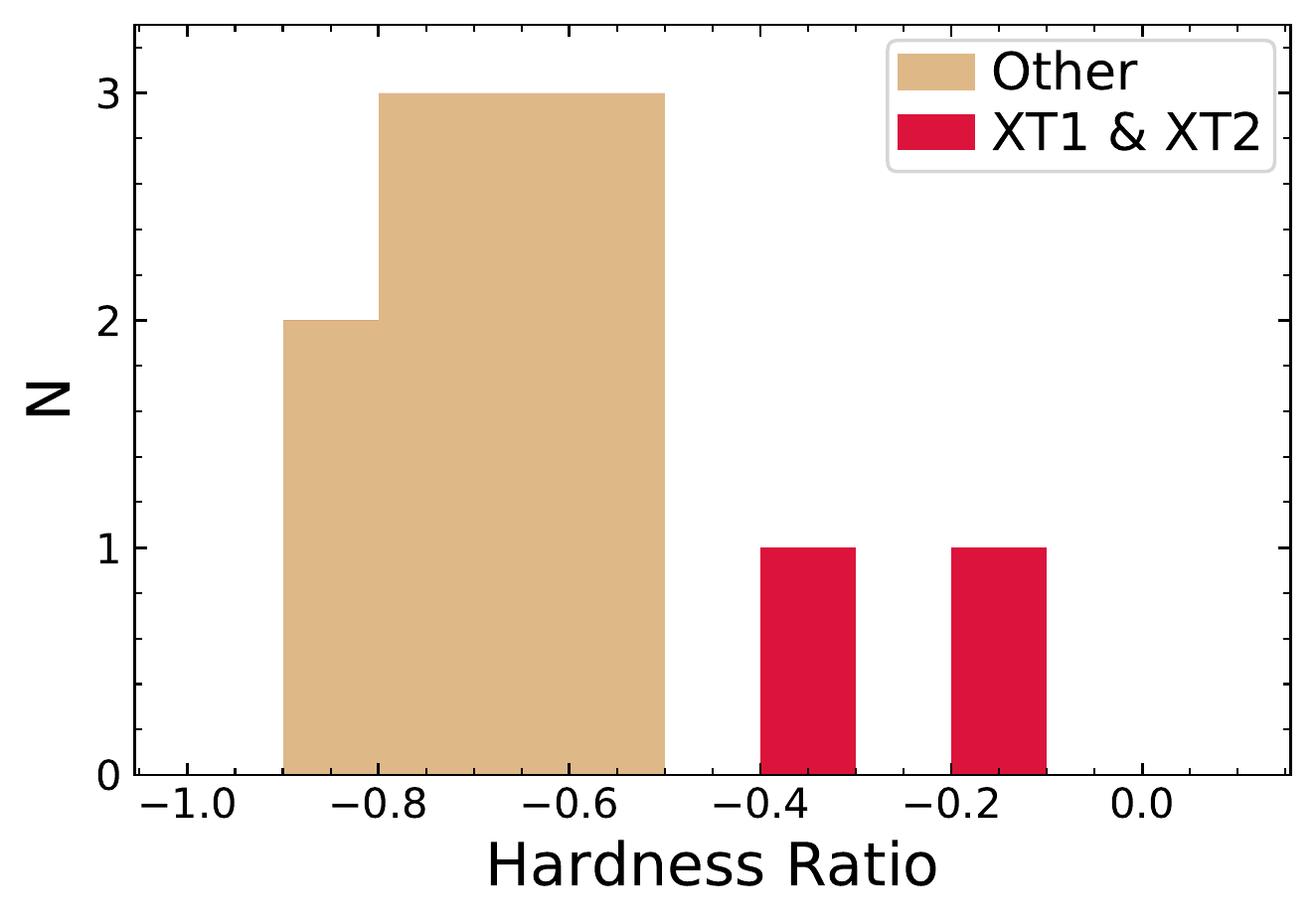}
\caption[]{Hardness-ratio distribution of our selected {transient candidates}. 
\cdfs\ XT1 and XT2 are highlighted with the red color.
The spectral shapes of XT1 and XT2 are harder than most of the other transients.
}
\label{fig:hr_hist}
\end{figure}

The \xray\ properties of the {13} transient candidates are listed in 
Table~\ref{tab:xray_prop}. 
ID1 and ID2 are \cdfs\ XT1 and XT2, respectively.
Their successful selection indicates that our method of transient searching
is effective for selecting {\cdfs\ XT-like transients (\S\ref{sec:sim_res})}.
For each {transient candidate}, we calculate the hardness ratio for the 
observation where the transient is identified.
Here, hardness ratio is defined as $(H-S)/(H+S)$, where $H$ and $S$ are hard-band 
(\hbox{2--7 keV}) and soft-band (\hbox{0.5--2 keV}) net counts, respectively.
The ${1\sigma}$ uncertainty is calculated with {\sc behr}, a Bayesian 
code for hardness ratio estimation \citep{park06}.
The results are listed in Table.~\ref{tab:xray_prop}.
In Fig~\ref{fig:hr_hist}, we show the distribution of hardness ratios.
The spectral shapes of XT1 and XT2 are harder than for other 
{transient candidates}.

\begin{figure*}
\includegraphics[width=0.45\linewidth]{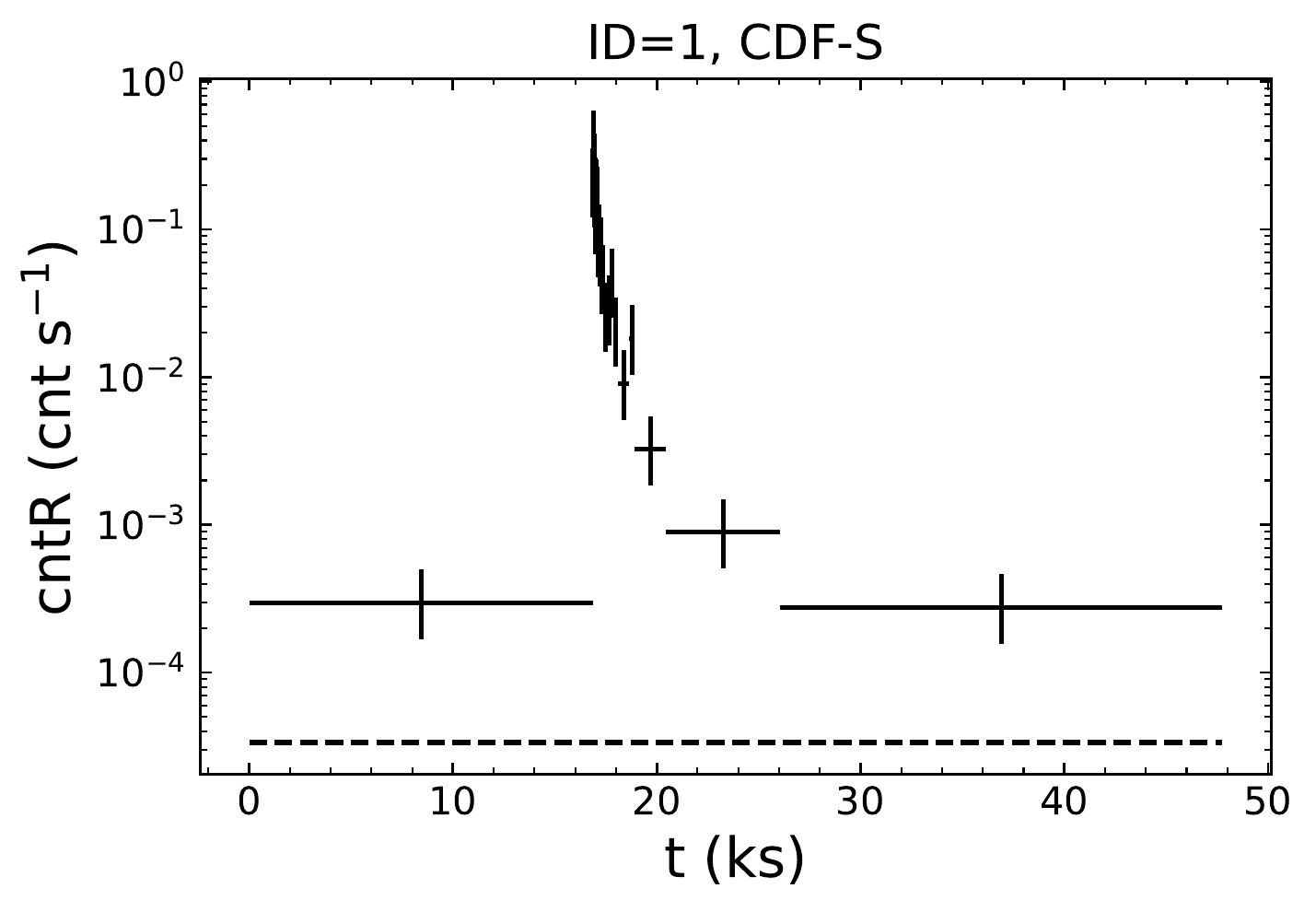}
\includegraphics[width=0.45\linewidth]{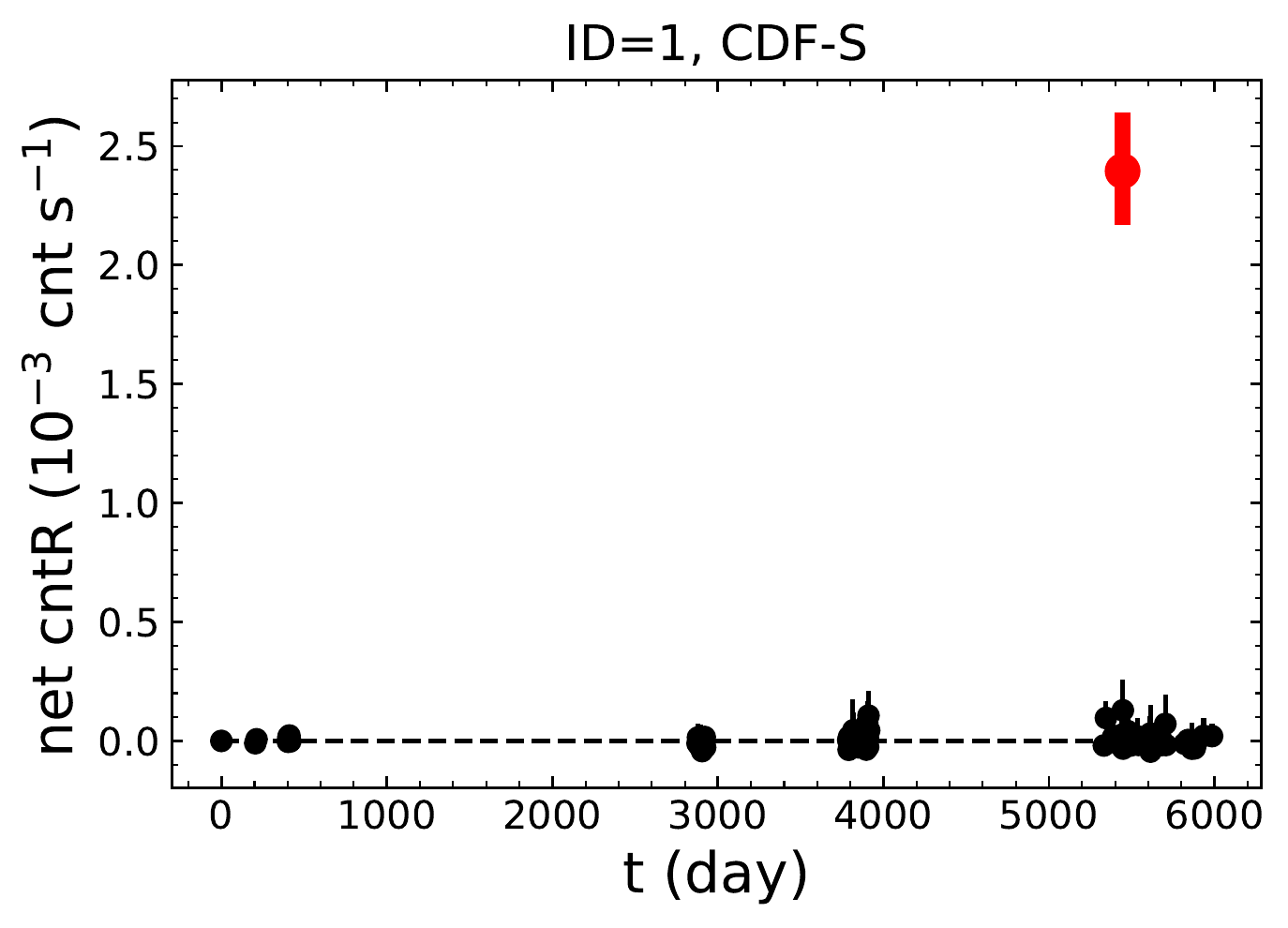} \\
\includegraphics[width=0.45\linewidth]{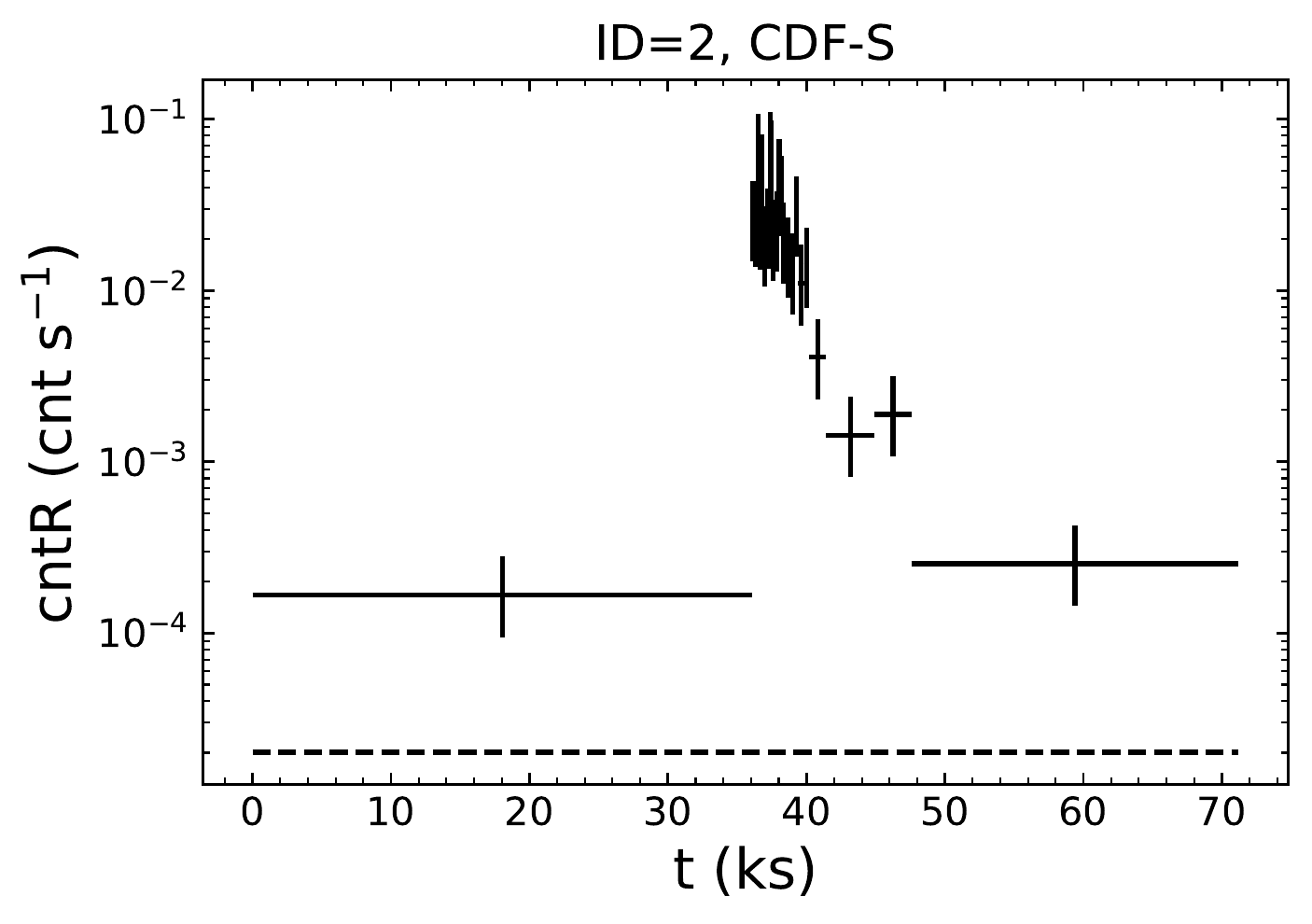}
\includegraphics[width=0.45\linewidth]{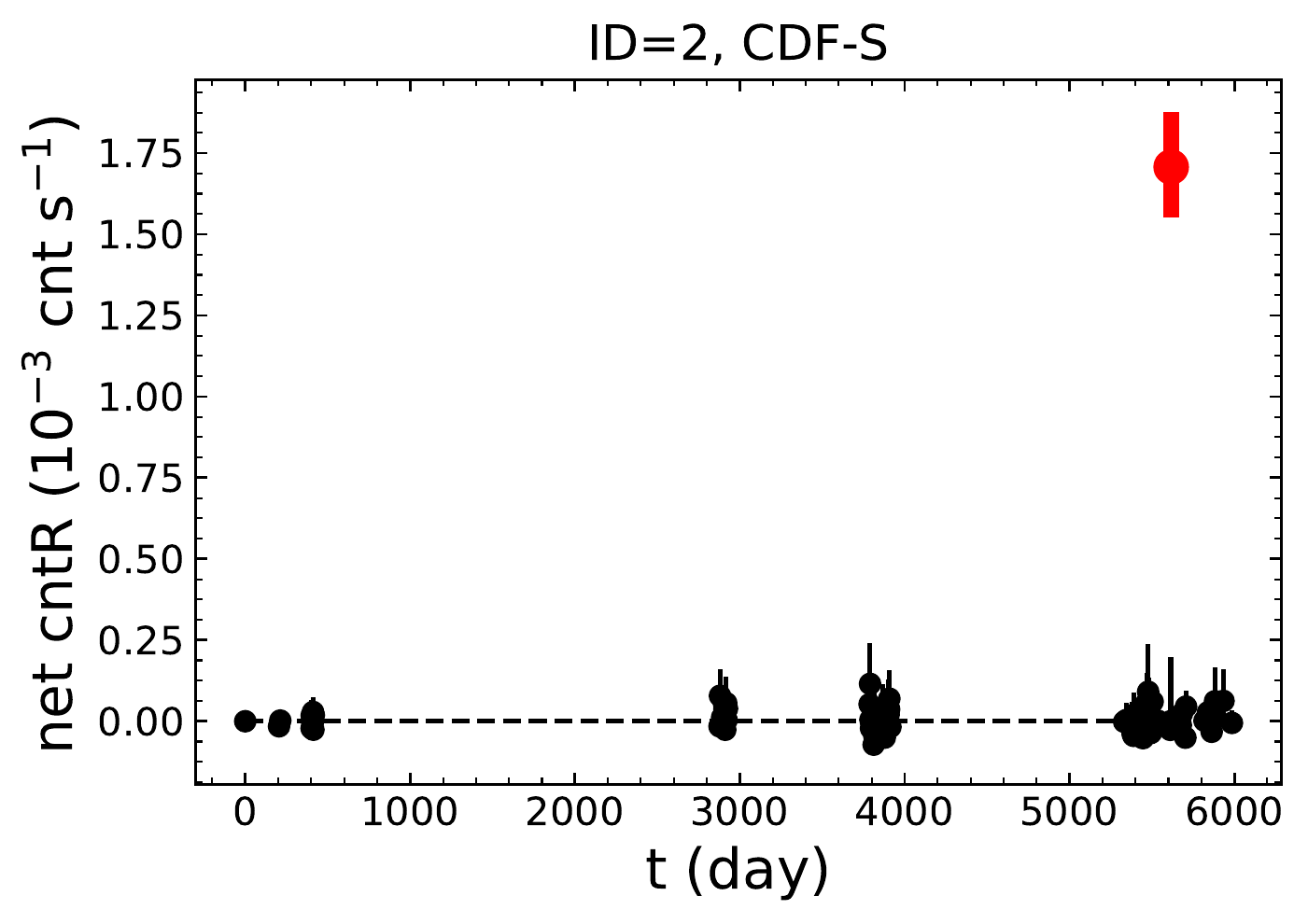} \\
\includegraphics[width=0.45\linewidth]{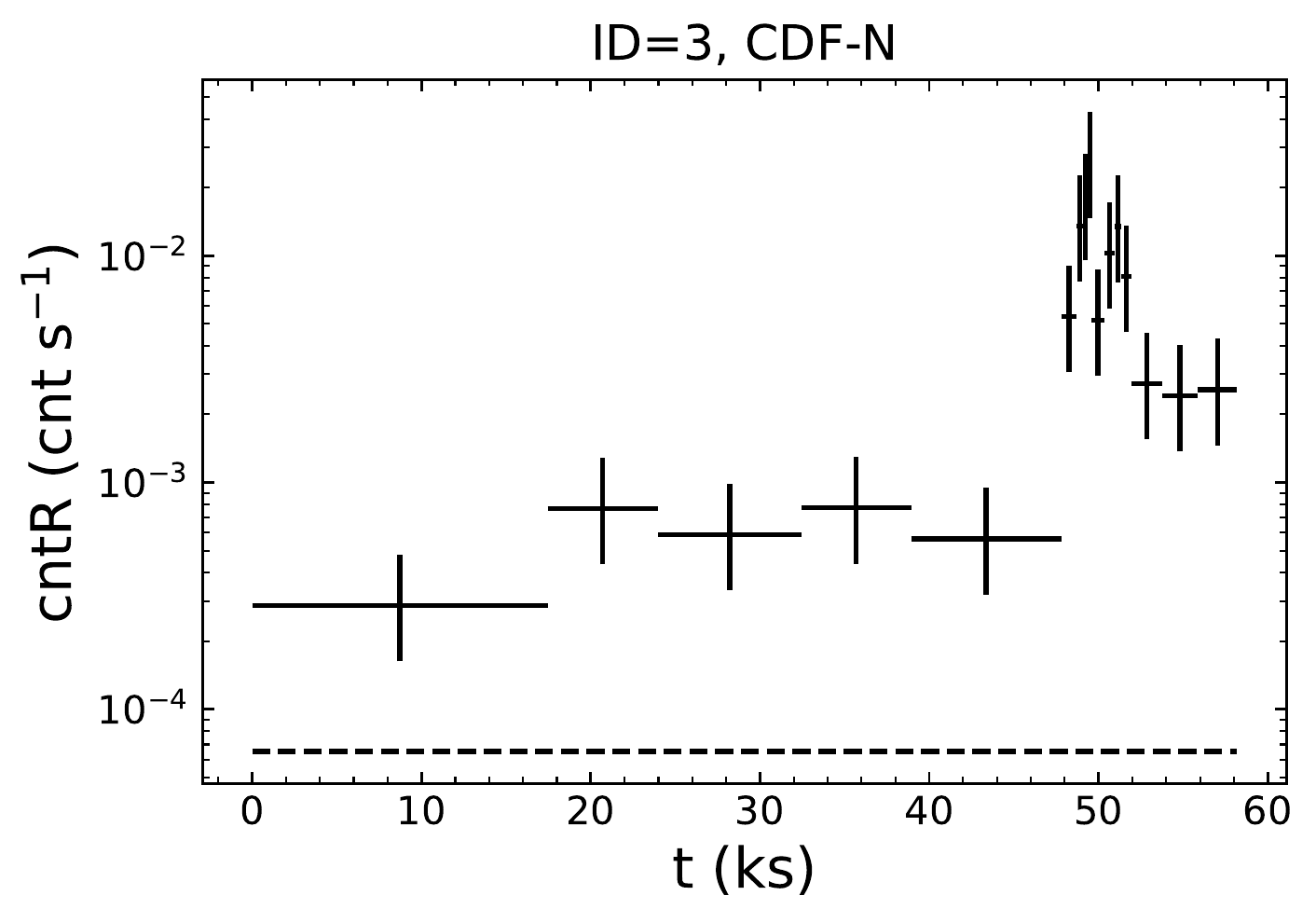}
\includegraphics[width=0.45\linewidth]{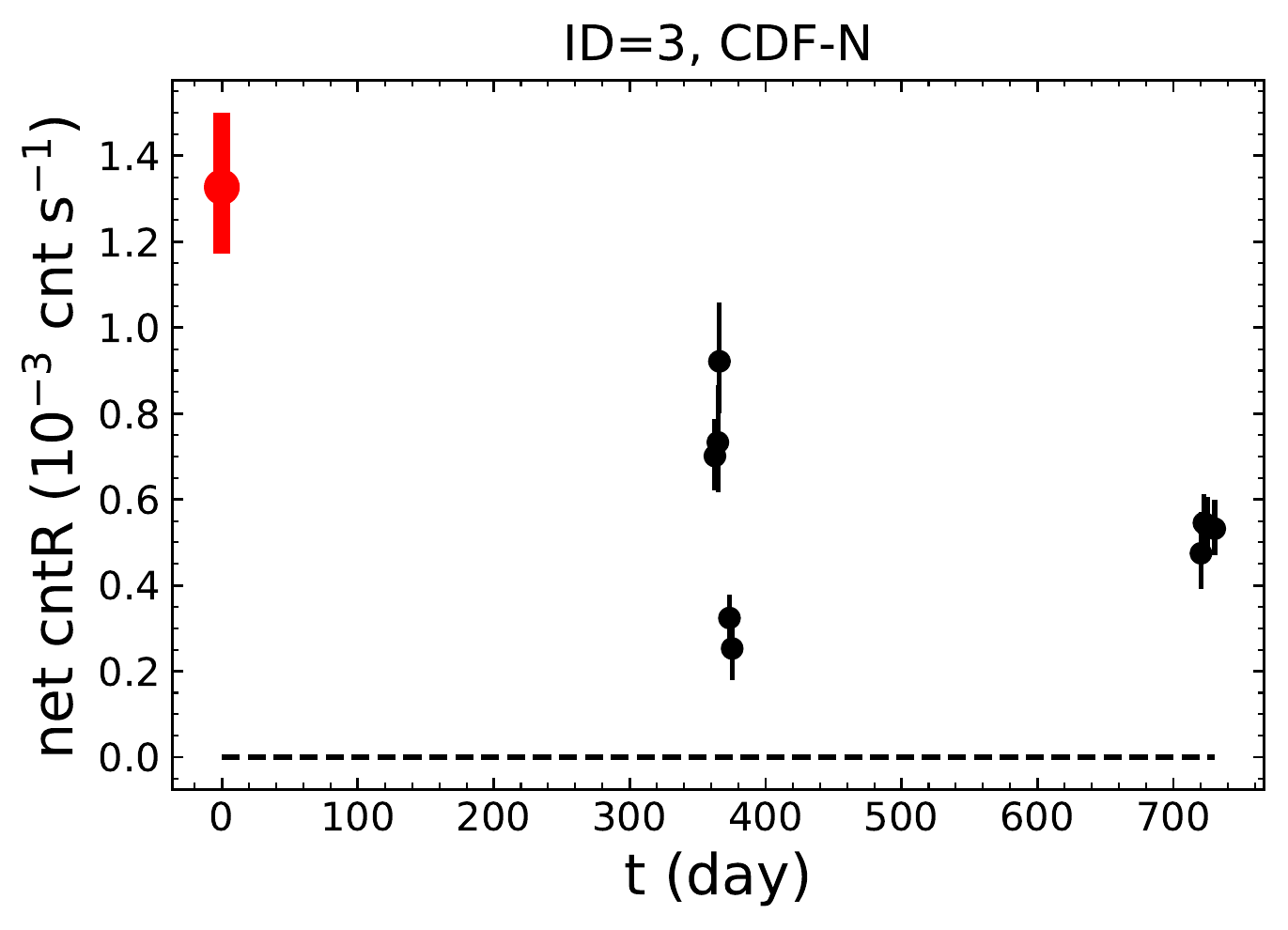} \\
\includegraphics[width=0.45\linewidth]{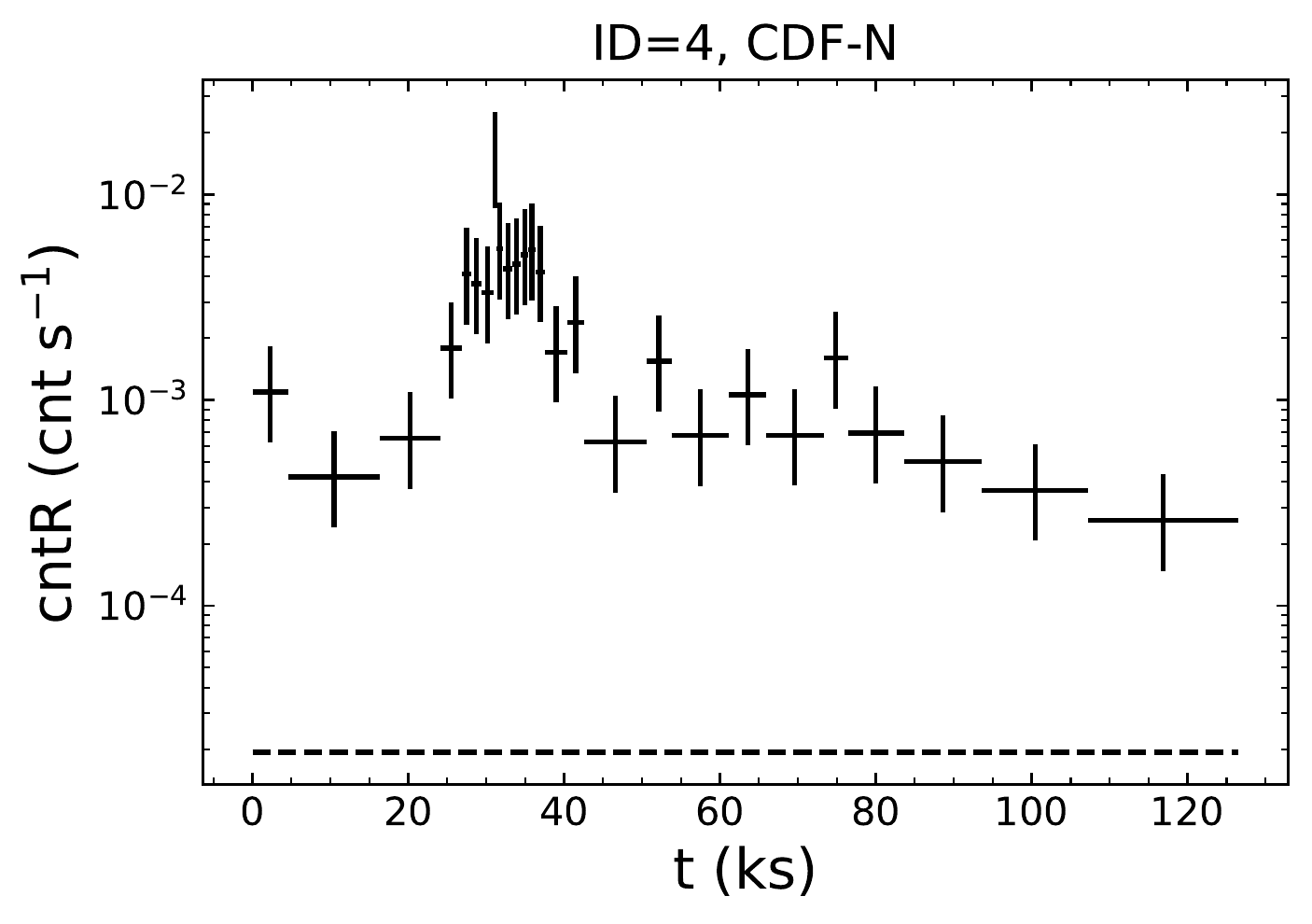}
\includegraphics[width=0.45\linewidth]{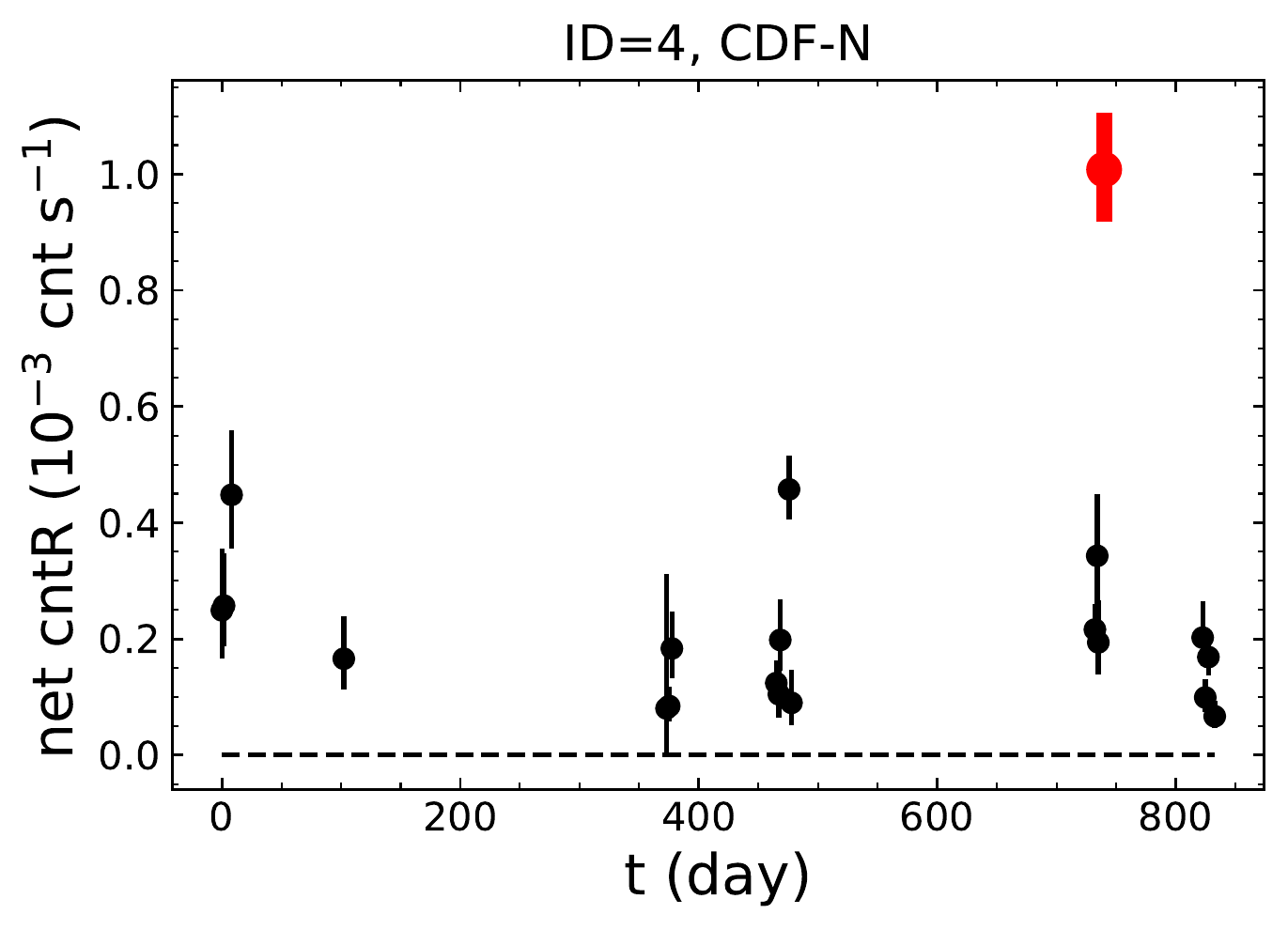} \\
\caption{Light curves for each {transient candidate}.
The left panels are light curves for the observation with the 
transient, with each bin including 5 counts.
The horizontal dashed lines indicate the estimated average 
background count rates.
The right panels are long-term light curves with each data point
representing a \chandra\ observation.
The transient observation is highlighted in red color.
The horizontal dashed lines indicate a net count rate of zero.
}
\label{fig:lc}
\end{figure*}

In Fig.~\ref{fig:lc} (left), we show the light curves of the {transient candidates}
during the observation when the transient happens.
The light curves are derived from the \xray\ events extracted in \S\ref{sec:select},
and are binned by 5-count intervals.
The data points in these light curves indicate total count rates, including 
contributions from the source and background.
The estimated average background count rate is marked as the dashed line 
in each panel of Fig.~\ref{fig:lc} (left).
The durations of XT1 and XT2 tend to be shorter than for other 
{transient candidates} (Fig.~\ref{fig:lc} left).
The $T_{90}$ values of XT1 and XT2 are $5.0^{+4.2}_{-0.3}$~ks and 
$11.1^{+0.4}_{-0.6}$~ks, respectively (see \hbox{\citealt{bauer17}} and \hbox{\citealt{xue19}}
for details).
We do not derive $T_{90}$ for other sources, because $T_{90}$ cannot be 
derived for many transients that extend beyond the \chandra\ exposures 
(e.g.\ ID3 and ID9 in Fig.~\ref{fig:lc} left).
Also, unlike XT1 and XT2, many of the other {transient candidates} 
have non-zero fluxes in the quiescent states, and thus their $T_{90}$ calculation 
requires careful subtraction of the quiescent fluxes, which is beyond the 
scope of this work.

We plot the long-term light curves in Fig.~\ref{fig:lc} (right), where each 
\chandra\ observation is represented by a data point.
These data points indicate net count rates, which are background-subtracted.
As expected, the transient observation generally has a count rate much higher 
than other observations.
However, unlike the \cdfs\ XT1 and XT2 events, most of the other {transient candidates}
have detectable signals in some of the non-transient observations.
Also, \cdfs\ XT1 and XT2 tend to have higher hardness ratios than the rest
of the selected {transient candidates} (Fig.~\ref{fig:hr_hist}).
These differences indicate that most of the new {transient candidates} are 
physically distinct from \cdfs\ XT1 and XT2 (see \S\ref{sec:cp}).

\begin{table*}
\centering
\caption{\xray\ Properties of {Transient Candidates}}
\label{tab:xray_prop}
\begin{tabular}{cccccccccc}
\hline\hline
ID & Survey & RA & DEC & Pos.\ Unc. & Obs.\ ID & Off.\ Ang. & HR & $\log F_{\rm peak}$ & Method \\
(1) & (2) & (3) & (4) & (5) & (6) & (7) & (8) & (9) & (10) \\
\hline
1 & CDF-S & 53.16156 & $-27.85934$ & 0.32$\arcsec$ &   16454 & 4.3$\arcmin$ &   $-0.13^{+0.09}_{-0.10}$ &  $-11.41$ & 1,2 \\ 
2 & CDF-S & 53.07648 & $-27.87339$ & 0.31$\arcsec$ &   16453 & 4.1$\arcmin$ &   $-0.32^{+0.08}_{-0.09}$ &  $-12.18$ & 1,2 \\ 
3 & CDF-N & 189.02046 & $62.33728$ & 0.20$\arcsec$ &   957 & 6.6$\arcmin$ &   $-0.54^{+0.08}_{-0.12}$ &  $-12.59$ & 1 \\ 
4 & CDF-N & 189.10587 & $62.23467$ & 0.10$\arcsec$ &   3389 & 3.2$\arcmin$ &   $-0.82^{+0.05}_{-0.06}$ &  $-12.82$ & 1 \\ 
5 & DEEP2 & 215.07414 & $53.10650$ & 0.36$\arcsec$ &   9875 & 6.7$\arcmin$ &   $-0.72^{+0.08}_{-0.12}$ &  $-12.53$ & 1 \\ 
6 & DEEP2 & 214.96015 & $52.74344$ & 0.26$\arcsec$ &   9456 & 6.6$\arcmin$ &   $-0.63^{+0.12}_{-0.14}$ &  $-12.97$ & 1 \\ 
7 & DEEP2 & 214.61007 & $52.54347$ & 0.20$\arcsec$ &   9735 & 4.8$\arcmin$ &   $-0.83^{+0.04}_{-0.17}$ &  $-12.75$ & 1,2 \\ 
8 & DEEP2 & 214.66798 & $52.66658$ & 0.11$\arcsec$ &   5849 & 3.0$\arcmin$ &   $-0.77^{+0.08}_{-0.10}$ &  $-13.21$ & 2 \\ 
9 & DEEP2 & 252.12761 & $34.96337$ & 0.53$\arcsec$ &   8636 & 7.5$\arcmin$ &   $-0.62^{+0.09}_{-0.12}$ &  $-12.46$ & 1 \\ 
10 & UDS & 34.48317 & $-5.09118$ & 0.96$\arcsec$ &   17305 & 0.7$\arcmin$ &   $-0.53^{+0.15}_{-0.18}$ &  $-13.14$ & 1 \\ 
11 & COSMOS & 149.75403 & $2.14188$ & 0.30$\arcsec$ &   8021 & 4.0$\arcmin$ &   $-0.80^{+0.09}_{-0.14}$ &  $-13.38$ & 1 \\ 
12 & COSMOS & 149.82641 & $2.71812$ & 0.30$\arcsec$ &   15214 & 5.9$\arcmin$ &   $-0.58^{+0.07}_{-0.09}$ &  $-12.59$ & 1 \\ 
13 & COSMOS & 149.99794 & $2.77972$ & 0.90$\arcsec$ &   15211 & 6.5$\arcmin$ &   $-0.69^{+0.12}_{-0.14}$ &  $-12.66$ & 2 \\ 
\hline
\hline
\end{tabular}
\begin{flushleft}
{\sc Note.} ---
(1) {Transient-candidate} ID in this work.
(2) \xray\ survey name.
(3), (4), and (5) \xray\ source position and positional error from the corresponding survey catalog.
The positional error is taken from the survey catalog, and is calculated based on all observations 
that cover the source (not only the observation in Column~6). 
For example, ID6 has a lower positional uncertainty than ID15, because the former has more total net 
counts than the latter ($\approx$~100 vs.\ $\approx$~25).
(6) \chandra\ ID of the observation where the transient is identified.
(7) Off-axis angle of the transient in the observation.
(8) Hardness ratio based on the observation in Column~6. 
{The uncertainties are at the $1\sigma$ level and are calculated with
{\sc behr} (\S\ref{sec:select})}.
(9) Logarithmic \hbox{0.5--7 keV} peak flux converted from the peak count rate in Fig.~\ref{fig:lc} 
with the method in \S\ref{sec:conf}.
(10) The Method(s) responsible for identifying the transient candidate.
\end{flushleft}
\end{table*}

\renewcommand{\thefigure}{\arabic{figure} (Continued)}
\addtocounter{figure}{-1}
\begin{figure*}
\includegraphics[width=0.45\linewidth]{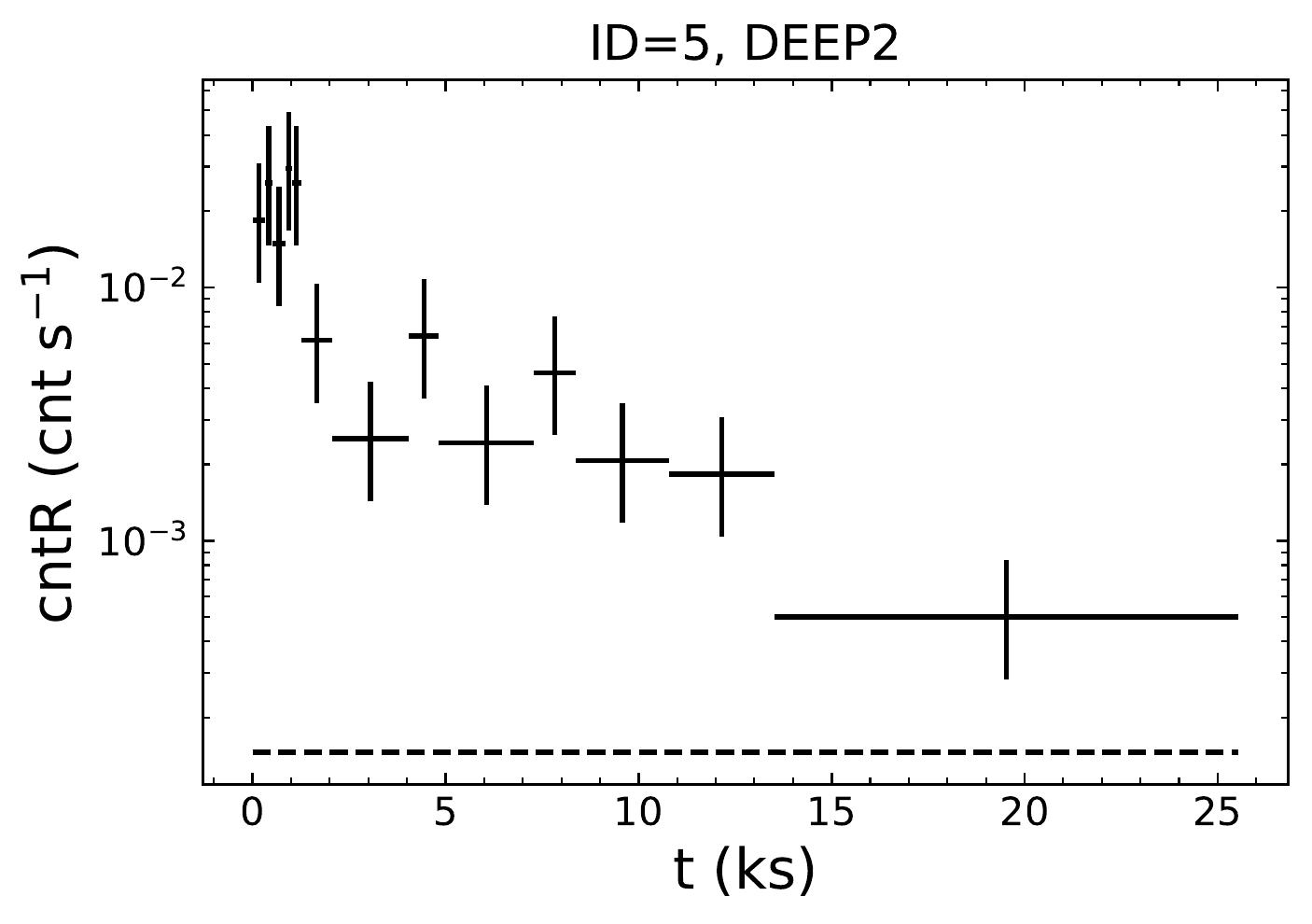}
\includegraphics[width=0.45\linewidth]{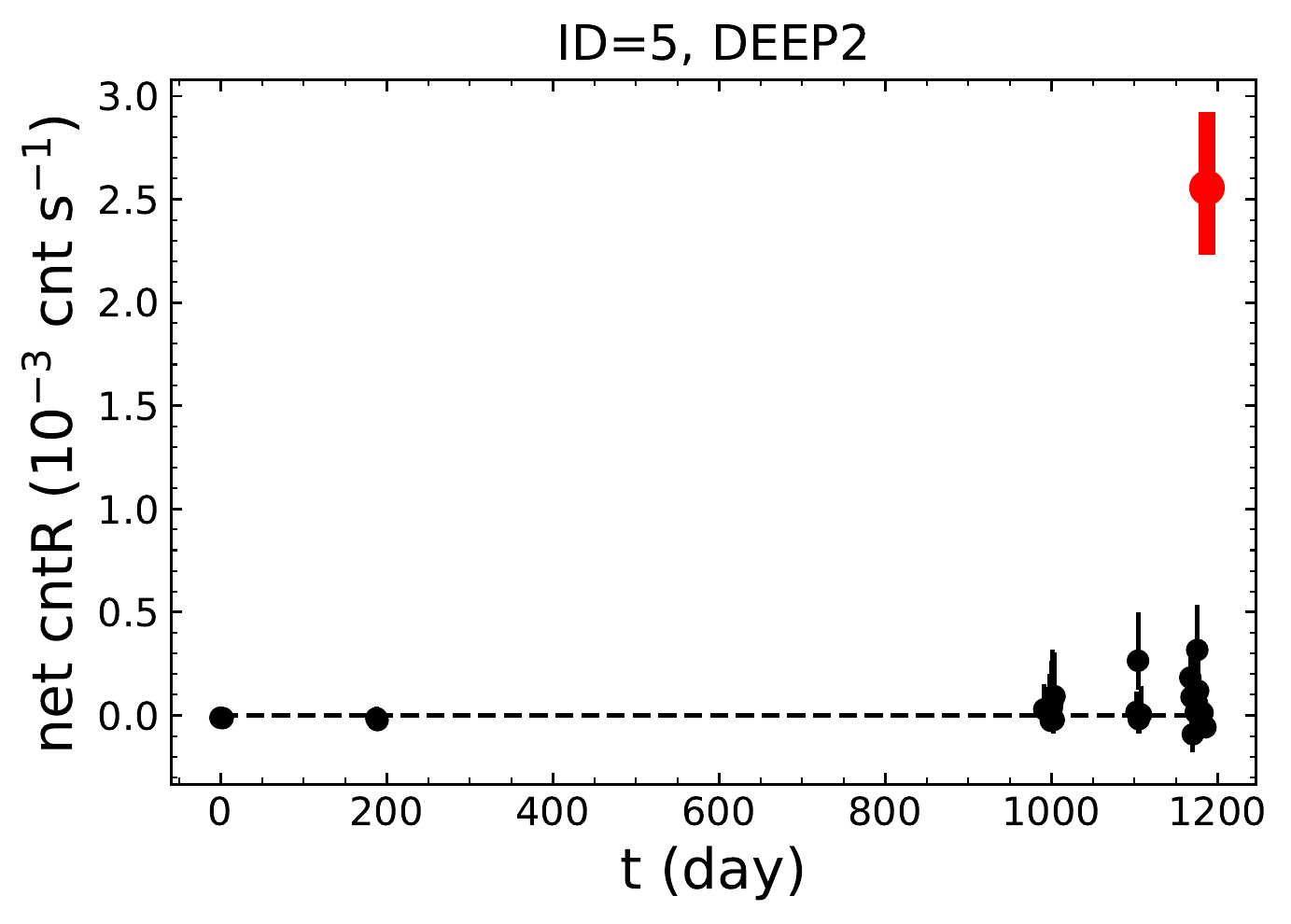} \\
\includegraphics[width=0.45\linewidth]{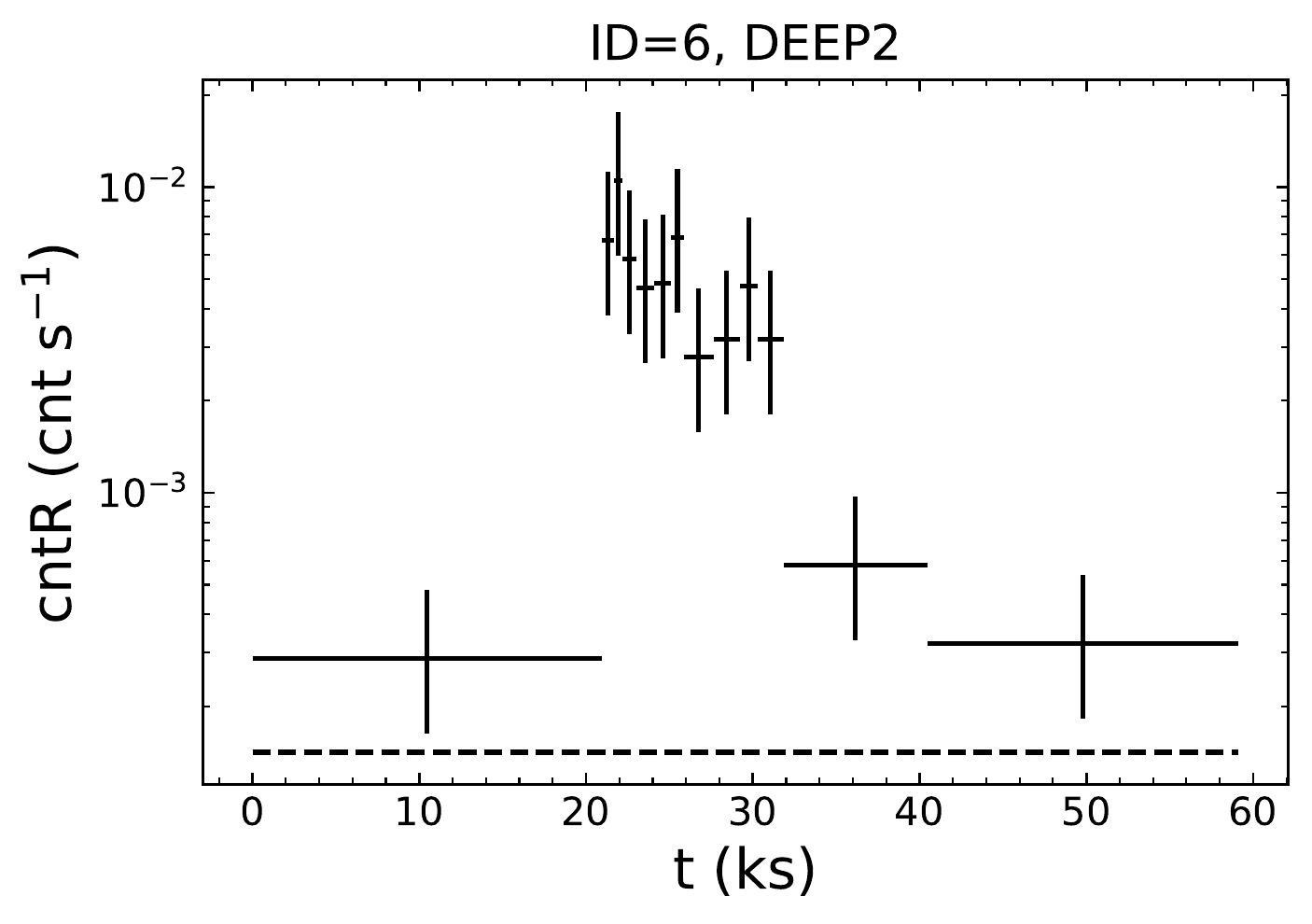}
\includegraphics[width=0.45\linewidth]{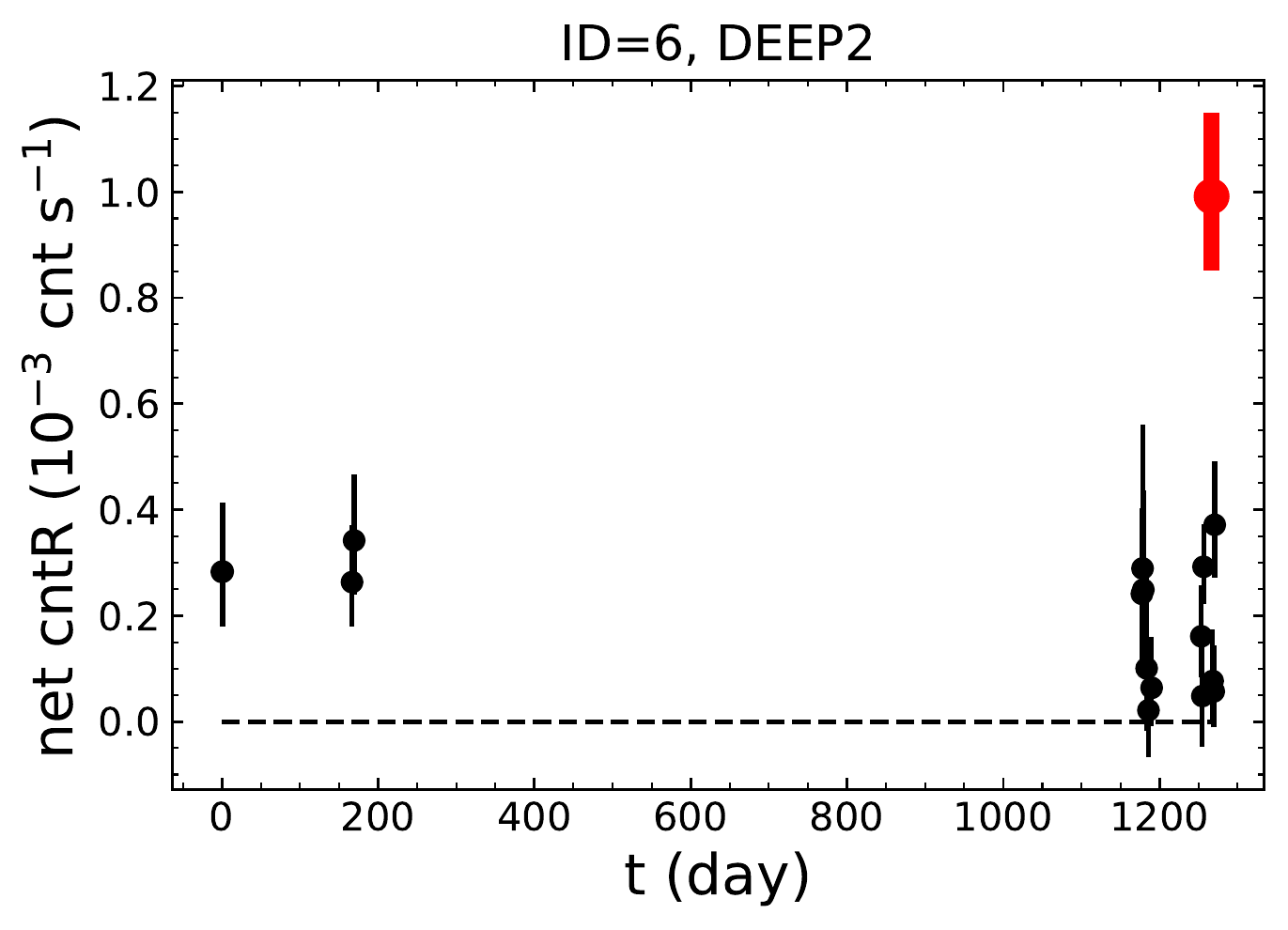} \\
\includegraphics[width=0.45\linewidth]{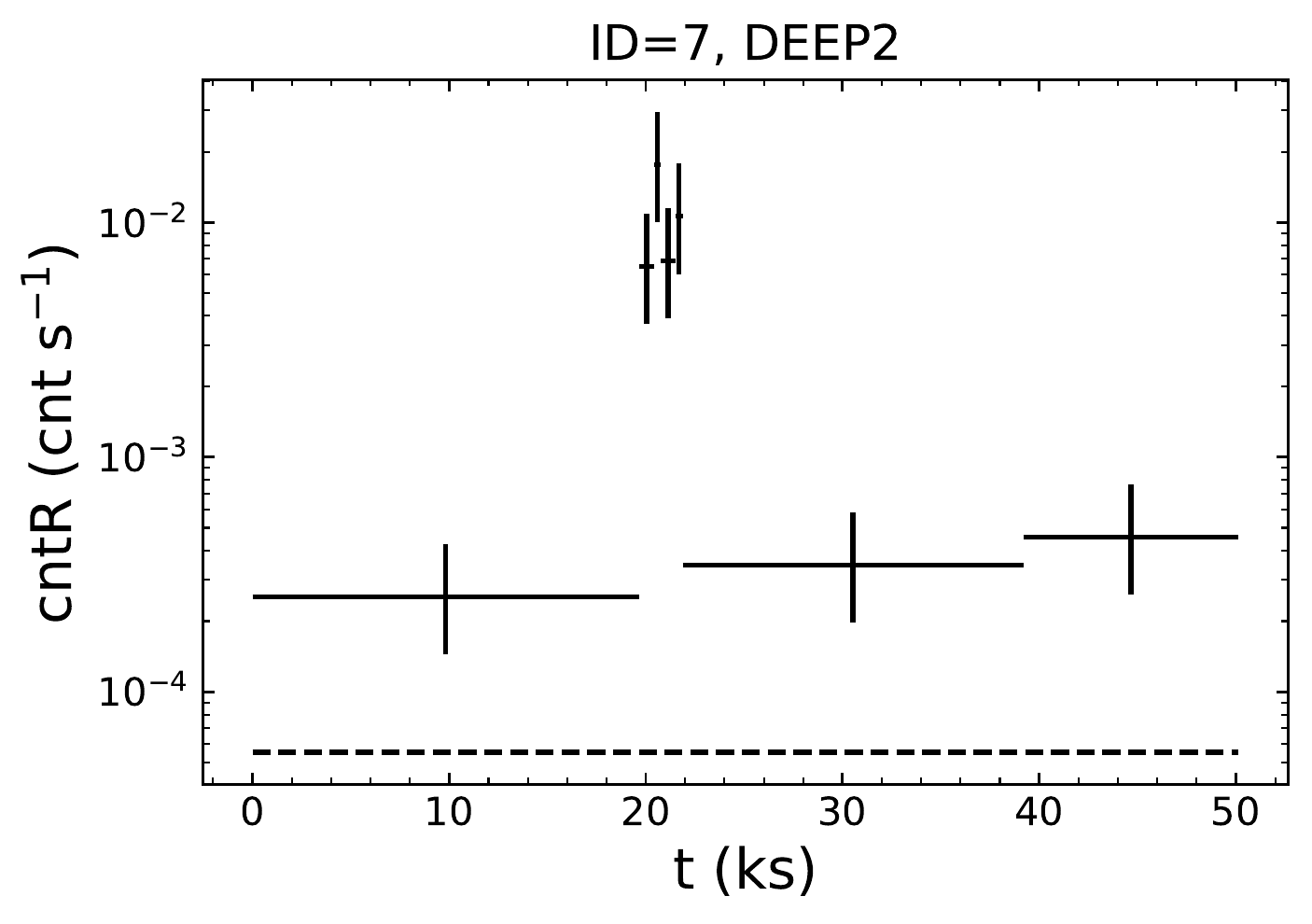}
\includegraphics[width=0.45\linewidth]{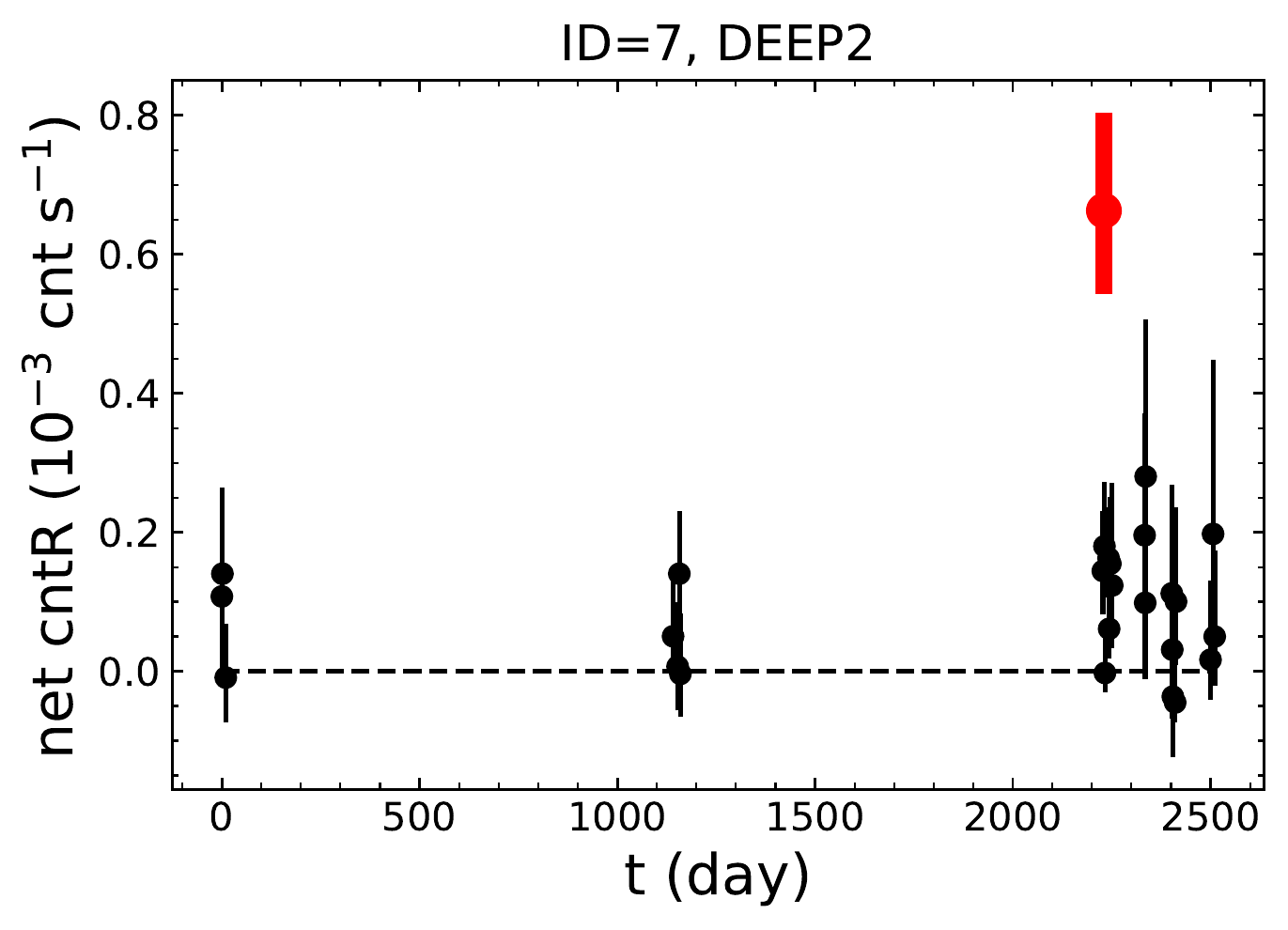} \\
\includegraphics[width=0.45\linewidth]{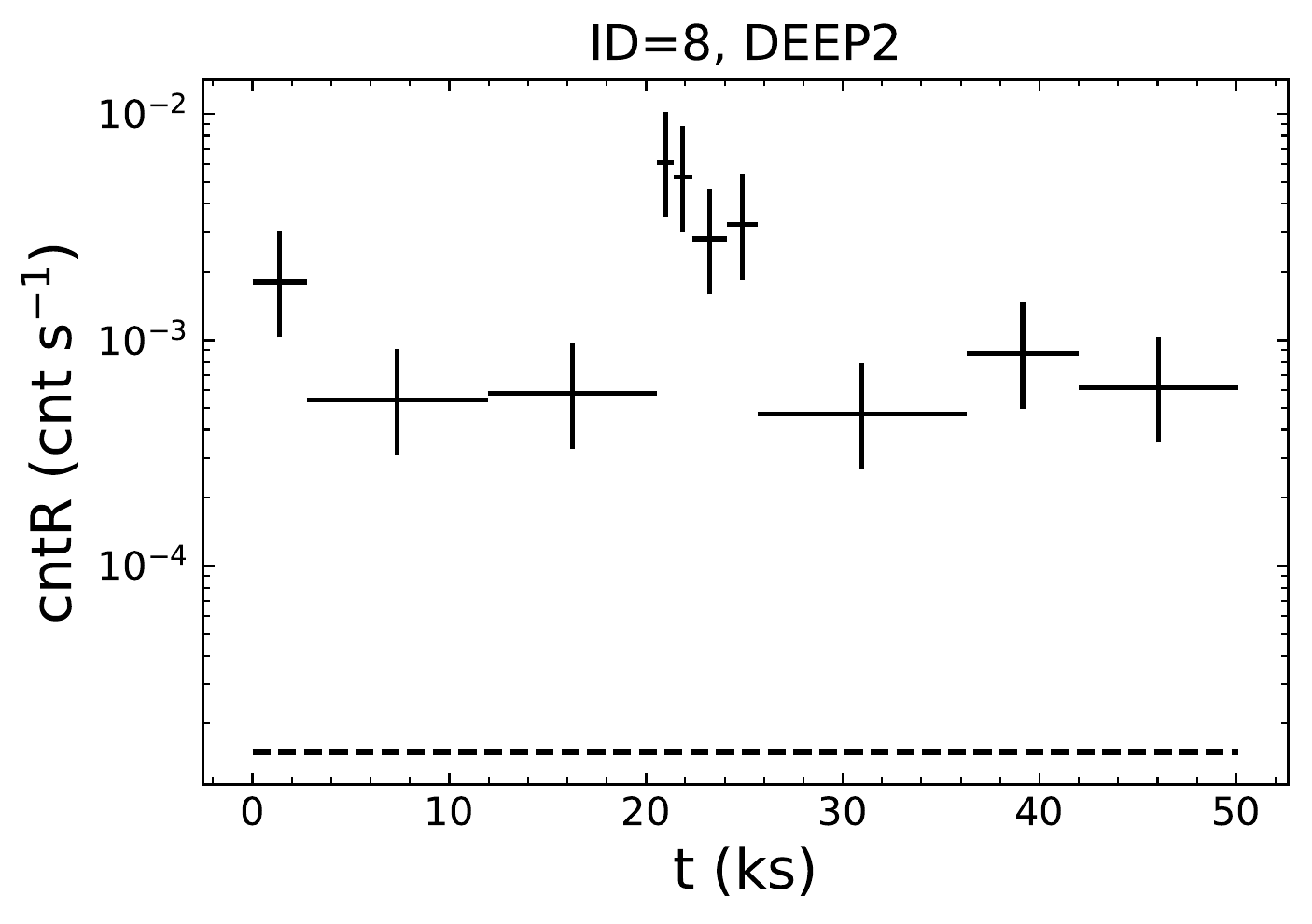}
\includegraphics[width=0.45\linewidth]{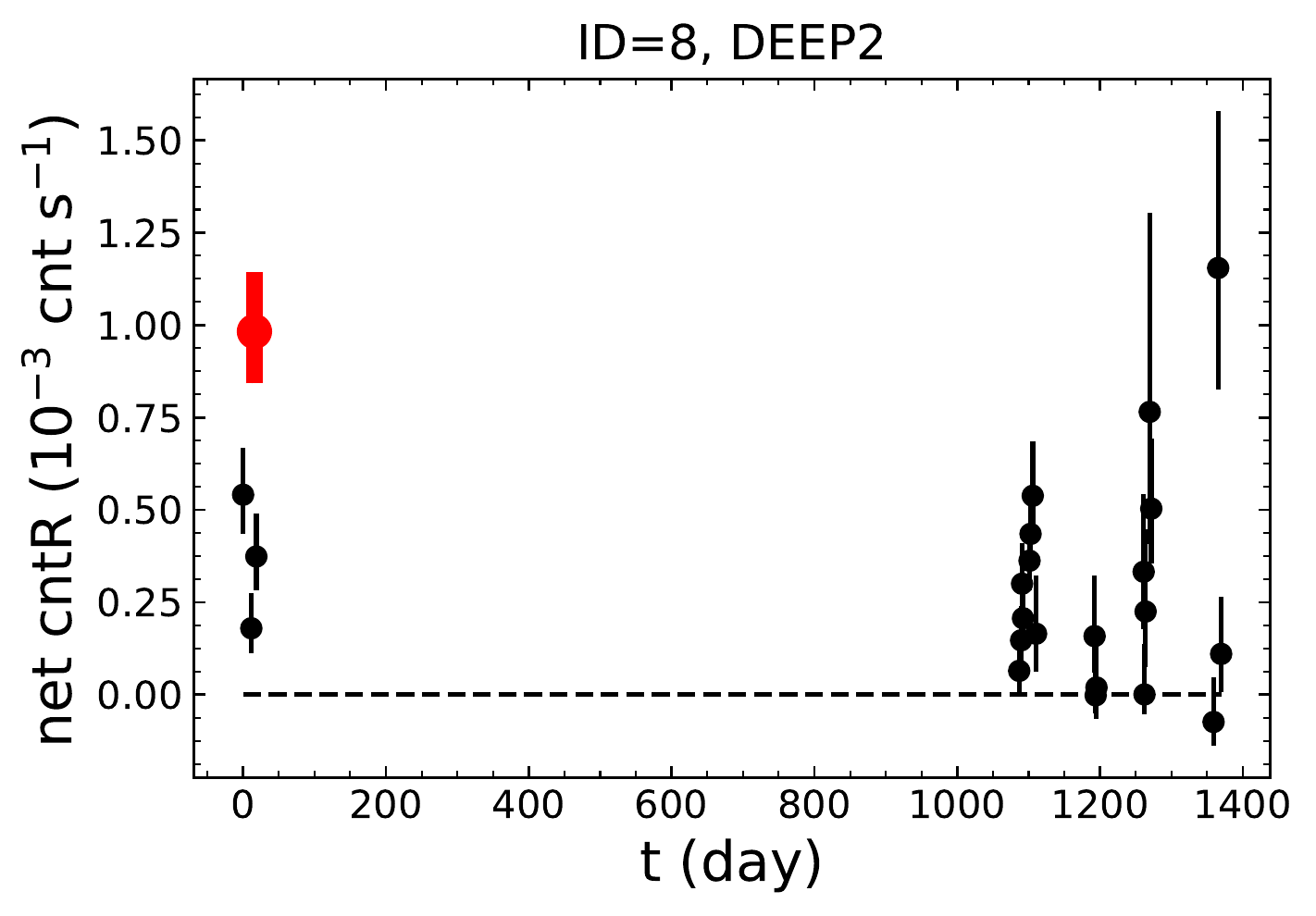} \\
\caption{}
\end{figure*}
\renewcommand{\thefigure}{\arabic{figure}}

\renewcommand{\thefigure}{\arabic{figure} (Continued)}
\addtocounter{figure}{-1}
\begin{figure*}
\includegraphics[width=0.45\linewidth]{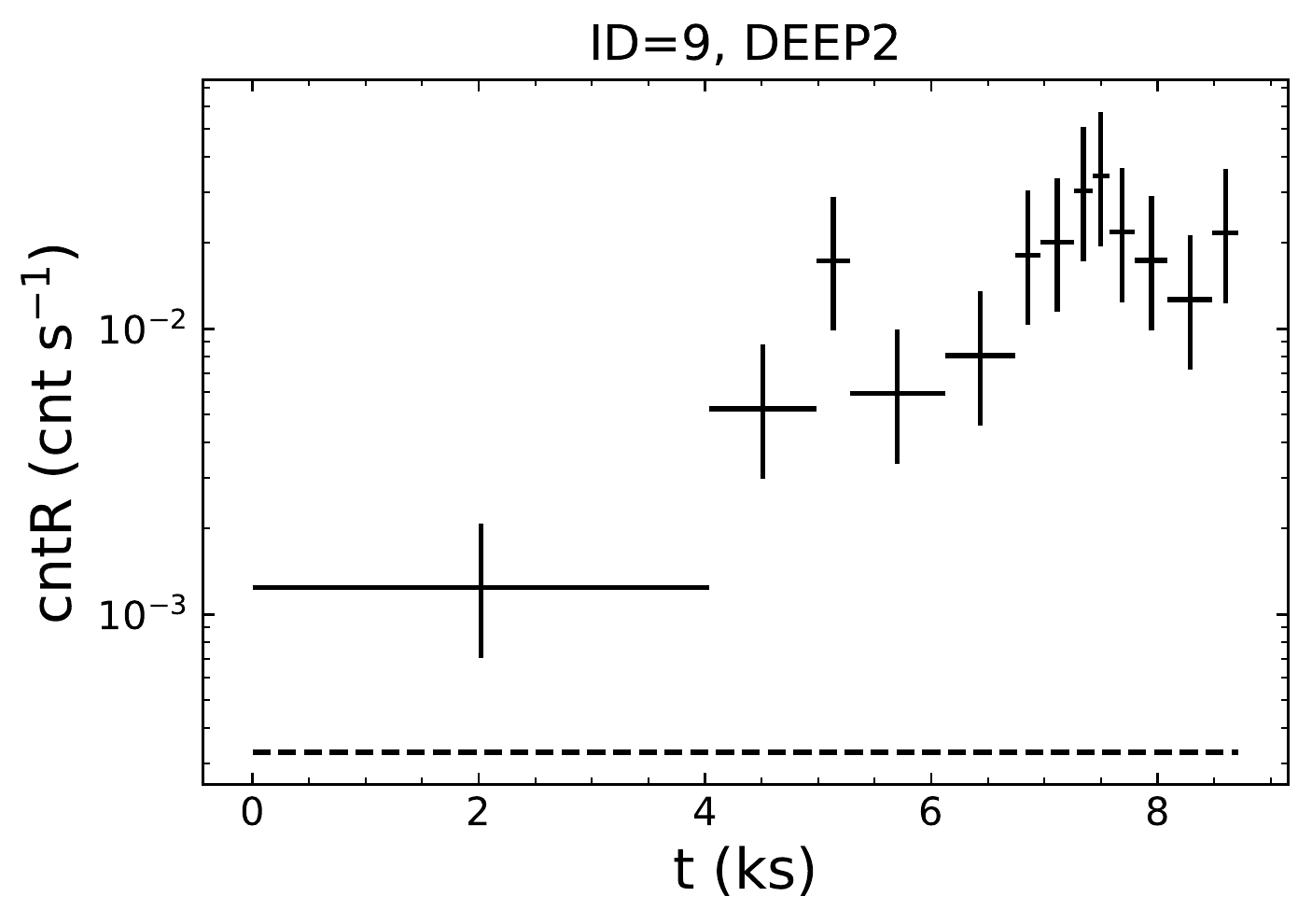}
\includegraphics[width=0.45\linewidth]{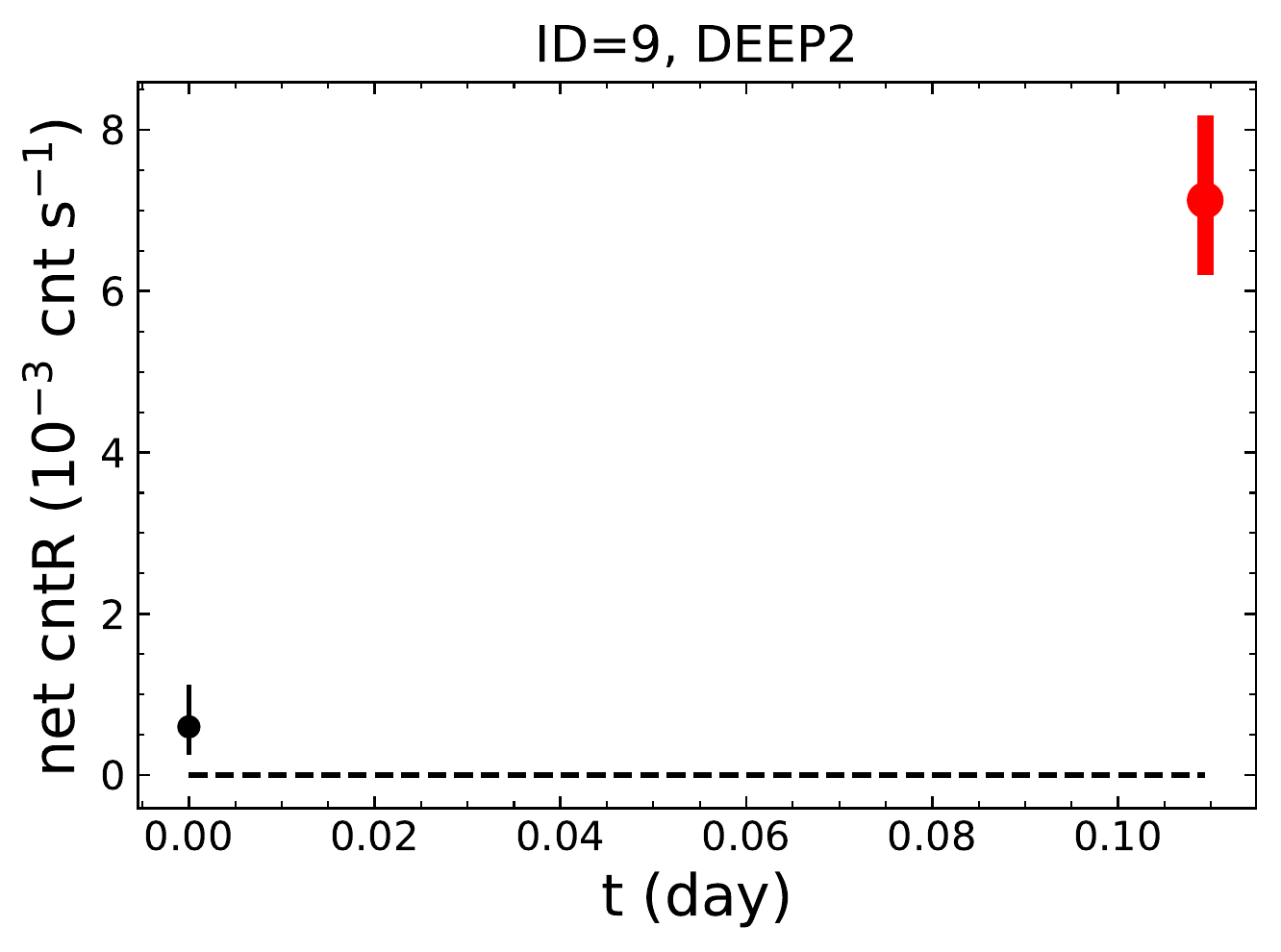} \\
\includegraphics[width=0.45\linewidth]{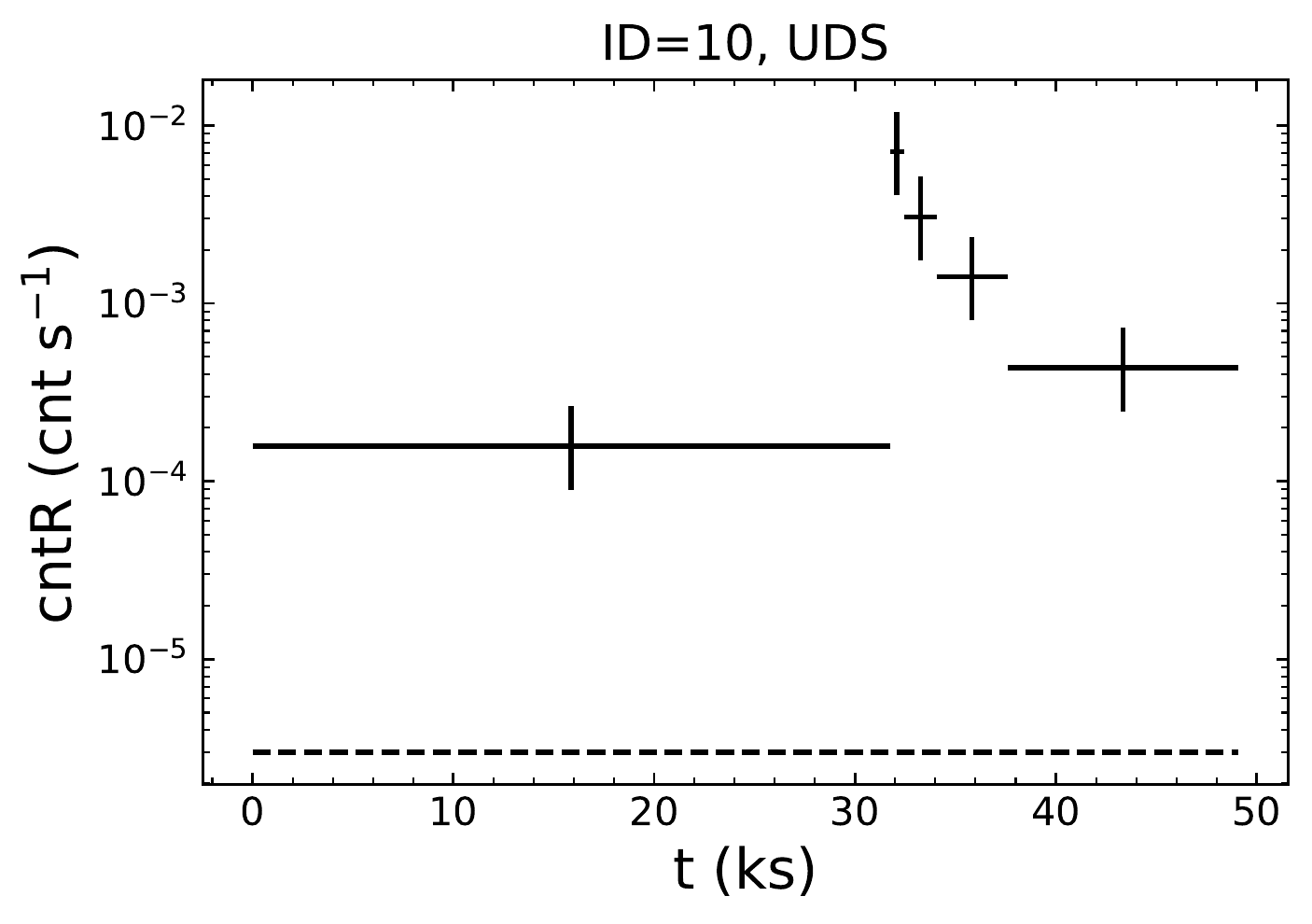}
\includegraphics[width=0.45\linewidth]{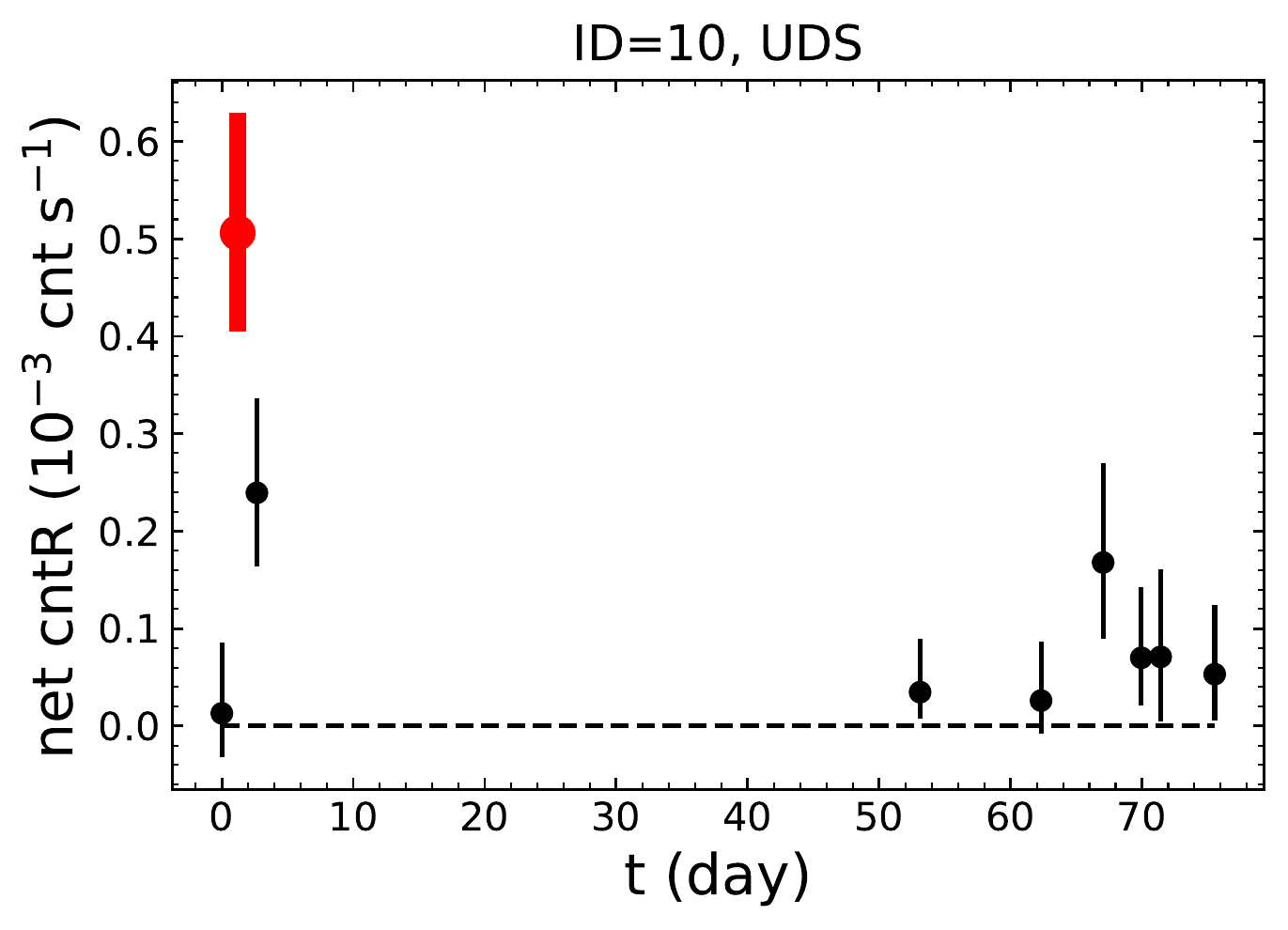} \\
\includegraphics[width=0.45\linewidth]{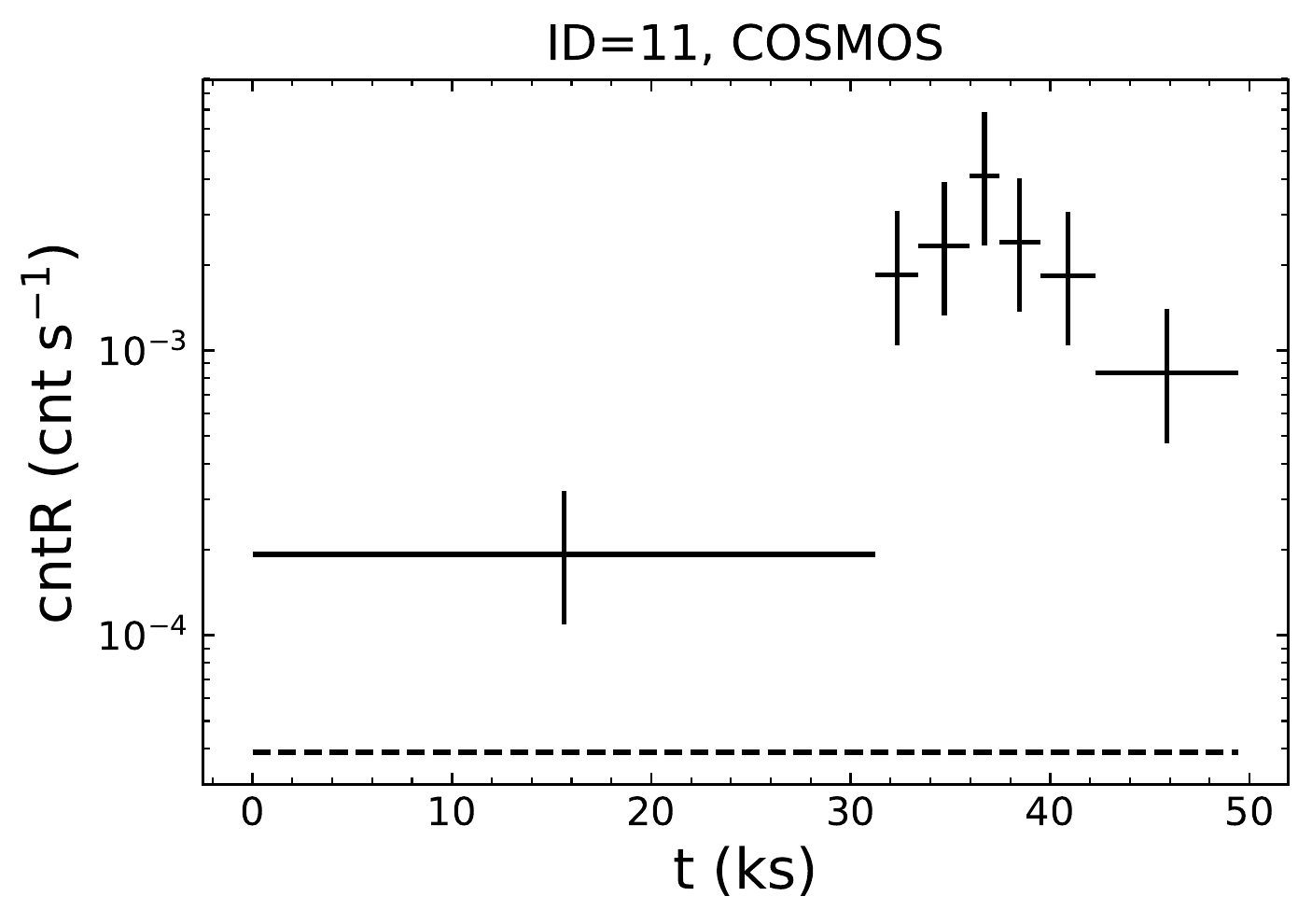}
\includegraphics[width=0.45\linewidth]{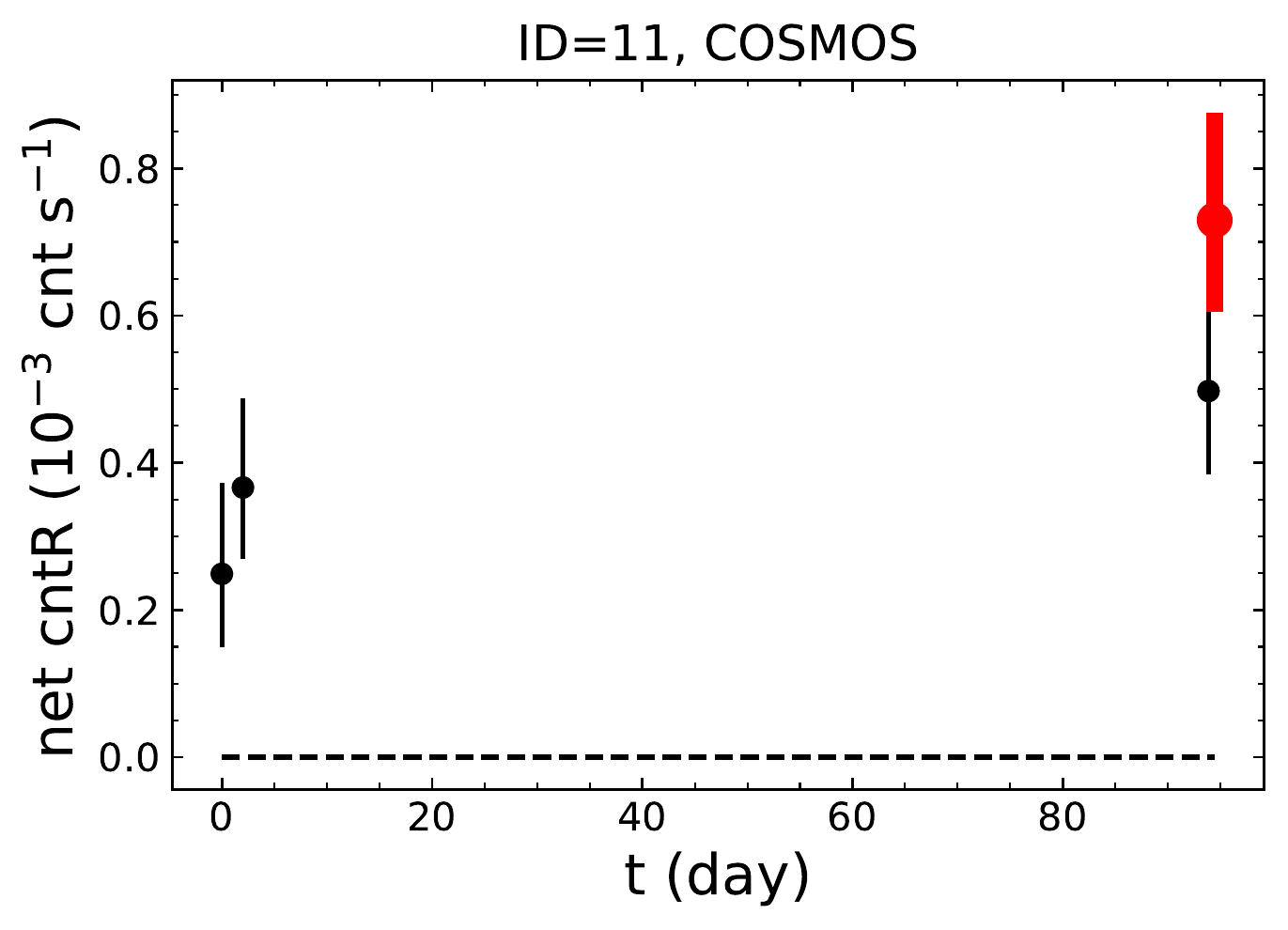} \\
\includegraphics[width=0.45\linewidth]{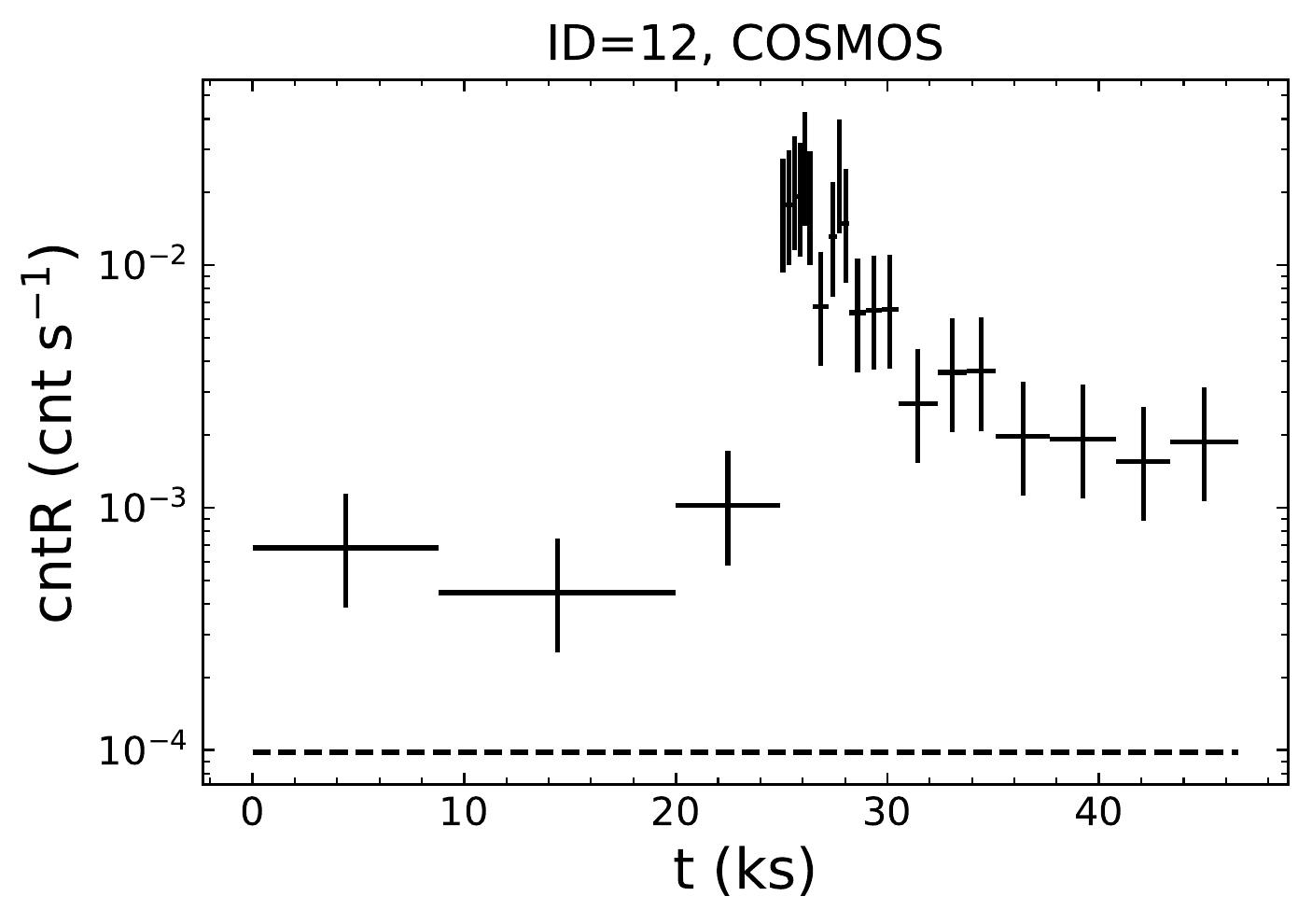}
\includegraphics[width=0.45\linewidth]{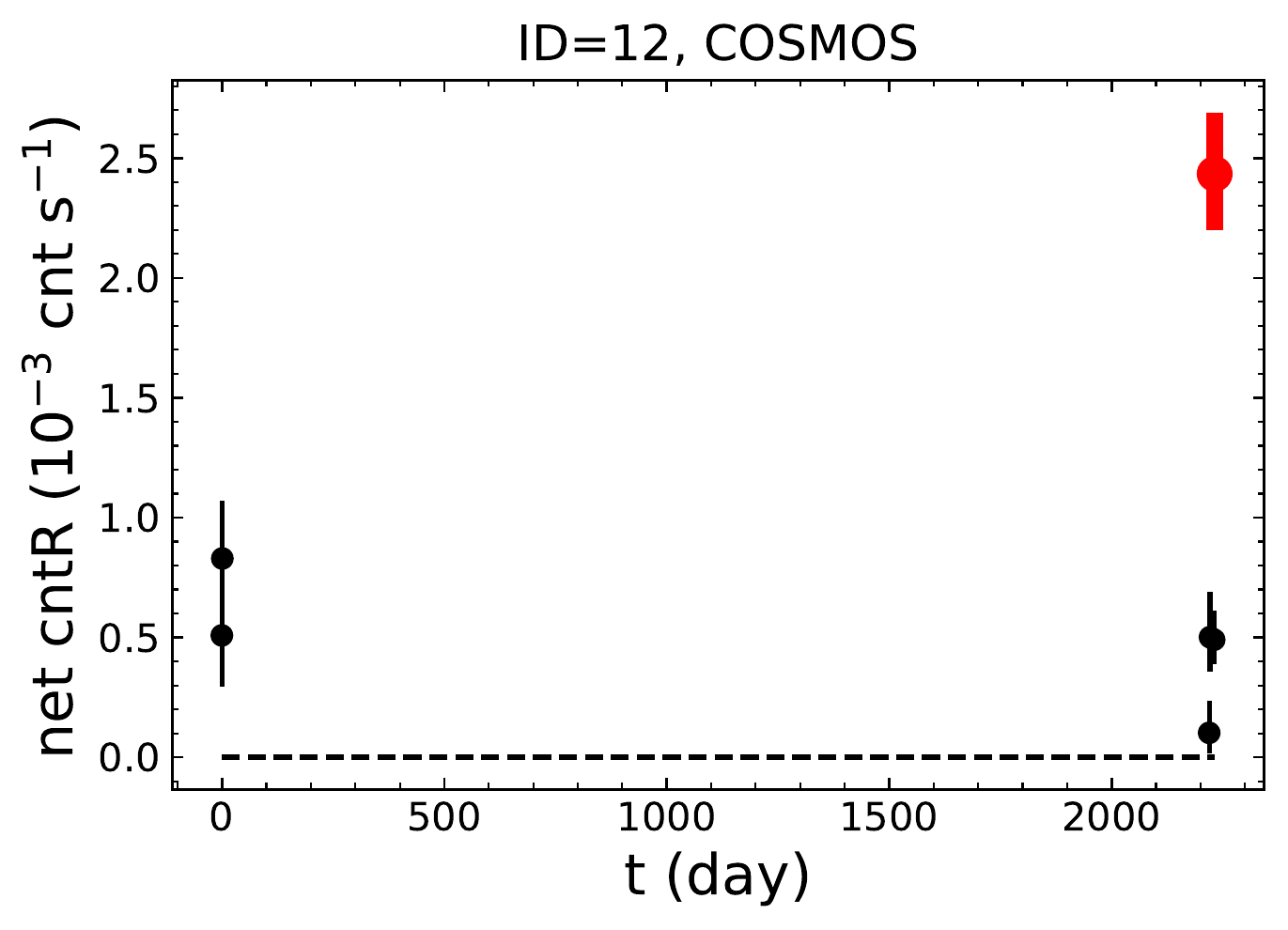} \\
\caption{}
\end{figure*}
\renewcommand{\thefigure}{\arabic{figure}}

\renewcommand{\thefigure}{\arabic{figure} (Continued)}
\addtocounter{figure}{-1}
\begin{figure*}
\includegraphics[width=0.45\linewidth]{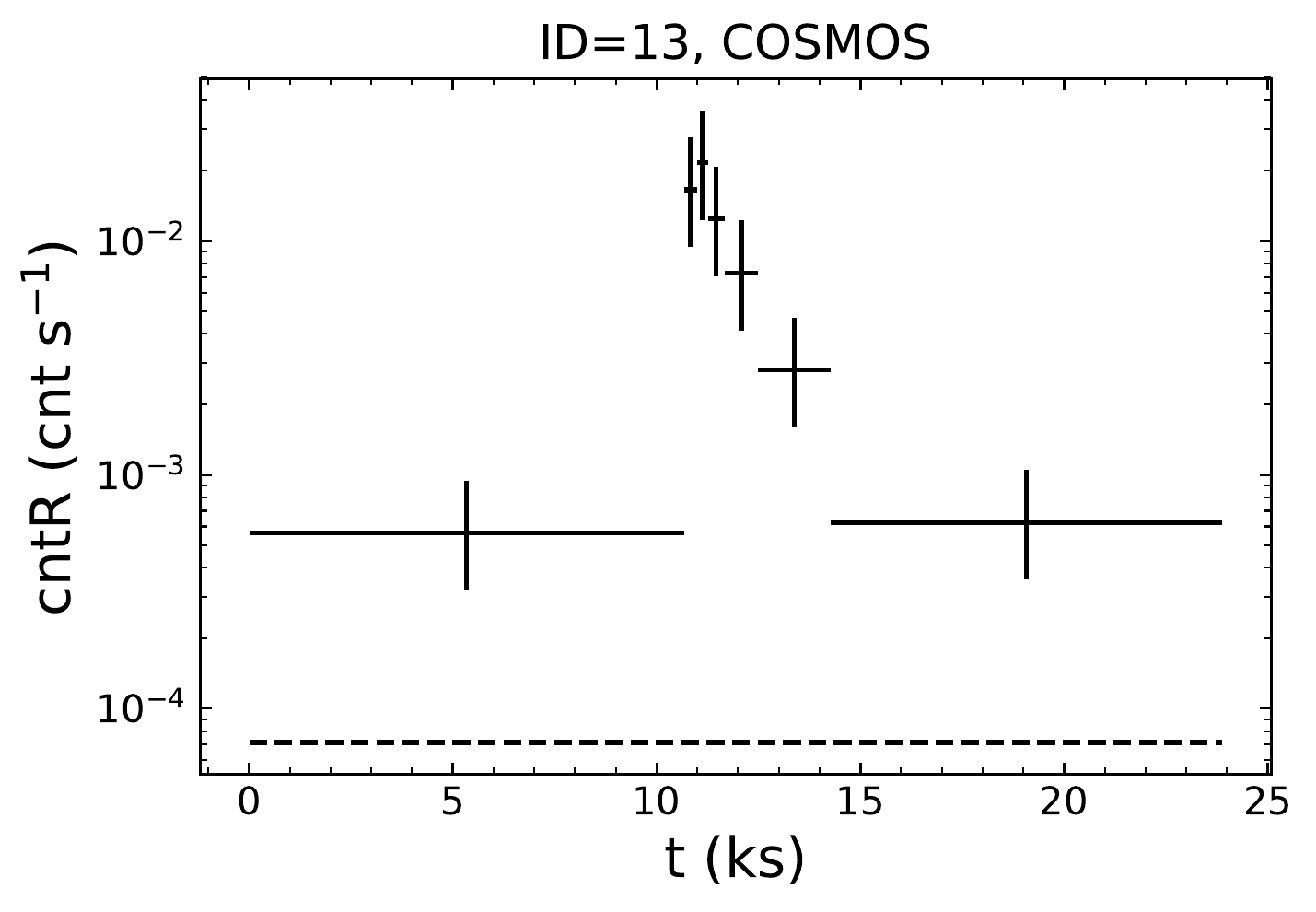}
\includegraphics[width=0.45\linewidth]{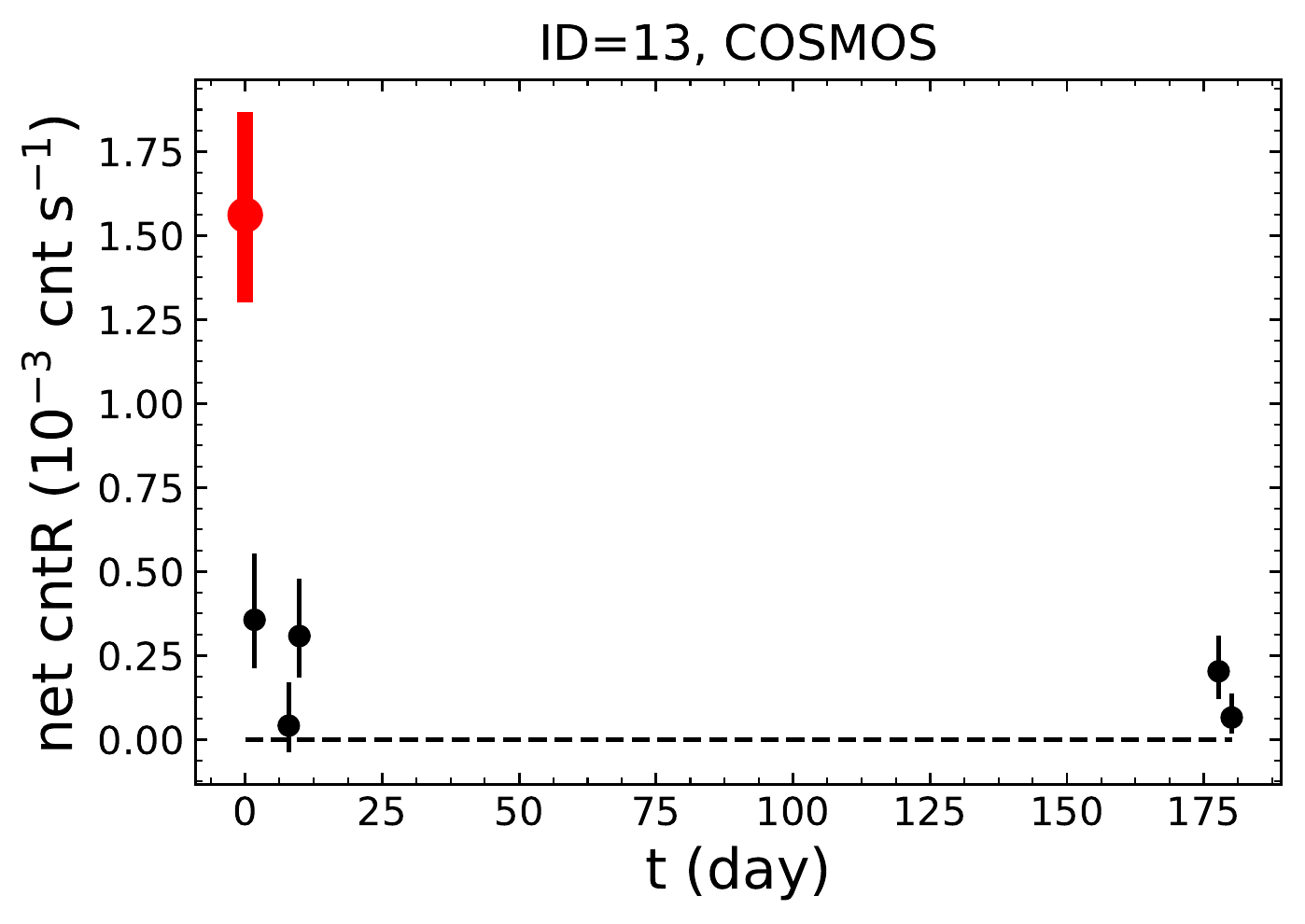} \\
\caption{}
\end{figure*}
\renewcommand{\thefigure}{\arabic{figure}}

\subsection{Optical/NIR Counterparts}
\label{sec:cp}
We have compiled the likelihood counterpart matching results from the 
survey catalogs (Table~\ref{tab:surv_prop}).
All the {transient candidates} have optical/NIR counterparts.
The counterpart properties are presented in Table~\ref{tab:cp_prop}.
We also match the counterparts with the \gaia\ catalog \citep{gaia18} 
using a 1$\arcsec$ matching radius, and mark the sources with non-zero 
parallax and/or proper motion as ``star'' in Table~\ref{tab:cp_prop}.

\begin{figure*}
\includegraphics[width=\linewidth]{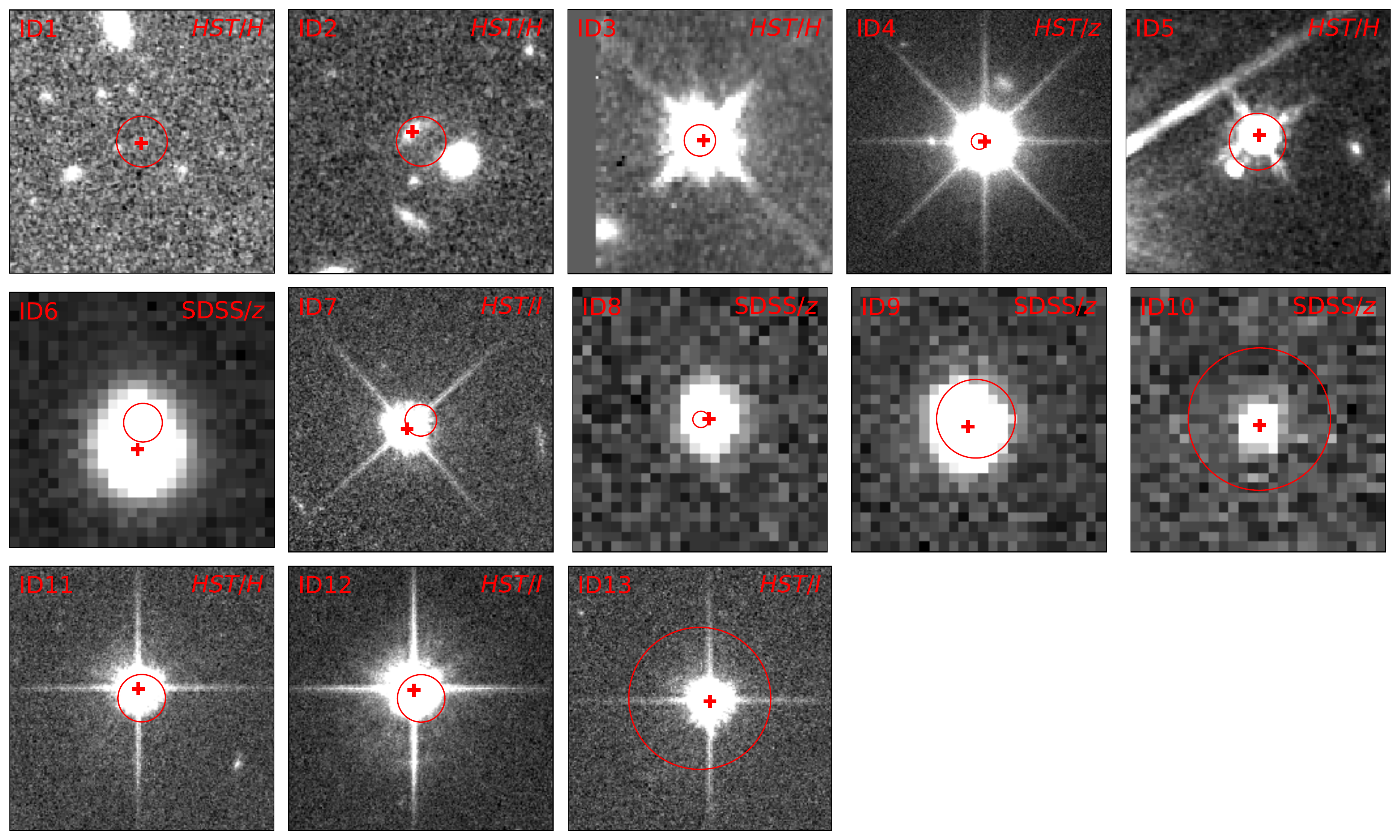}
\caption[]{Optical/NIR image $10\arcsec \times 10\arcsec$ cutouts of the transients.
Each cutout is centered at the \xray\ position. 
The central red circle denotes the \xray\ positional uncertainty, and has a radius 
$3\times \Delta X$, where $\Delta X$ is the $1\sigma$ \xray\ positional 
error listed in Table~\ref{tab:xray_prop}.
{The red cross marks the position of the optical/NIR counterpart 
(Table~\ref{tab:cp_prop}).}
The cutouts are from the \hst\ bands (as labelled) when available or the SDSS~$z$ band.
The \hst\ cutouts are from \cite{koekemoer07} and the \textit{Hubble} Legacy Archive 
(https://hla.stsci.edu/).
{The \xray\ and optical/NIR positions are generally consistent with each other.}
}
\label{fig:cutout}
\end{figure*}

We show the optical/IR image cutouts in Fig.~\ref{fig:cutout}.
{From Fig.~\ref{fig:cutout}, the optical positions\footnote{{The 
positional errors of the optical/NIR counterparts are not provided in the 
corresponding catalogs. 
Estimating the optical/NIR positional errors requires addressing factors
such as CCD saturation and seeing (for ground-based telescopes), 
which are beyond the scope of this work.}}
are within (or marginally outside, i.e. ID6 and ID7) the 
${3\sigma}$ \xray\ 
positional errors, indicating that the \xray\ and optical/NIR positions are 
generally consistent with each other.
For ID6 and ID7, in the image cutouts nearby the \xray\ positions, 
there appear to be no other optical/NIR sources except the counterparts, and 
thus the counterparts are likely the same physical objects as the \xray\ 
sources.
}

From Table~\ref{tab:cp_prop}, ID1 and ID2 (namely \cdfs\ XT1 and XT2) 
are likely of extragalactic origin and have already been discussed 
in detail (\hbox{\citealt{bauer17}}; \hbox{\citealt{xue19}}).
The other transients are relatively bright ($\mathrm{mag}_z < 20$), 
and all of them are reliably identified as stellar objects by optical/NIR 
spectroscopy and/or \gaia.  
Therefore, all the new {transient candidates} (aside from 
\cdfs\ XT1 and XT2) are stellar flares. 
{These stellar objects have different variability properties, e.g.
some have significant non-zero fluxes detected in the non-bursting observations 
(e.g. ID3 and ID4; see Fig.~\ref{fig:lc}) while others do not (e.g. ID5 
and ID9).
However, since the main scope of this paper is to study extragalactic 
transients similar to \cdfs\ XT1 and XT2, we do not further classify 
the stellar objects into, e.g. ``transient stars'' vs.\ ``variable stars''.
}

{
Since our algorithm is optimized for selecting \cdfs\ XT-like transients
(see \S\ref{sec:eff}), the fact that only two such transients are found 
indicates such events are relatively rare. 
We further estimate the \cdfs\ XT-like event rate in \S\ref{sec:evtR}.
The prevalence of stars among our transient candidates is likely because 
stellar flares are intrinsically more common than \cdfs\ XT-like extragalactic 
transients, and it does not necessarily indicate that our algorithm is more 
sensitive in selecting stellar flares.
There should be even more stellar flares in the survey data not identified 
by our algorithm, which is designed to select XT-like transients rather than
stellar flares. 
In fact, we have tested adjusting our algorithm slightly, and the resulting stellar 
sample changes while the extragalactic sample remains the same.
For example, if we chop the exposures to ${t_{\rm exp}<70}$~ks 
instead of ${t_{\rm exp}<50}$~ks (\S\ref{sec:alg}), \cdfs\ 
XT1 and XT2 will be still identified. 
However, this change will select 6 new stellar flares while missing 3 old 
stellar flares.
}


\begin{table*}
\centering
\caption{Counterpart Properties of {Transient Candidates}}
\label{tab:cp_prop}
\begin{tabular}{ccccccccc}
\hline\hline
ID & Source &  RAc & DECc & Offset & Mag$_z$ & $z$ & $z$~type & \gaia \\
(1) & (2) & (3) & (4) & (5) & (6) & (7) & (8) & (9) \\
\hline
1 & CANDELS &   53.16157 & $-27.85936$ & 0.07$\arcsec$ &   27.9 & 2.14 & phot    & n/a \\
2 & CANDELS &   53.07659 & $-27.87329$ & 0.50$\arcsec$ &   24.5 & 0.74 & spec    & n/a \\
3 & WIRCam &   189.02037 & $62.33728$ & 0.14$\arcsec$ &   16.8 & 0.00 & spec    & star \\
4 & CANDELS &   189.10575 & $62.23467$ & 0.21$\arcsec$ &   16.4 & 0.00 & spec    & star \\
5 & DEEP2-1 &   215.07411 & $53.10657$ & 0.26$\arcsec$ &   19.6 & 0.00 & spec    & n/a \\
6 & DEEP2-1 &   214.95966 & $52.74351$ & 1.10$\arcsec$ &   14.1 & n/a & n/a    & star \\
7 & DEEP2-1 &   214.61031 & $52.54338$ & 0.62$\arcsec$ &   17.1 & 0.00 & spec    & n/a \\
8 & DEEP2-1 &   214.66805 & $52.66666$ & 0.33$\arcsec$ &   16.8 & n/a & n/a    & star \\
9 & DEEP2-2 &   252.12746 & $34.96339$ & 0.45$\arcsec$ &   15.8 & 0.00 & spec    & star \\
10 & HSC &   34.48311 & $-5.09118$ & 0.25$\arcsec$ &   18.1 & 0.00 & spec    & star \\
11 & UltraVISTA &   149.75412 & $2.14183$ & 0.38$\arcsec$ &   16.7 & 0.00 & spec    & star \\
12 & UltraVISTA &   149.82649 & $2.71803$ & 0.41$\arcsec$ &   15.8 & 0.00 & spec    & star \\
13 & UltraVISTA &   149.99794 & $2.77960$ & 0.40$\arcsec$ &   16.5 & 0.00 & spec    & n/a \\
\hline
\end{tabular}
\begin{flushleft}
{\sc Note.} ---
(1) Transient ID in this work.
(2) Source of the counterpart: CANDELS \citep{grogin11, koekemoer11}, 
    WIRCam \citep{wang_w10}, DEEP2 \citep{coil04}, HSC \citep{aihara18}, and UltraVISTA 
    \citep{laigle16}.
(3) and (4) The position of the optical/NIR counterpart.
(5) The distance between the \xray\ position and the counterpart. 
(6) {${z}$-band AB magnitude of the counterpart. 
    For ID1 and ID2, the ${z}$-band filter refers to \hst\ F850LP; 
    for other sources, the filter refers to SDSS~${z}$.}
(7) and (8) redshift and its type. 
    ``0.00'' means stellar object. 
    ``n/a'' means redshift unavailable.
    $z=0.00$ and $z\ \rm{type}= \rm{phot}$ mean the source's SED prefers a stellar template rather than a 
    quasar/galaxy template.
    For ID7, we adopt the redshift from SDSS, since redshift information is not provided 
    in the \xray\ catalog \citep{goulding12}.
(9) \gaia\ classification. 
``star'' indicates the source has non-zero parallax and/or proper motion ($\rm S/N>5$) measured from \gaia;
otherwise, ``n/a'' is listed. 
\end{flushleft}
\end{table*}

\section{Event Rate and Future Prospects}\label{sec:evtR}
Our transient-search algorithm is able to find {\cdfs\ XT-like} transients 
with ${\log F_{\rm peak}\gtrsim -12.6}$ (cgs) 
effectively (\S\ref{sec:sim_res}).
{We remind that the limiting peak flux here is estimated for a typical 
off-axis angle of ${5'}$ (see \S\ref{sec:method}).
For an off-axis angle of ${0.5'}$ (nearly on-axis) and ${8'}$ 
(the maximum value accepted by our algorithm; \S\ref{sec:alg}), the limiting flux 
changes slightly (${\approx 0.1}$~dex; see Appendix~\ref{sec:oth_ang}).
}
However, we do not find any new extragalactic transients {that are similar 
to \cdfs\ XT1 and XT2}, despite searching \chandra\ observations totaling 
19~Ms exposure (\S\ref{sec:cp}).
{Based on this search result, we estimate the event rate of \cdfs\ XT-like
transients in \S\ref{sec:evtR_est}.
From the estimated event rate, we discuss the prospects of future missions 
(\athena\ and \ep) in detecting \cdfs\ XT-like transients.
}


\subsection{Event-Rate Estimation}\label{sec:evtR_est}
{
Since our simulations in \S\ref{sec:sim_res} show that the efficiency of our transient 
selection in short \chandra\ exposures (${t_{\rm exp} \lesssim 8}$~ks) is low, 
we do not include exposures shorter than 8~ks in when estimating the event rate below.
These short exposures only add up to 0.022~Ms of observation time in total, which is 
negligible compared to the total observation time analyzed (19.3~Ms).
}

{
For a set of \chandra\ observations, the expected number of transients brighter than
the flux limit (${\log F_{\rm peak}\gtrsim -12.6}$, cgs) can be written as
\begin{equation}\label{eq:evt1}
{
\mathcal{N} = \sum_{i} R_i \Omega_i t_i,
}
\end{equation}
where ${R_i}$ is the event rate; ${\Omega_i}$ and ${t_i}$ 
are the field of view (FOV) and exposure 
time, respectively; the subscript (${i}$) denotes different exposures.
In general, ${R_i}$ is a function of the sky coordinate of the telescope pointing.
However, considering that our focus is extragalactic transients and the Universe is largely 
isotropic, we assume that ${R_i}$ is a constant and denote it as ${R}$.
${\Omega_i}$ depends on the instrument used. 
All of our analyzed survey data are from \chandra/ACIS-I imaging observations, and 
thus ${\Omega_i}$ is a constant and we denote it as 
${\Omega}={\pi\times(8')^2=201}$~arcmin$^2$.
Eq.~\ref{eq:evt1} can then be simplified as 
\begin{equation}\label{eq:evt2}
{
\begin{split}
\mathcal{N} = \sum_{i} R \Omega t_i 
            = R \Omega \sum_{i} t_i,
\end{split}
}
\end{equation}
i.e. $\mathcal{N}$ only depends on the total exposure time of these observations.
In other words, it does not matter whether our analyzed 19~Ms of data are from a
single sky zone or multiple sky zones. 
From Eq.~\ref{eq:evt2}, the event rate $R$ can be calculated as 
\begin{equation}\label{eq:evt2}
{
\begin{split}
R &= \frac{\mathcal{N}}{\Omega \sum_{i} t_i}
\end{split}
}
\end{equation}
Based on the fact that 2 events are detected in 19~Ms of data, we estimate 
${R\approx 59^{+77}_{-38}\ \rm{evt\ yr^{-1} deg^{-2}}}$,
where the uncertainties are Poisson ${1\sigma}$ errors, calculated with 
the {\sc astropy.stats} package.
We stress that the event rate estimated throughout this paper refers to that of 
a particular type of transients (i.e. similar to \cdfs\ XT1 and XT2 with 
${\log F_{\rm peak}\gtrsim -12.6}$, cgs) rather than general 
extragalactic transients.
}

Given the event rate estimated above, we can estimate the number of \cdfs\ XT-like 
transients potentially existing in the \chandra\ archive.
As of March 2019, there are {95~Ms and 94~Ms} of ACIS-I and ACIS-S archival 
imaging observations {(excluding $< 8$~ks exposures)}
at Galactic latitudes of ${|b|>20^{\circ}}$.\footnote{Here, we do not consider the 7~Ms 
of observations performed by HRC, because the sensitivities and thereby flux limits of 
HRC and ACIS are different. 
{We also do not include ACIS subarray-mode observations to avoid complexity 
in the calculation of FOV. 
Such observations only contribute 1\% and 17\% of the exposure time for ACIS-I and 
ACIS-S, respectively.
Accounting for these observations is technically challenging, but would 
only affect our estimated transient number by a few percent at most.
}}
{ACIS-I and ACIS-S consist of 4 and 6 CCD chips, respectively.
For ACIS-I, all the chips are front-illuminated (FI); 
for ACIS-S, 4 and 2 chips are FI and back-illuminated (BI), respectively.
The BI chips have a slightly higher ($\approx$~10\%) flux-to-counts 
conversion factor than the FI chips.\footnote{{http://cxc.harvard.edu/proposer/POG/html/chap6.html}} 
The former have a higher background ($\approx$~2 times) than the 
latter, but still at a low level (only $\approx$~6 counts for a 50~ks 
exposure).
After considering these differences in conversion factor and background
in our simulations (\S\ref{sec:eff}), we find the flux limits of our 
transient detection are similar for FI and BI chips 
($\log F_{\rm lim} \approx -12.6$ for both).
Therefore, the differences between the FI and BI chips should not affect our 
estimation of the transient number in \chandra\ archival observations below.
}

As for ACIS-I, we only account for the CCD area with off-axis 
angle ${<8'}$ for ACIS-S, {which covers the S2, S3, and S4 CCD chips.}
{However, unlike the case for ACIS-I generally, 
ACIS-S may have some chips turned off 
during an observation (S3 is always on as it covers the aimpoint).
When one (S3), two (S3$+$S4 or S2$+$S3), and three (S2$+$S3$+$S4) relevant 
chips are on, the CCD areas are $\Omega_{\rm S, 1} \approx 69$~arcmin$^2$, 
$\Omega_{\rm S, 2} \approx 96$~arcmin$^2$, and 
$\Omega_{\rm S,3} \approx 123$~arcmin$^2$, respectively.
The total exposure times for the three cases are 
$T_{\rm S, 1} \approx 16$~Ms, $T_{\rm S, 1} \approx 25$~Ms, 
$T_{\rm S, 1} \approx 53$~Ms, respectively.
}
Therefore, we can estimate the total number of \cdfs\ XT-like transients in these 
archival observations as 
\begin{equation}
{
    \mathcal{N} = R(\Omega_{\rm I} T_{\rm I} +  
                    \Omega_{\rm S,1} T_{\rm S,1} + \Omega_{\rm S,2} T_{\rm S,2} + \
                    \Omega_{\rm S,2} T_{\rm S,3}) 
                = {15^{+20}_{-10}}, 
}
\end{equation}
where ${\Omega_{\rm I}}$ (${\Omega_{\rm S}}$) 
and ${T_{\rm I}}$ (${T_{\rm S}}$) are the 
FOV and total exposure time (${|b|>20^{\circ}}$) of ACIS-I (ACIS-S) in the \chandra\ archive.
We note that, at ${|b|>20^{\circ}}$, Galactic absorption is 
typically low, with column density of 
${N_{\rm H} \lesssim 10^{21}}$~cm${^{-2}}$ 
\citep[e.g.][]{stark92}, and such absorption only reduces the observed flux by 
${\lesssim 10\%}$ (estimated with {\sc pimms}).
Therefore, Galactic absorption is unlikely to significantly affect the estimated
number of transients above. 

We will perform an extensive \chandra\ archival search in a separate paper 
(Quirola V\'asquez et al.\ in prep.).
From our results (\S\ref{sec:cp}), the stellar objects found in archival 
data are likely to have bright optical/NIR counterparts ($z$-band magnitudes 
$\lesssim 20$), 
and thus their stellar nature can be largely determined with current wide-field 
surveys, e.g.\ SDSS, UKIDSS, and \textit{Gaia}.
In contrast, the counterparts of extragalactic transients will likely be faint in 
the optical/NIR, and follow-up observations with large ground-based telescopes will 
be helpful to study their properties such as redshift and host-galaxy stellar mass.
These counterparts may also be studied with future deep wide-field
surveys such as LSST \citep{ivezic19} and \textit{Euclid} \citep{euclid11}.
{\xmm\ has had a similar operational time as \chandra, 
and it notably has a larger effective area and FOV but also higher background than \chandra.}
Future work could also search \xmm\ archival data for \cdfs\ XT-like transients 
(e.g.\ the EXTraS project; \citealt{de_luca16}). 


\subsection{The Perspectives for Future Missions}\label{sec:future}
{Future \xray\ missions such as \athena\ and \ep\ should be able to discover 
a large number of extragalactic transients similar to \cdfs\ XT1 and XT2. 
Now, we estimate the sample sizes of transients that will be potentially detected 
by \athena\ and \ep.
As a first-order approximation, we assume that the event-rate density 
(event rate per dex of flux) is a power-law function, i.e.
\begin{equation}\label{eq:dN}
{
    \frac{dR}{d\log F_{\rm peak}} \propto F_{\rm peak}^{-\gamma}
}
\end{equation}
Here, the power-law index (${\gamma}$) is positive, because otherwise the 
event rate above a given ${F_{\rm peak}}$ would be divergent.
By integrating Eq.~\ref{eq:dN} from $\log F_{\rm lim}$ (limiting peak flux of 
the mission)\footnote{{Here, we integrate from $\log F_{\rm lim}$ 
rather than $-12.6$ (the limiting flux of \chandra; \S\ref{sec:sim_res}).
This is because, in this Section, our goal is to estimate the number of 
XT-like transients detectable by future missions (i.e. sources with 
$F_{\rm peak}$ above $F_{\rm lim}$ of these missions).
Therefore, the integration lower limit should be $F_{\rm lim}$ of the mission 
of interest.
}} 
to ${\infty}$ and applying Eq.~\ref{eq:evt1}, we can estimate 
the number of detected \cdfs\ XT-like transients as
\begin{equation}\label{eq:N}
{
\begin{split}
\mathcal{N} &\propto F_{\rm lim}^{-\gamma} \Omega T \\
    &\propto F_{\rm lim}^{-\gamma+1} \frac{\Omega}{F_{\rm lim}} T \\
    &\propto A^{\gamma-1} (\Omega A)  T \\ 
    &\propto A^{\gamma-1} G T, \\ 
\end{split}
}
\end{equation}
where ${A}$ and ${G}$ are the effective area and grasp 
(defined as ${\Omega \times A}$) of the mission.
In Eq.~\ref{eq:N}, we adopt the approximation of ${F_{\rm lim} \propto A^{-1}}$.
If further assuming ${\gamma = 1}$ and ${T}$ is similar for different missions, 
we have ${N \propto G}$.
Since both \athena\ and \ep\ have ${G}$ values ${\sim 200}$~times larger than 
that of \chandra\ (e.g. \hbox{\citealt{nandra13}}; \hbox{\citealt{burrows18}}; \hbox{\citealt{yuan18}}) 
which can detect ${\sim 15}$ transients (see above), we expect that 
\athena\ and \ep\ will each detect ${\sim 3000}$ sources if they operate for 
${\approx 20}$~years. 
These samples will be sufficiently large for detailed sample studies.
Note that the estimated sample sizes depend on the assumption that ${\gamma =1}$.
If ${\gamma >1}$, \athena\ (\ep) will detect more (fewer) transients; 
if ${0<\gamma <1}$, the situation is the opposite. 
}

{Our estimation above is based on the assumption of a power-law 
function of event-rate density (Eq.~\ref{eq:dN}) with $\gamma =1$.
A natural prediction of this power-law function is that there are more 
faint sources than bright sources in general.
One might be concerned that this prediction contradicts our results, i.e.
the 19~Ms of \chandra\ data only contains two relatively bright sources 
(XT1 and XT2, both having $\log F_{\rm peak} > -12.2$; see 
Table~\ref{tab:xray_prop}) but no fainter sources.
We now test whether this apparent inconsistency is statistically 
significant or not.
Assuming there are two transients above the \chandra\ flux limit 
($\log F_{\rm lim}=-12.6$) detected in the 19~Ms of data, we estimate the 
chance for these two both to have $\log F_{\rm peak} > -12.2$.
From Eq.~\ref{eq:dN} ($\gamma =1$), the probability for one detected 
transient to be bright ($\log F_{\rm peak} > -12.2$) is
\begin{equation}\label{eq:P_one}
\begin{split}
    P_\mathrm{bright} &= \frac{\int_{-12.2}^{\infty} F_\mathrm{peak}^{-1} d \log F_\mathrm{peak}}
                          {\int_{-12.6}^{\infty} F_\mathrm{peak}^{-1} d \log F_\mathrm{peak}} \\
                   &= 0.40.
\end{split}
\end{equation}
Then, according to the binomial distribution, the probability ($p$-value) 
for both sources to be bright ($\log F_{\rm peak} > -12.2$) is 
$0.40^2=0.16$, only corresponding to $1.4\sigma$ significance.
Therefore, the assumption of Eq.~\ref{eq:dN} ($\gamma =1$) does not 
contradict our results significantly.
Actually, we find Eq.~\ref{eq:dN} is always consistent with our 
results at a $3\sigma$ level, as long as $0<\gamma<3.1$.
}



%

\section{Summary}\label{sec:sum}
We have performed a systematic search for {\cdfs\ XT-like} extragalactic transients 
in 19~Ms of \chandra\ surveys, including \cdfs, \cdfn, DEEP2, UDS, COSMOS, and \hbox{E-CDF-S}.
Our main results are summarized below.
\begin{enumerate}[wide=0pt, widest=99, leftmargin=\parindent, labelsep=*]

\item We developed a method to select transients within a \chandra\ observation 
(\S\ref{sec:method}).
From simulations, we show that our method is efficient in discovering transients
with \hbox{0.5\text{--}7 keV} peak flux
$\log F_{\rm peak}\gtrsim -12.6$ (erg~cm$^{-2}$~s$^{-1}$). 

\item Our selection yields {13 transient candidates} (\S\ref{sec:analyses}), 
including \cdfs\ XT1 and XT2 which have been reported in previous works 
(\hbox{\citealt{bauer17}}; \hbox{\citealt{xue19}}).
All the {candidates} have optical/NIR counterparts (\S\ref{sec:cp}). 
Except for \cdfs\ XT1 and XT2, all other sources are stellar objects. 

\item The lack of new {\cdfs\ XT-like} transients in our search indicates that 
such objects are rare (\S\ref{sec:evtR}). 
We estimate an event rate of ${59^{+77}_{-38}}\ \rm{evt\ yr^{-1} deg^{-2}}$, 
corresponding to a total of ${15^{+20}_{-10}}$ events in \chandra\ archival 
observations at $|b|>20^\circ$.
Future \xray\ missions such as \athena\ and the \ep\  with 
large {grasps} might be able to find {thousands} of extragalactic 
transients, and sample studies will be feasible then.

\end{enumerate}


\section*{Acknowledgements}
We thank the referee for helpful feedback that improved this work.
We thank David Burrows, Qingling Ni, John Timlin, and Fabio Vito 
for helpful discussions. 
GY, WNB, and SFZ acknowledge support from CXC grant AR8-19016X, 
CXC grant AR8-19011X, and NASA ADP grant 80NSSC18K0878.
FEB acknowledges support from CONICYT-Chile (Basal AFB-170002, FONDO ALMA 31160033) 
and the Ministry of Economy, Development, and Tourism's Millennium Science Initiative 
through grant IC120009, awarded to The Millennium Institute of Astrophysics, MAS.
YQX acknowledges support from the 973 Program (2015CB857004), NSFC 
(11890693, 11421303), and the CAS Frontier Science Key Research 
Program (QYZDJ-SSW-SLH006).
The Guaranteed Time Observations (GTO) for the CDF-N included here were
selected by the ACIS Instrument Principal Investigator, Gordon P.
Garmire, currently of the Huntingdon Institute for X-ray Astronomy, LLC,
which is under contract to the Smithsonian Astrophysical Observatory;
Contract SV2-82024.
This project uses Astropy (a Python package; see 
\citealt{astropy}).




\bibliographystyle{mnras}
\bibliography{all.bib} 

\begin{thebibliography}{}
\makeatletter
\relax
\def\mn@urlcharsother{\let\do\@makeother \do\$\do\&\do\#\do\^\do\_\do\%\do\~}
\def\mn@doi{\begingroup\mn@urlcharsother \@ifnextchar [ {\mn@doi@}
  {\mn@doi@[]}}
\def\mn@doi@[#1]#2{\def\@tempa{#1}\ifx\@tempa\@empty \href
  {http://dx.doi.org/#2} {doi:#2}\else \href {http://dx.doi.org/#2} {#1}\fi
  \endgroup}
\def\mn@eprint#1#2{\mn@eprint@#1:#2::\@nil}
\def\mn@eprint@arXiv#1{\href {http://arxiv.org/abs/#1} {{\tt arXiv:#1}}}
\def\mn@eprint@dblp#1{\href {http://dblp.uni-trier.de/rec/bibtex/#1.xml}
  {dblp:#1}}
\def\mn@eprint@#1:#2:#3:#4\@nil{\def\@tempa {#1}\def\@tempb {#2}\def\@tempc
  {#3}\ifx \@tempc \@empty \let \@tempc \@tempb \let \@tempb \@tempa \fi \ifx
  \@tempb \@empty \def\@tempb {arXiv}\fi \@ifundefined
  {mn@eprint@\@tempb}{\@tempb:\@tempc}{\expandafter \expandafter \csname
  mn@eprint@\@tempb\endcsname \expandafter{\@tempc}}}

\bibitem[\protect\citeauthoryear{{Aihara} et~al.,}{{Aihara}
  et~al.}{2018}]{aihara18}
{Aihara} H.,  et~al., 2018, \mn@doi [\pasj] {10.1093/pasj/psx081}, \href
  {http://ads.nao.ac.jp/abs/2018PASJ...70S...8A} {70, S8}

\bibitem[\protect\citeauthoryear{{Astropy Collaboration} et~al.,}{{Astropy
  Collaboration} et~al.}{2018}]{astropy}
{Astropy Collaboration} et~al., 2018, preprint, \href
  {http://ui.adsabs.harvard.edu/abs/2018arXiv180102634T} {} (\mn@eprint {arXiv}
  {1801.02634})

\bibitem[\protect\citeauthoryear{{Bauer} et~al.,}{{Bauer}
  et~al.}{2017}]{bauer17}
{Bauer} F.~E.,  et~al., 2017, \mn@doi [\mnras] {10.1093/mnras/stx417}, \href
  {http://ui.adsabs.harvard.edu/abs/2017MNRAS.467.4841B} {467, 4841}

\bibitem[\protect\citeauthoryear{{Belloni} \& {Stella}}{{Belloni} \&
  {Stella}}{2014}]{belloni14}
{Belloni} T.~M.,  {Stella} L.,  2014, \mn@doi [\ssr]
  {10.1007/s11214-014-0076-0}, \href
  {https://ui.adsabs.harvard.edu/#abs/2014SSRv..183...43B} {183, 43}

\bibitem[\protect\citeauthoryear{{Brandt} \& {Alexander}}{{Brandt} \&
  {Alexander}}{2015}]{brandt15}
{Brandt} W.~N.,  {Alexander} D.~M.,  2015, \mn@doi [\aapr]
  {10.1007/s00159-014-0081-z}, \href
  {http://ui.adsabs.harvard.edu/abs/2015A%26ARv..23....1B} {23, 1}

\bibitem[\protect\citeauthoryear{{Burrows} et~al.,}{{Burrows}
  et~al.}{2018}]{burrows18}
{Burrows} D.~N.,  et~al., 2018, in Space Telescopes and Instrumentation 2018:
  Ultraviolet to Gamma Ray. p. 106991J (\mn@eprint {arXiv} {1808.02883}),
  \mn@doi{10.1117/12.2312785}

\bibitem[\protect\citeauthoryear{{Civano} et~al.,}{{Civano}
  et~al.}{2016}]{civano16}
{Civano} F.,  et~al., 2016, \mn@doi [\apj] {10.3847/0004-637X/819/1/62}, \href
  {http://ui.adsabs.harvard.edu/abs/2016ApJ...819...62C} {819, 62}

\bibitem[\protect\citeauthoryear{{Coil}, {Newman}, {Kaiser}, {Davis}, {Ma},
  {Kocevski}  \& {Koo}}{{Coil} et~al.}{2004}]{coil04}
{Coil} A.~L.,  {Newman} J.~A.,  {Kaiser} N.,  {Davis} M.,  {Ma} C.-P.,
  {Kocevski} D.~D.,   {Koo} D.~C.,  2004, \mn@doi [\apj] {10.1086/425676},
  \href {http://ui.adsabs.harvard.edu/abs/2004ApJ...617..765C} {617, 765}

\bibitem[\protect\citeauthoryear{{De Luca}, {Salvaterra}, {Tiengo},
  {D'Agostino}, {Watson}, {Haberl}  \& {Wilms}}{{De Luca}
  et~al.}{2016}]{de_luca16}
{De Luca} A.,  {Salvaterra} R.,  {Tiengo} A.,  {D'Agostino} D.,  {Watson}
  M.~G.,  {Haberl} F.,   {Wilms} J.,  2016, in {Napolitano} N.~R.,  {Longo} G.,
   {Marconi} M.,  {Paolillo} M.,   {Iodice} E.,  eds,  Vol. 42, The Universe of
  Digital Sky Surveys. p.~291 (\mn@eprint {arXiv} {1503.01497}),
  \mn@doi{10.1007/978-3-319-19330-4_46}

\bibitem[\protect\citeauthoryear{{Evans} et~al.,}{{Evans}
  et~al.}{2010}]{evans10}
{Evans} I.~N.,  et~al., 2010, \mn@doi [\apjs] {10.1088/0067-0049/189/1/37},
  \href {http://ui.adsabs.harvard.edu/abs/2010ApJS..189...37E} {189, 37}

\bibitem[\protect\citeauthoryear{{Gaia Collaboration} et~al.,}{{Gaia
  Collaboration} et~al.}{2018}]{gaia18}
{Gaia Collaboration} et~al., 2018, \mn@doi [\aap]
  {10.1051/0004-6361/201833051}, \href
  {https://ui.adsabs.harvard.edu/\#abs/2018A&A...616A...1G} {616, A1}

\bibitem[\protect\citeauthoryear{{Gallo}}{{Gallo}}{2018}]{gallo18}
{Gallo} L.,  2018, in Revisiting narrow-line Seyfert 1 galaxies and their place
  in the Universe. 9-13 April 2018. Padova Botanical Garden. p.~34 (\mn@eprint
  {arXiv} {1807.09838})

\bibitem[\protect\citeauthoryear{{Glennie}, {Jonker}, {Fender}, {Nagayama}  \&
  {Pretorius}}{{Glennie} et~al.}{2015}]{glennie15}
{Glennie} A.,  {Jonker} P.~G.,  {Fender} R.~P.,  {Nagayama} T.,   {Pretorius}
  M.~L.,  2015, \mn@doi [\mnras] {10.1093/mnras/stv801}, \href
  {http://ui.adsabs.harvard.edu/abs/2015MNRAS.450.3765G} {450, 3765}

\bibitem[\protect\citeauthoryear{{Goulding} et~al.,}{{Goulding}
  et~al.}{2012}]{goulding12}
{Goulding} A.~D.,  et~al., 2012, \mn@doi [\apjs] {10.1088/0067-0049/202/1/6},
  \href {http://ui.adsabs.harvard.edu/abs/2012ApJS..202....6G} {202, 6}

\bibitem[\protect\citeauthoryear{{Grogin} et~al.,}{{Grogin}
  et~al.}{2011}]{grogin11}
{Grogin} N.~A.,  et~al., 2011, \mn@doi [\apjs] {10.1088/0067-0049/197/2/35},
  \href {http://ui.adsabs.harvard.edu/abs/2011ApJS..197...35G} {197, 35}

\bibitem[\protect\citeauthoryear{{G{\"u}del} \& {Naz{\'e}}}{{G{\"u}del} \&
  {Naz{\'e}}}{2009}]{gudel09}
{G{\"u}del} M.,  {Naz{\'e}} Y.,  2009, \mn@doi [Astronomy and Astrophysics
  Review] {10.1007/s00159-009-0022-4}, \href
  {https://ui.adsabs.harvard.edu/\#abs/2009A&ARv..17..309G} {17, 309}

\bibitem[\protect\citeauthoryear{{Haisch}, {Strong}  \& {Rodono}}{{Haisch}
  et~al.}{1991}]{haisch91}
{Haisch} B.,  {Strong} K.~T.,   {Rodono} M.,  1991, \mn@doi [\araa]
  {10.1146/annurev.aa.29.090191.001423}, \href
  {http://ui.adsabs.harvard.edu/abs/1991ARA%26A..29..275H} {29, 275}

\bibitem[\protect\citeauthoryear{{Ivezi{\'c}} et~al.,}{{Ivezi{\'c}}
  et~al.}{2019}]{ivezic19}
{Ivezi{\'c}} {\v Z}.,  et~al., 2019, \mn@doi [\apj] {10.3847/1538-4357/ab042c},
  \href {http://adsabs.harvard.edu/abs/2019ApJ...873..111I} {873, 111}

\bibitem[\protect\citeauthoryear{{Kara}, {Miller}, {Reynolds}  \& {Dai}}{{Kara}
  et~al.}{2016}]{kara16}
{Kara} E.,  {Miller} J.~M.,  {Reynolds} C.,   {Dai} L.,  2016, \mn@doi [\nat]
  {10.1038/nature18007}, \href
  {http://adsabs.harvard.edu/abs/2016Natur.535..388K} {535, 388}

\bibitem[\protect\citeauthoryear{{Kocevski} et~al.,}{{Kocevski}
  et~al.}{2018}]{kocevski18}
{Kocevski} D.~D.,  et~al., 2018, \mn@doi [\apjs] {10.3847/1538-4365/aab9b4},
  \href {http://ui.adsabs.harvard.edu/abs/2018ApJS..236...48K} {236, 48}

\bibitem[\protect\citeauthoryear{{Koekemoer} et~al.,}{{Koekemoer}
  et~al.}{2007}]{koekemoer07}
{Koekemoer} A.~M.,  et~al., 2007, \mn@doi [\apjs] {10.1086/520086}, \href
  {http://ui.adsabs.harvard.edu/abs/2007ApJS..172..196K} {172, 196}

\bibitem[\protect\citeauthoryear{{Koekemoer} et~al.,}{{Koekemoer}
  et~al.}{2011}]{koekemoer11}
{Koekemoer} A.~M.,  et~al., 2011, \mn@doi [\apjs] {10.1088/0067-0049/197/2/36},
  \href {http://ui.adsabs.harvard.edu/abs/2011ApJS..197...36K} {197, 36}

\bibitem[\protect\citeauthoryear{{Komossa}}{{Komossa}}{2015}]{komossa15}
{Komossa} S.,  2015, \mn@doi [Journal of High Energy Astrophysics]
  {10.1016/j.jheap.2015.04.006}, \href
  {https://ui.adsabs.harvard.edu/\#abs/2015JHEAp...7..148K} {7, 148}

\bibitem[\protect\citeauthoryear{Krishnamoorthy \& Thomson}{Krishnamoorthy \&
  Thomson}{2004}]{krishnamoorthy04}
Krishnamoorthy K.,  Thomson J.,  2004, \mn@doi [Journal of Statistical Planning
  and Inference] {https://doi.org/10.1016/S0378-3758(02)00408-1}, 119, 23

\bibitem[\protect\citeauthoryear{{Laigle} et~al.,}{{Laigle}
  et~al.}{2016}]{laigle16}
{Laigle} C.,  et~al., 2016, \mn@doi [\apjs] {10.3847/0067-0049/224/2/24}, \href
  {http://ui.adsabs.harvard.edu/abs/2016ApJS..224...24L} {224, 24}

\bibitem[\protect\citeauthoryear{{Laureijs} et~al.,}{{Laureijs}
  et~al.}{2011}]{euclid11}
{Laureijs} R.,  et~al., 2011, arXiv e-prints, \href
  {https://ui.adsabs.harvard.edu/\#abs/2011arXiv1110.3193L} {p.
  arXiv:1110.3193}

\bibitem[\protect\citeauthoryear{{Lawrence} et~al.,}{{Lawrence}
  et~al.}{2007}]{lawrence07}
{Lawrence} A.,  et~al., 2007, \mn@doi [\mnras]
  {10.1111/j.1365-2966.2007.12040.x}, \href
  {http://ui.adsabs.harvard.edu/abs/2007MNRAS.379.1599L} {379, 1599}

\bibitem[\protect\citeauthoryear{{Luo} et~al.,}{{Luo} et~al.}{2017}]{luo17}
{Luo} B.,  et~al., 2017, \mn@doi [\apjs] {10.3847/1538-4365/228/1/2}, \href
  {http://ui.adsabs.harvard.edu/abs/2017ApJS..228....2L} {228, 2}

\bibitem[\protect\citeauthoryear{{Marchesi} et~al.,}{{Marchesi}
  et~al.}{2016}]{marchesi16}
{Marchesi} S.,  et~al., 2016, \mn@doi [\apj] {10.3847/0004-637X/817/1/34},
  \href {http://ui.adsabs.harvard.edu/abs/2016ApJ...817...34M} {817, 34}

\bibitem[\protect\citeauthoryear{{Markowitz} et~al.,}{{Markowitz}
  et~al.}{2003a}]{markowitz03b}
{Markowitz} A.,  et~al., 2003a, \mn@doi [\apj] {10.1086/375330}, \href
  {http://ui.adsabs.harvard.edu/abs/2003ApJ...593...96M} {593, 96}

\bibitem[\protect\citeauthoryear{{Markowitz}, {Edelson}  \&
  {Vaughan}}{{Markowitz} et~al.}{2003b}]{markowitz03}
{Markowitz} A.,  {Edelson} R.,   {Vaughan} S.,  2003b, \mn@doi [\apj]
  {10.1086/379103}, \href
  {http://ui.adsabs.harvard.edu/abs/2003ApJ...598..935M} {598, 935}

\bibitem[\protect\citeauthoryear{{Nandra} et~al.,}{{Nandra}
  et~al.}{2013}]{nandra13}
{Nandra} K.,  et~al., 2013, arXiv e-prints, \href
  {https://ui.adsabs.harvard.edu/\#abs/2013arXiv1306.2307N} {p.
  arXiv:1306.2307}

\bibitem[\protect\citeauthoryear{{Nandra} et~al.,}{{Nandra}
  et~al.}{2015}]{nandra15}
{Nandra} K.,  et~al., 2015, \mn@doi [\apjs] {10.1088/0067-0049/220/1/10}, \href
  {http://ui.adsabs.harvard.edu/abs/2015ApJS..220...10N} {220, 10}

\bibitem[\protect\citeauthoryear{{Paolillo} et~al.,}{{Paolillo}
  et~al.}{2017}]{paolillo17}
{Paolillo} M.,  et~al., 2017, \mn@doi [\mnras] {10.1093/mnras/stx1761}, \href
  {http://ui.adsabs.harvard.edu/abs/2017MNRAS.471.4398P} {471, 4398}

\bibitem[\protect\citeauthoryear{{Park}, {Kashyap}, {Siemiginowska}, {van Dyk},
  {Zezas}, {Heinke}  \& {Wargelin}}{{Park} et~al.}{2006}]{park06}
{Park} T.,  {Kashyap} V.~L.,  {Siemiginowska} A.,  {van Dyk} D.~A.,  {Zezas}
  A.,  {Heinke} C.,   {Wargelin} B.~J.,  2006, \mn@doi [\apj] {10.1086/507406},
  \href {http://ui.adsabs.harvard.edu/abs/2006ApJ...652..610P} {652, 610}

\bibitem[\protect\citeauthoryear{{Pooley}, {Kumar}, {Wheeler}  \&
  {Grossan}}{{Pooley} et~al.}{2018}]{pooley18}
{Pooley} D.,  {Kumar} P.,  {Wheeler} J.~C.,   {Grossan} B.,  2018, \mn@doi
  [\apj] {10.3847/2041-8213/aac3d6}, \href
  {https://ui.adsabs.harvard.edu/\#abs/2018ApJ...859L..23P} {859, L23}

\bibitem[\protect\citeauthoryear{{Ricci} et~al.,}{{Ricci}
  et~al.}{2016}]{ricci16}
{Ricci} C.,  et~al., 2016, \mn@doi [\apj] {10.3847/0004-637X/820/1/5}, \href
  {http://ui.adsabs.harvard.edu/abs/2016ApJ...820....5R} {820, 5}

\bibitem[\protect\citeauthoryear{{Stark}, {Gammie}, {Wilson}, {Bally}, {Linke},
  {Heiles}  \& {Hurwitz}}{{Stark} et~al.}{1992}]{stark92}
{Stark} A.~A.,  {Gammie} C.~F.,  {Wilson} R.~W.,  {Bally} J.,  {Linke} R.~A.,
  {Heiles} C.,   {Hurwitz} M.,  1992, \mn@doi [\apjs] {10.1086/191645}, \href
  {http://ui.adsabs.harvard.edu/abs/1992ApJS...79...77S} {79, 77}

\bibitem[\protect\citeauthoryear{{Vito} et~al.,}{{Vito} et~al.}{2016}]{vito16}
{Vito} F.,  et~al., 2016, \mn@doi [\mnras] {10.1093/mnras/stw1998}, \href
  {http://ui.adsabs.harvard.edu/abs/2016MNRAS.463..348V} {463, 348}

\bibitem[\protect\citeauthoryear{{Wang}, {Cowie}, {Barger}, {Keenan}  \&
  {Ting}}{{Wang} et~al.}{2010}]{wang_w10}
{Wang} W.-H.,  {Cowie} L.~L.,  {Barger} A.~J.,  {Keenan} R.~C.,   {Ting} H.-C.,
   2010, \mn@doi [\apjs] {10.1088/0067-0049/187/1/251}, \href
  {http://ads.bao.ac.cn/abs/2010ApJS..187..251W} {187, 251}

\bibitem[\protect\citeauthoryear{{Xue}}{{Xue}}{2017}]{xue17}
{Xue} Y.~Q.,  2017, \mn@doi [\nar] {10.1016/j.newar.2017.09.002}, \href
  {http://ui.adsabs.harvard.edu/abs/2017NewAR..79...59X} {79, 59}

\bibitem[\protect\citeauthoryear{{Xue}, {Luo}, {Brandt}, {Alexander}, {Bauer},
  {Lehmer}  \& {Yang}}{{Xue} et~al.}{2016}]{xue16}
{Xue} Y.~Q.,  {Luo} B.,  {Brandt} W.~N.,  {Alexander} D.~M.,  {Bauer} F.~E.,
  {Lehmer} B.~D.,   {Yang} G.,  2016, \mn@doi [\apjs]
  {10.3847/0067-0049/224/2/15}, \href
  {http://ui.adsabs.harvard.edu/abs/2016ApJS..224...15X} {224, 15}

\bibitem[\protect\citeauthoryear{{Xue} et~al.,}{{Xue} et~al.}{2019}]{xue19}
{Xue} Y.~Q.,  et~al., 2019, \mn@doi [\nat] {10.1038/s41586-019-1079-5}, \href
  {https://ui.adsabs.harvard.edu/abs/2019Natur.568..198X} {568, 198}

\bibitem[\protect\citeauthoryear{{Yang} et~al.,}{{Yang} et~al.}{2016}]{yang16}
{Yang} G.,  et~al., 2016, \mn@doi [\apj] {10.3847/0004-637X/831/2/145}, \href
  {http://ui.adsabs.harvard.edu/abs/2016ApJ...831..145Y} {831, 145}

\bibitem[\protect\citeauthoryear{{York} et~al.,}{{York} et~al.}{2000}]{york00}
{York} D.~G.,  et~al., 2000, \mn@doi [\aj] {10.1086/301513}, \href
  {http://ui.adsabs.harvard.edu/abs/2000AJ....120.1579Y} {120, 1579}

\bibitem[\protect\citeauthoryear{{Yuan} et~al.,}{{Yuan} et~al.}{2018}]{yuan18}
{Yuan} W.,  et~al., 2018, \mn@doi [Scientia Sinica Physica, Mechanica and
  Astronomica] {10.1360/SSPMA2017-00297}, \href
  {http://ui.adsabs.harvard.edu/abs/2018SSPMA..48c9502Y} {48, 039502}

\bibitem[\protect\citeauthoryear{{Zheng} et~al.,}{{Zheng}
  et~al.}{2017}]{zheng17}
{Zheng} X.~C.,  et~al., 2017, \mn@doi [\apj] {10.3847/1538-4357/aa9378}, \href
  {http://ui.adsabs.harvard.edu/abs/2017ApJ...849..127Z} {849, 127}

\bibitem[\protect\citeauthoryear{{van der Klis}}{{van der
  Klis}}{1989}]{van_der_klis89}
{van der Klis} M.,  1989, \mn@doi [Annual Review of Astronomy and Astrophysics]
  {10.1146/annurev.aa.27.090189.002505}, \href
  {https://ui.adsabs.harvard.edu/#abs/1989ARA&A..27..517V} {27, 517}

\makeatother
\end{thebibliography}



\appendix

\section{Efficiency of the Selection Algorithm for Different Transient Models}\label{sec:oth_lc}
{The simulations in \S\ref{sec:eff} are based on a fiducial transient model similar to 
the \cdfs\ XTs. 
The employment of this fiducial model is driven by the main aim of this paper, i.e. 
investigating \cdfs\ XT-like transients in \chandra\ surveys. 
However, our algorithm might also be able to identify other types of transients as a ``bonus''.
In this Appendix, we perform Monte~Carlo simulations for some other transient models
as examples, although pursuing them is not the main focus of our paper.
}

\begin{figure}
\includegraphics[width=\linewidth]{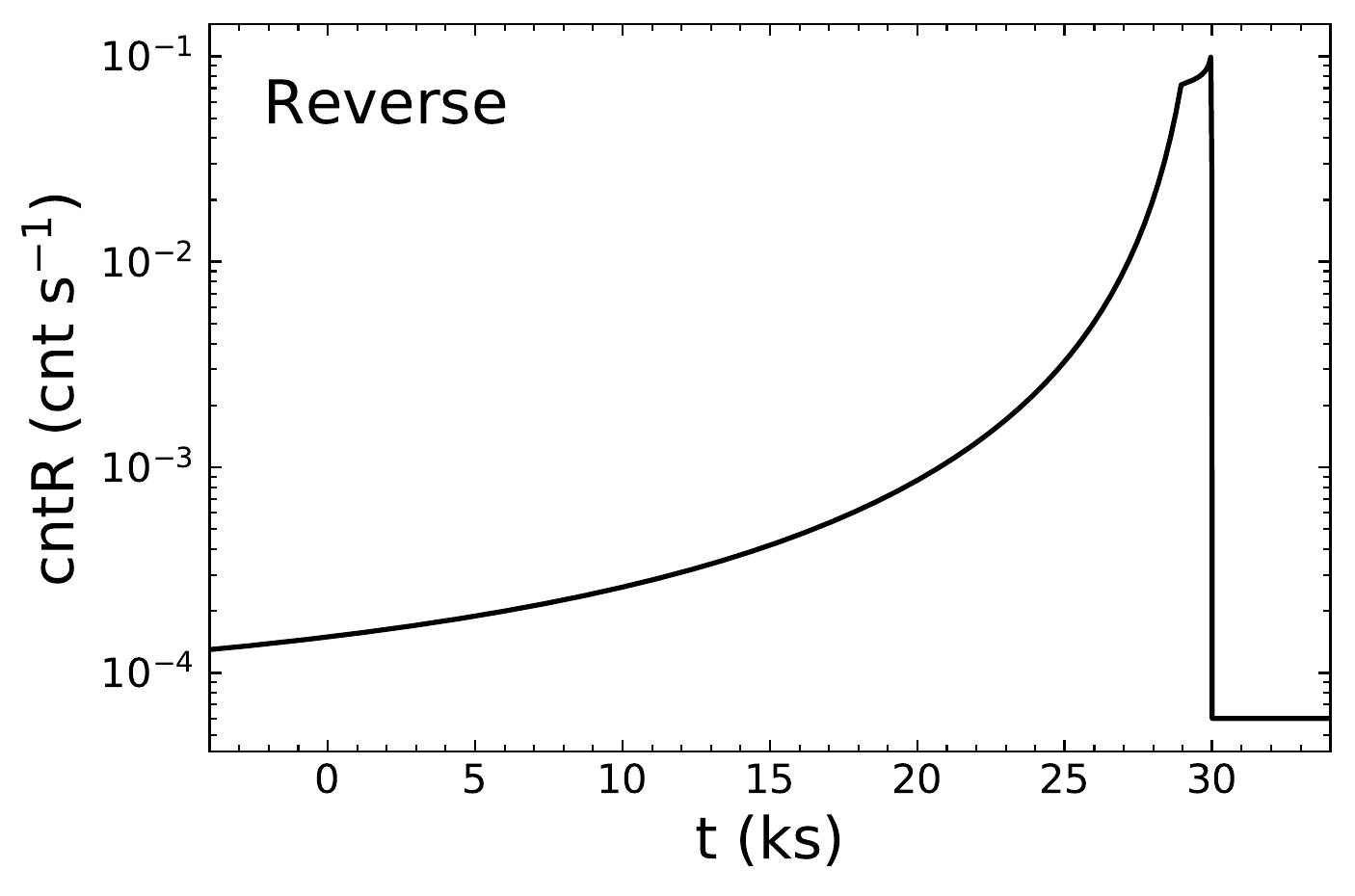}
\includegraphics[width=\linewidth]{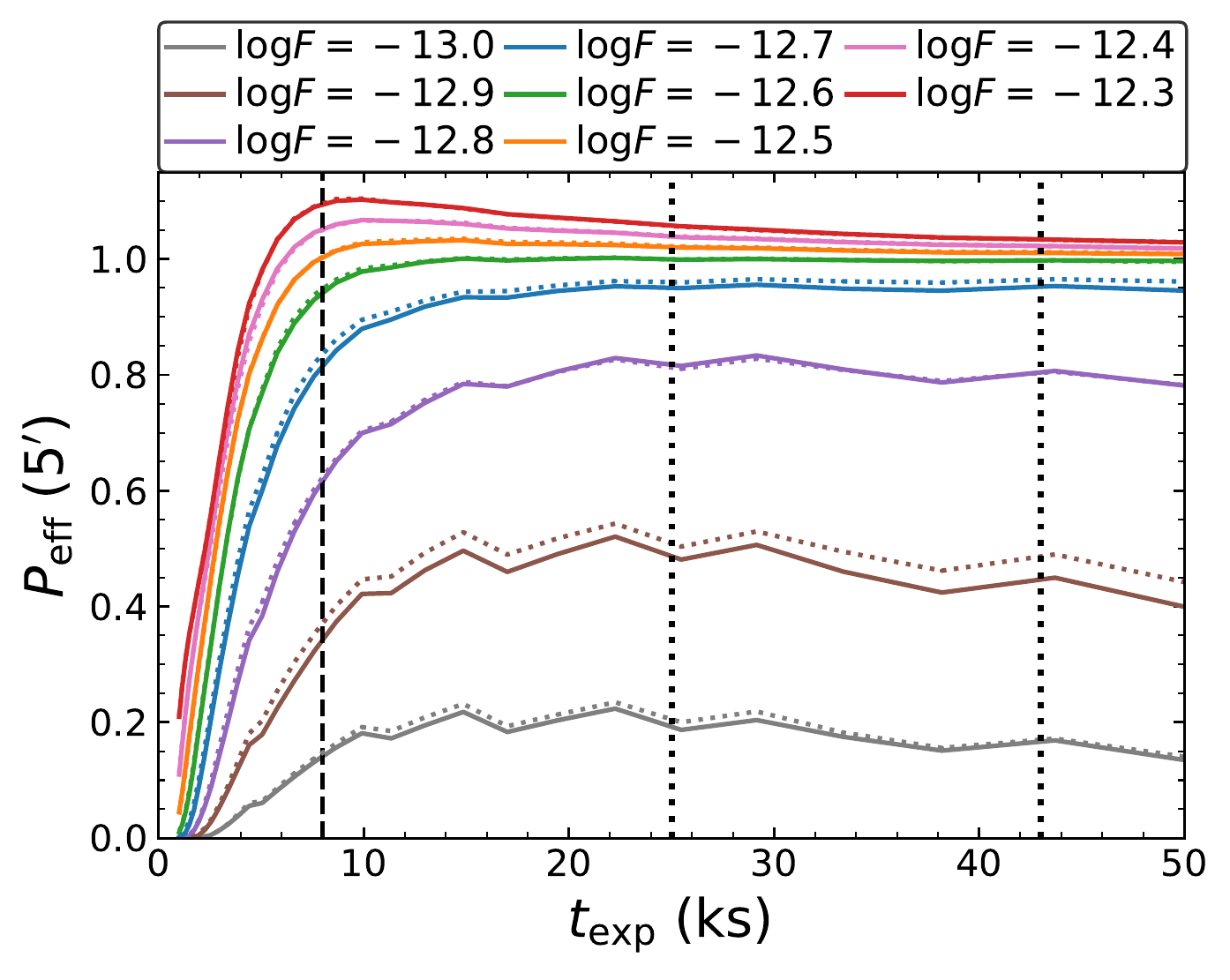}
\caption{{Top: Same format as Fig.~\ref{fig:sim_lc_model} but for a time-reversed 
fiducial model.
Bottom: Same format as Fig.~\ref{fig:Peff} but for the time-reversed model in the top 
panel.
For comparison, the ${P_{\rm eff}}$ for the fiducial model are also plotted as 
the dotted curves.
}}
\label{fig:sim_reverse}
\end{figure}

{
The first additional transient model we test is a ``time-reversed'' version of our fiducial model
(see Fig.~\ref{fig:sim_reverse} top for the light curve). 
The fiducial light curve has features of a fast rise and slow decline (Fig.~\ref{fig:sim_lc_model}), 
and thus the reverse has features of a slow rise and fast decline.
The reversed model has the same flux-to-counts conversion factor and timescale as 
the fiducial model.
We then apply the simulation process in \S\ref{sec:gau} to the reversed model, and 
show ${P_{\rm eff}}$ as a function of ${t_{\rm exp}}$ in 
Fig.~\ref{fig:sim_reverse} (bottom).
The simulation results are similar to those of the fiducial model, e.g.
for ${\log F_{\rm peak} \gtrsim -12.6}$~(cgs, corresponding to $\approx$~30 counts), 
${P_{\rm eff}}$ is $\approx$~1 for a wide range of 
${t_{\rm exp}=8\text{--}50}$~ks.
We have also tested some other light curves with different shapes but similar timescales, 
and found the sensitivity of our algorithm for these models is similar to the fiducial 
model.
These results indicate that our algorithm is also capable of detecting different types of 
transients with timescales similar to that of the \cdfs\ XTs.
}

\begin{figure}
\includegraphics[width=\linewidth]{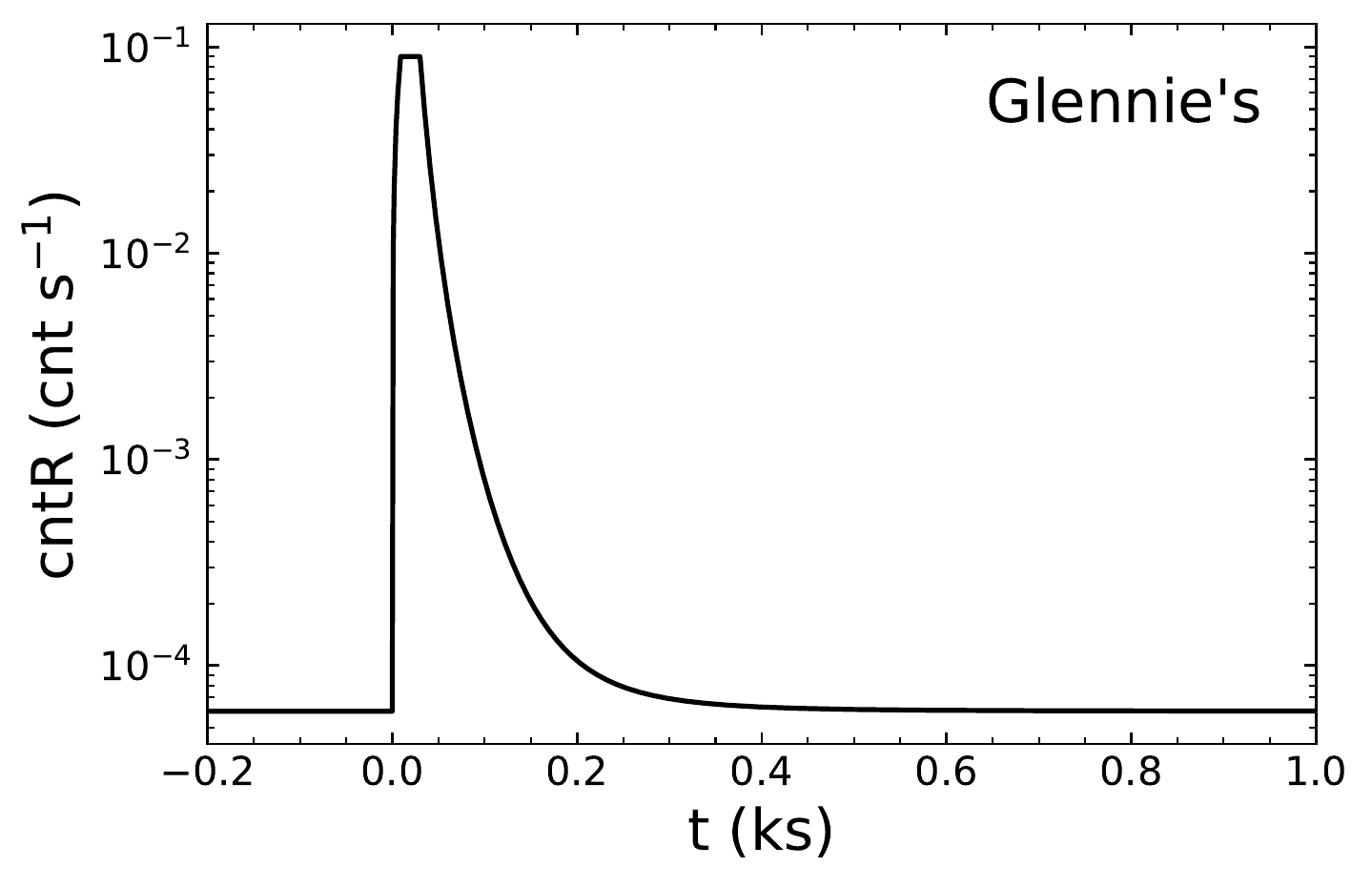}
\includegraphics[width=\linewidth]{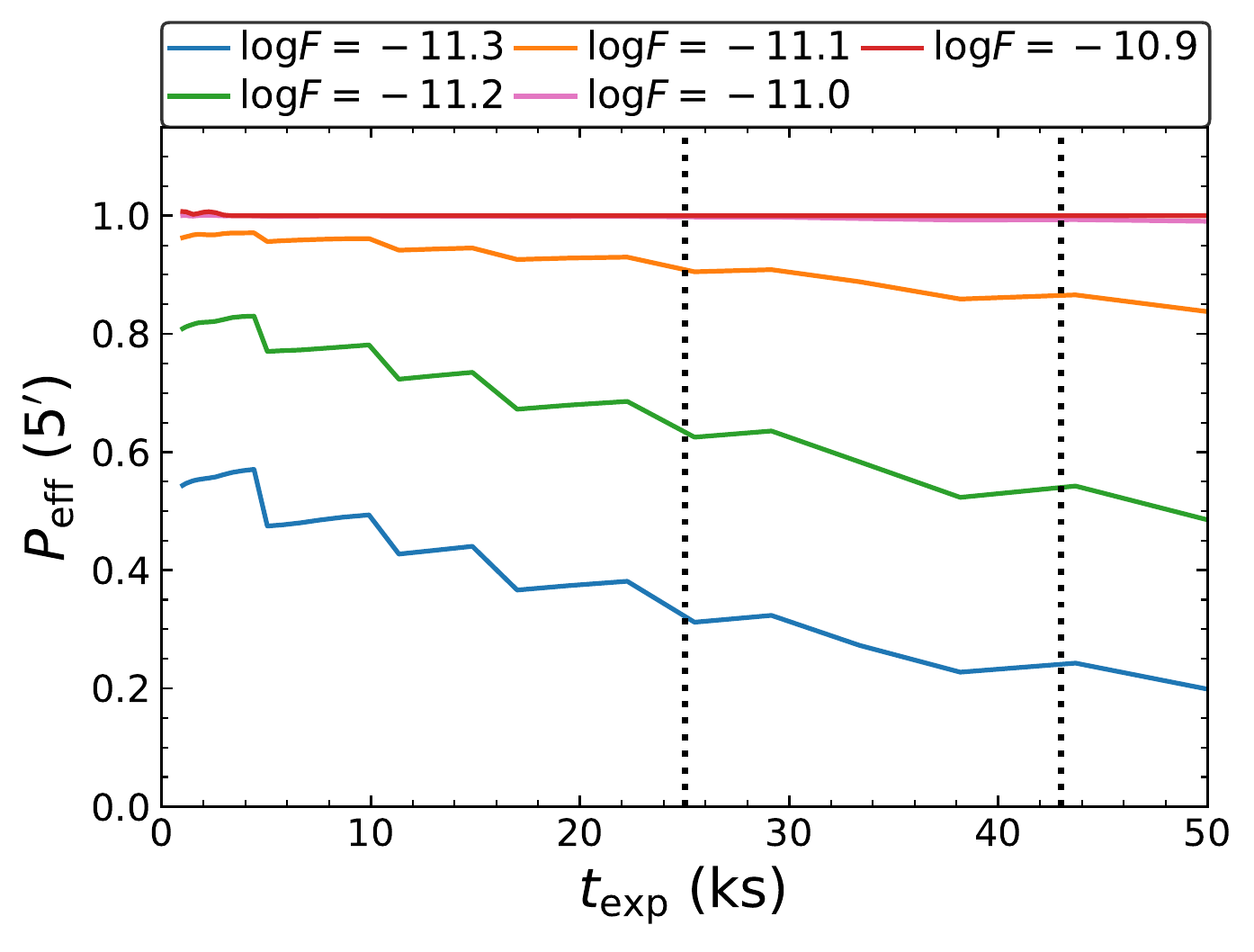}
\caption{{Top: Same format as Fig.~\ref{fig:sim_lc_model} but for an ultrafast transient 
model similar to Glennie's event.
Bottom: Same format as Fig.~\ref{fig:Peff} but for the ultrafast model in the top panel.
}}
\label{fig:sim_fast}
\end{figure}

{
Another additional transient model we test is based on the ultrafast transient discovered 
by \cite{glennie15}.
This transient lasts only ${\approx 100}$~s with ${\log F_{\rm peak}= -9.9}$ 
(cgs), and has a spectral shape of ${\Gamma \approx 1.4}$.
The nature of the transient remains unknown, as the optical/NIR counterpart has not been
found due to the lack of deep multiwavelength data (\S\ref{sec:intro}).
The light curve can also be approximated by the general formula in Eq.~\ref{eq:lc}, with
${(t_1, t_2, \alpha_1, \alpha_2) \approx (10\ \mathrm{s}, 30\ \mathrm{s}, 0, -4)}$.
This light-curve model is displayed in Fig.~\ref{fig:sim_fast} (top).
The flux-to-counts conversion factor (Eq.~\ref{eq:Nnet}) for this model is 
${3.2\times 10^{12}}$, and the ${T_{\rm 90}}$ is 47~s.
Here, the conversion factor is much lower than that in Eq.~\ref{eq:Nnet}. 
This is mainly because the ultrafast model has a timescale much shorter than the fiducial model, 
and to reach similar counts, the former must have a much higher peak flux than the latter. 
We show the simulation results in Fig.~\ref{fig:sim_fast} (bottom).
Unlike ${P_{\rm eff}}$ in Fig.~\ref{fig:sim_lc_model}, ${P_{\rm eff}}$ in 
Fig.~\ref{fig:sim_fast} does not drop below ${t_{\rm exp} \approx 8}$~ks.
The drop in Fig.~\ref{fig:sim_lc_model} is because, when the exposure time becomes 
shorter than the transient timescale, the observed light curve will be similar to a normal 
variable source (\S\ref{sec:sim_res}).
However, this is not the case in Fig.~\ref{fig:sim_fast}, since the ultrafast-transient timescale 
(${T_{\rm 90}=47}$~s) is even shorter than our shortest exposures 
(${3}$~ks).
In Fig.~\ref{fig:sim_fast}, for ${\log F_{\rm peak}\lesssim -11.1}$, ${P_{\rm eff}}$ 
declines toward high $t_{\rm exp}$ due to high background levels for long exposures 
(\S\ref{sec:alg}).
For ${\log F_{\rm peak}\gtrsim -11.0}$ (corresponding to $\approx$~30 counts), 
${P_{\rm eff}}$ is stable for different 
${t_{\rm exp}}$, because the \xray\ signal is dominated by the source rather than the 
background.
}

\begin{figure}
\includegraphics[width=\linewidth]{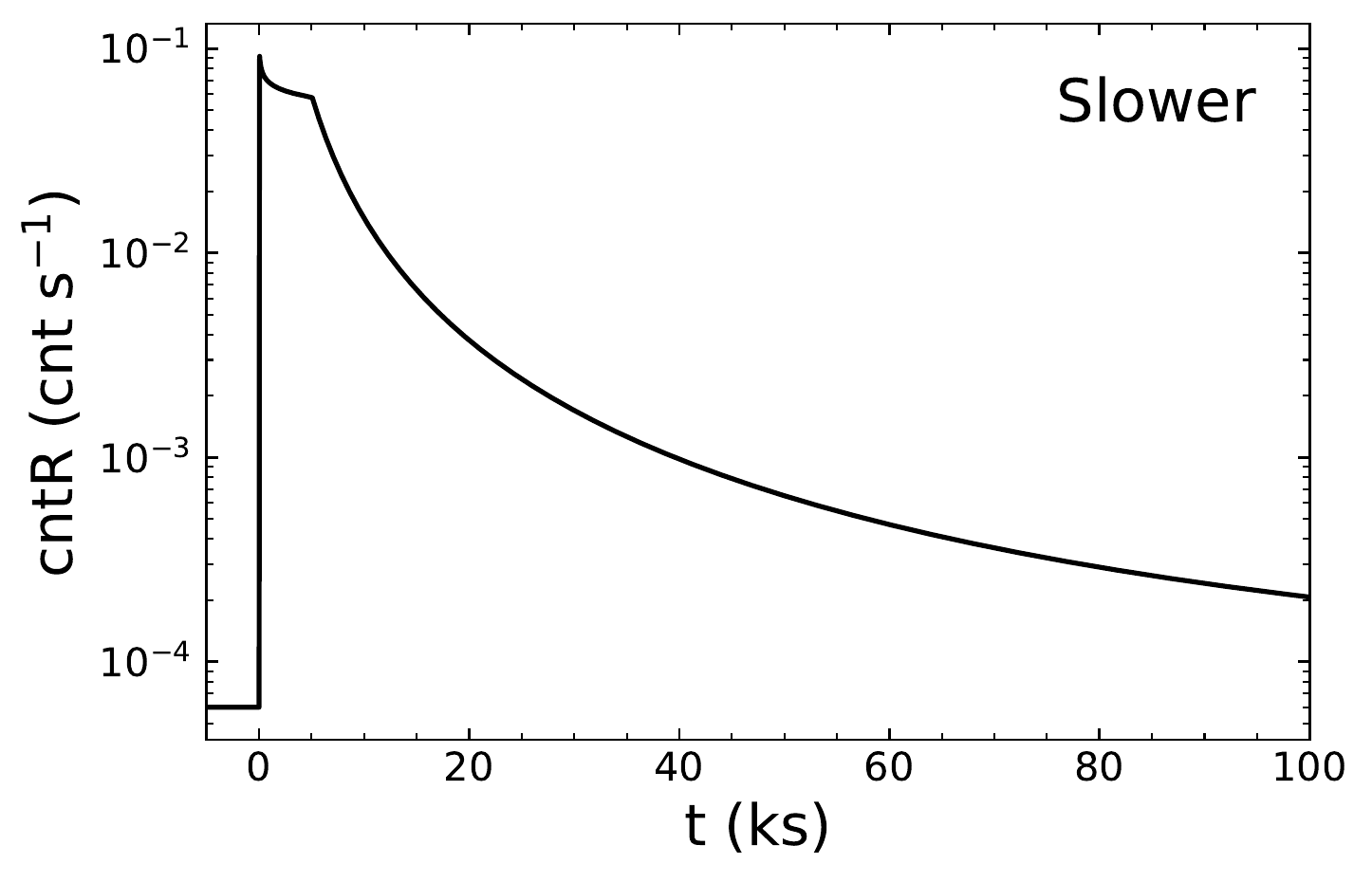}
\includegraphics[width=\linewidth]{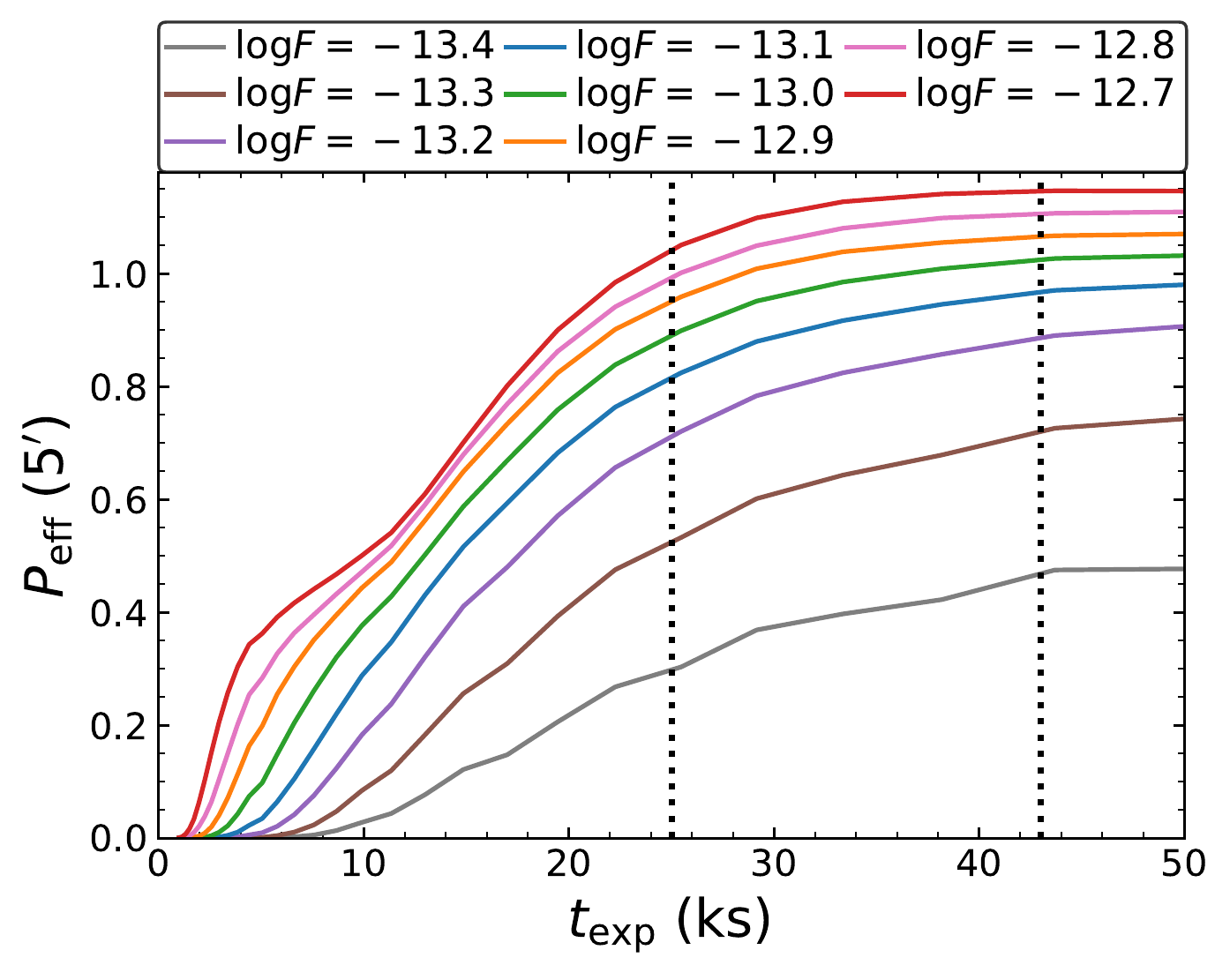}
\caption{{Top: Same format as Fig.~\ref{fig:sim_lc_model} but for a slower 
transient model with ${t_2=t_1 + 5}$~ks (Eq.~\ref{eq:lc}).
Bottom: Same format as Fig.~\ref{fig:Peff} but for the slower model in the top panel.
}}
\label{fig:sim_slow}
\end{figure}

{
Glennie's model tested above is faster than our fiducial model. 
Now, we test another transient model which is ``slower'' than the fiducial model.  
We extend the plateau phase of the fiducial model (\S\ref{sec:conf}) by setting 
${t_2=t_1 + 5}$~ks (Eq.~\ref{eq:lc}), while keeping the other parameters the same.
The light curve of this slower model is displayed in Fig.~\ref{fig:sim_slow} (top).
The flux-to-counts conversion factor (Eq.~\ref{eq:Nnet}) for this model is 
${6.0\times 10^{14}}$, and the $T_{\rm 90}$ is 16.7~ks.
The simulation results are displayed in Fig.~\ref{fig:sim_slow} (bottom).
For a given ${F_{\rm peak}}$, ${P_{\rm eff}}$ rises toward high 
${t_{\rm exp}}$ for the aforementioned reason, i.e. our algorithm may not
be able to differentiate the transient from normal variable sources when 
${t_{\rm exp} \lesssim}$ transient timescale.
Since most ($\approx$~90\%; \S\ref{sec:select}) of our exposures are longer than 
the timescale of the slower model, our algorithm is largely capable of detecting such 
transients in our data.
}

{
In summary, our algorithm can detect different types of transients with timescales similar 
to or below that of the \cdfs\ XTs, as long as $\gtrsim$~30 counts are available.
For transients with longer timescales, only observations with 
${t_{\rm exp}} \gtrsim$ transient timescale can have high detection probabilities.
Since 80\% of our exposures are longer than 25~ks (\S\ref{sec:select}), we are potentially 
able to detect transients with timescales shorter than $\approx$~25~ks in our data.
}

\section{Efficiency of Selection Algorithm at Different Off-Axis Angles}\label{sec:oth_ang}
{The simulations in \S\ref{sec:eff} are performed for a typical off-axis angle of 
${5'}$.
In this Appendix, we perform simulations at off-axis angles of ${0.5\arcmin}$ 
(nearly on-axis) and ${8\arcmin}$ (the maximum value accepted by our algorithm; 
\S\ref{sec:alg}).
In our simulation configurations (\S\ref{sec:conf}), there are two parameters dependent 
on off-axis angle, i.e. flux-to-counts conversion factor and background noise. 
The conversion factors (Eq.~\ref{eq:Nnet}) are ${\approx 1.7\times 10^{14}}$ and 
${\approx 1.5\times 10^{14}}$ (cgs) at ${0.5'}$ and ${8'}$, 
respectively;
the typical background count rates are ${5.9\times 10^{-6}}$~cnt~s$^{-1}$ 
to ${2.5\times 10^{-4}}$~cnt~s$^{-1}$.
}

\begin{figure}
\includegraphics[width=\linewidth]{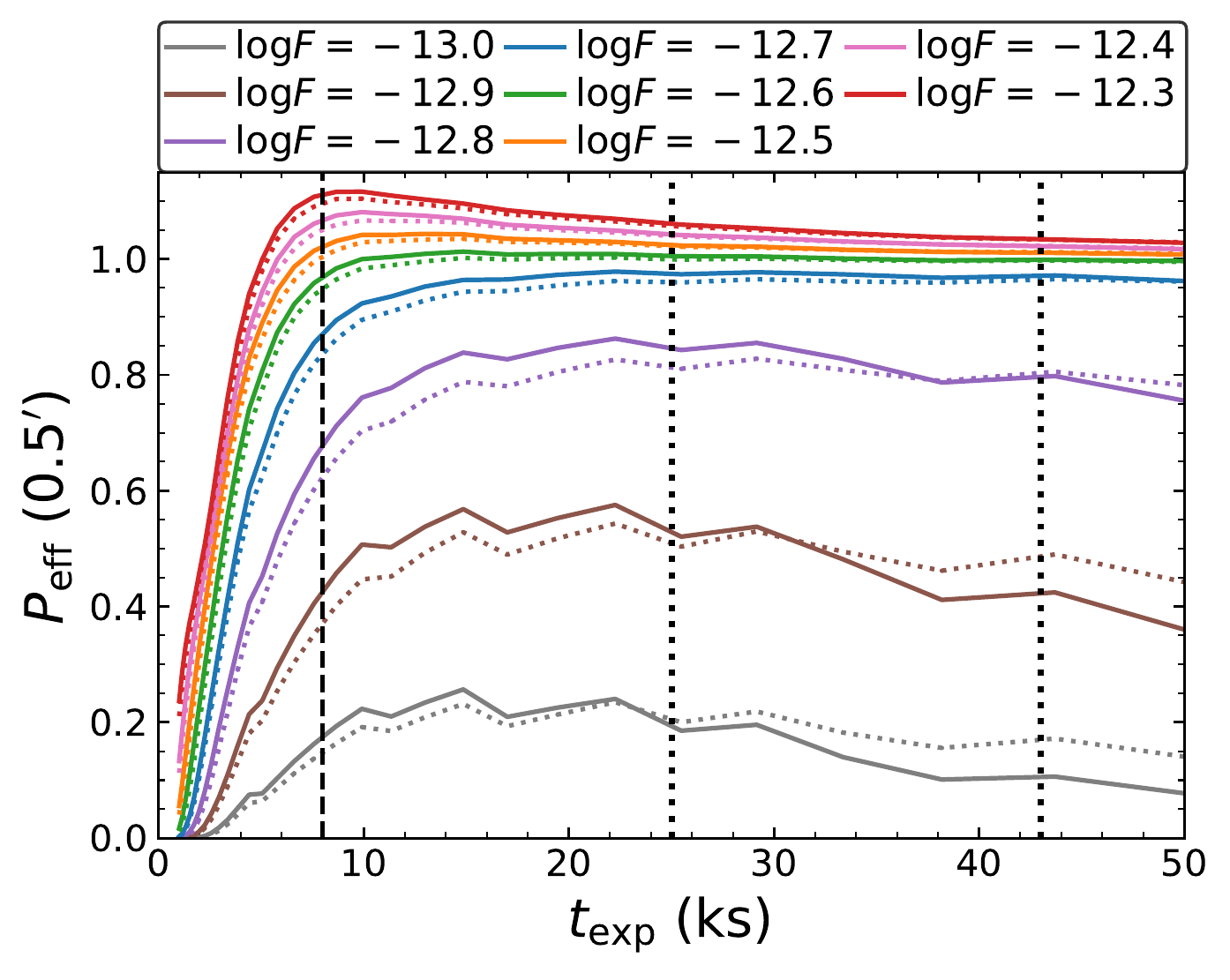}
\includegraphics[width=\linewidth]{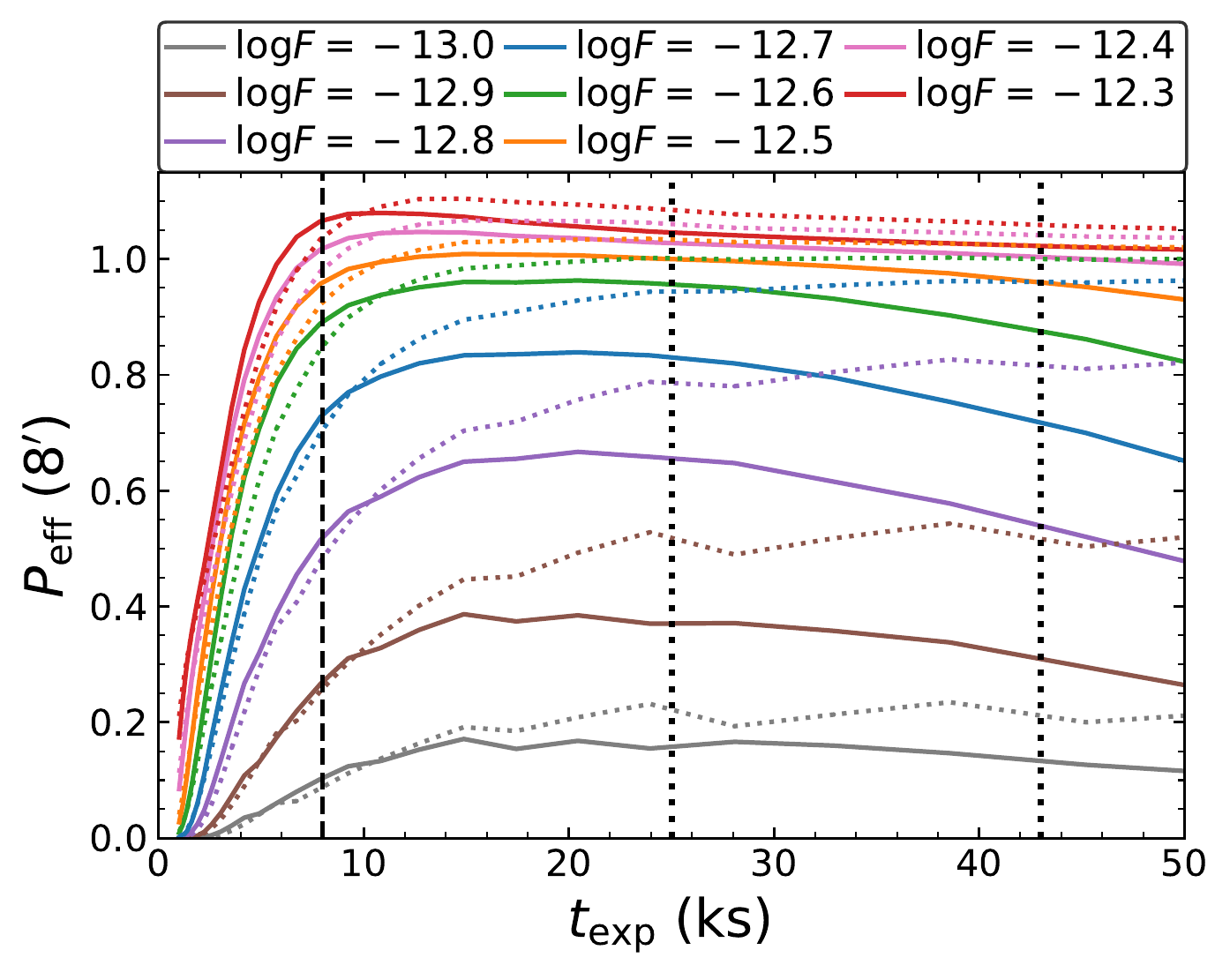}
\caption{{Same format as Fig.~\ref{fig:Peff} but for off-axis angles of ${0.5'}$ (top)
and ${8'}$ (bottom). 
For comparison, the ${P_{\rm eff}}$ for ${5'}$ are also plotted as the dotted 
curves.
}}
\label{fig:Peff_oth_ang}
\end{figure}

{We perform our simulations under these new configurations, and display the results
in Fig.~\ref{fig:Peff_oth_ang}.
Similar to the results for ${5'}$, ${P_{\rm eff}}$ drops significantly below 
${t_{\rm exp}\approx 8}$~ks, because short exposures cannot differentiate between variable 
sources and transients (\S\ref{sec:sim_res}).
Compared to that for ${5'}$, ${P_{\rm eff}}$ for 
${0.5'}$ (${8'}$) generally increases (decreases) 
for a given ${F_{\rm peak}}$ and ${t_{\rm exp}}$, as expected.
As a consequence, the peak-flux limit could change if using the simulation configurations 
for ${0.5'}$ (${8'}$).
In \S\ref{sec:sim_res}, we choose the peak-flux limit as the minimum flux above which 
${P_{\rm eff}}$ is ${\approx 1}$ for 
${t_{\rm exp} = 8\text{--}50}$~ks.
Applying the same criteria to Fig.~\ref{fig:Peff_oth_ang}, the peak-flux limits are
${\log F_{\rm peak} \approx -12.7}$ (${\approx -12.5}$) for ${0.5'}$ \
(${8'}$).
}

%


\bsp	
\label{lastpage}
\end{document}